Serguei A. Mokhov
Mashrur Mia
Petr Solodov
Kai Zhao
Jihed Halimi

Travel Search Engine

# A UI Design Case Study and a Prototype of a Travel Search Engine

# Project Report

*Concordia University, Montreal, Canada, 2003*

# Project Report

*Travel Search Engine*

**Contents:**

## Flight & Shopping Cart Scenarios

The following figures illustrate the prototype screenshots of the UI pages used to reserve flights for both general and corporate users. The goal is to allow the users to select one of the three common trip types: One Way, Round Trip and Multi City.

Many design guidelines were applied to the flight pages after using the mapping technique to generate them from the improved task models. Please note the use of colors, frames and proximity to highlight grouping as well as the use of interaction styles that would provide a better error prevention mechanism (e.g. country and city combo boxes). Due to the magnitude of the system, we made sure that the design of the pages within an individual task is consistent; however we did not propagate this model to the entire system.

These scenarios also demonstrate the shopping cart and its usage by the travel search engine as a central repository of the purchased items and how it offers a convenient shopping experience.

# Scenario1: General User Makes a Reservation

Mr. John Doe plans to travel from Montreal to New York. He heard about the Travel Search Engine and plans to use it to organize his trip. Let us follow Mr. Doe as he uses the system to accomplish his task (i.e. perform a round trip flight reservation from Montreal to New York).

### The Travel Search Engine Main Page

Since the main page will be the subject of a later section we will only look at the flight quick access section.

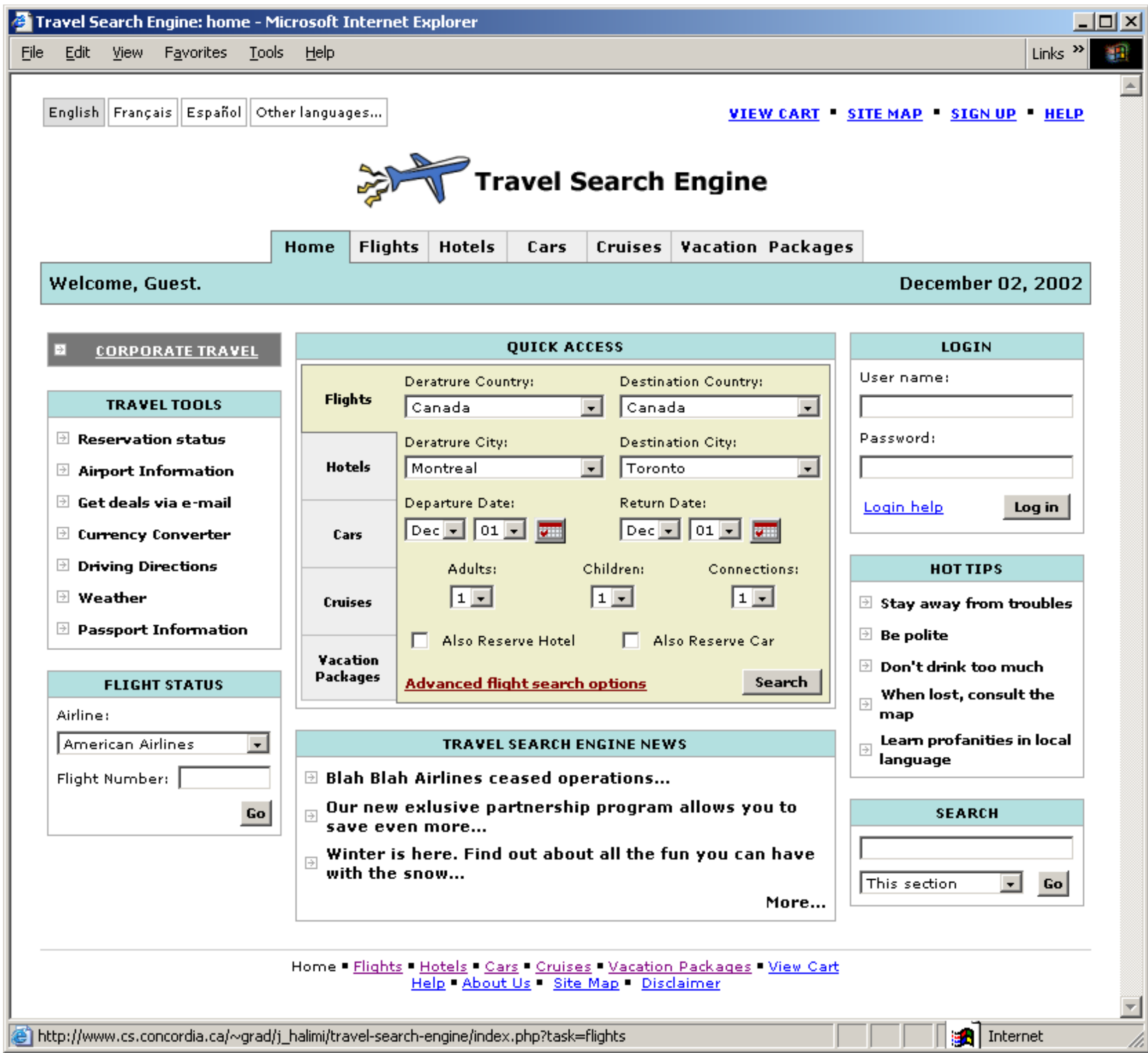

.

# The Advanced Flight Reservation Screen

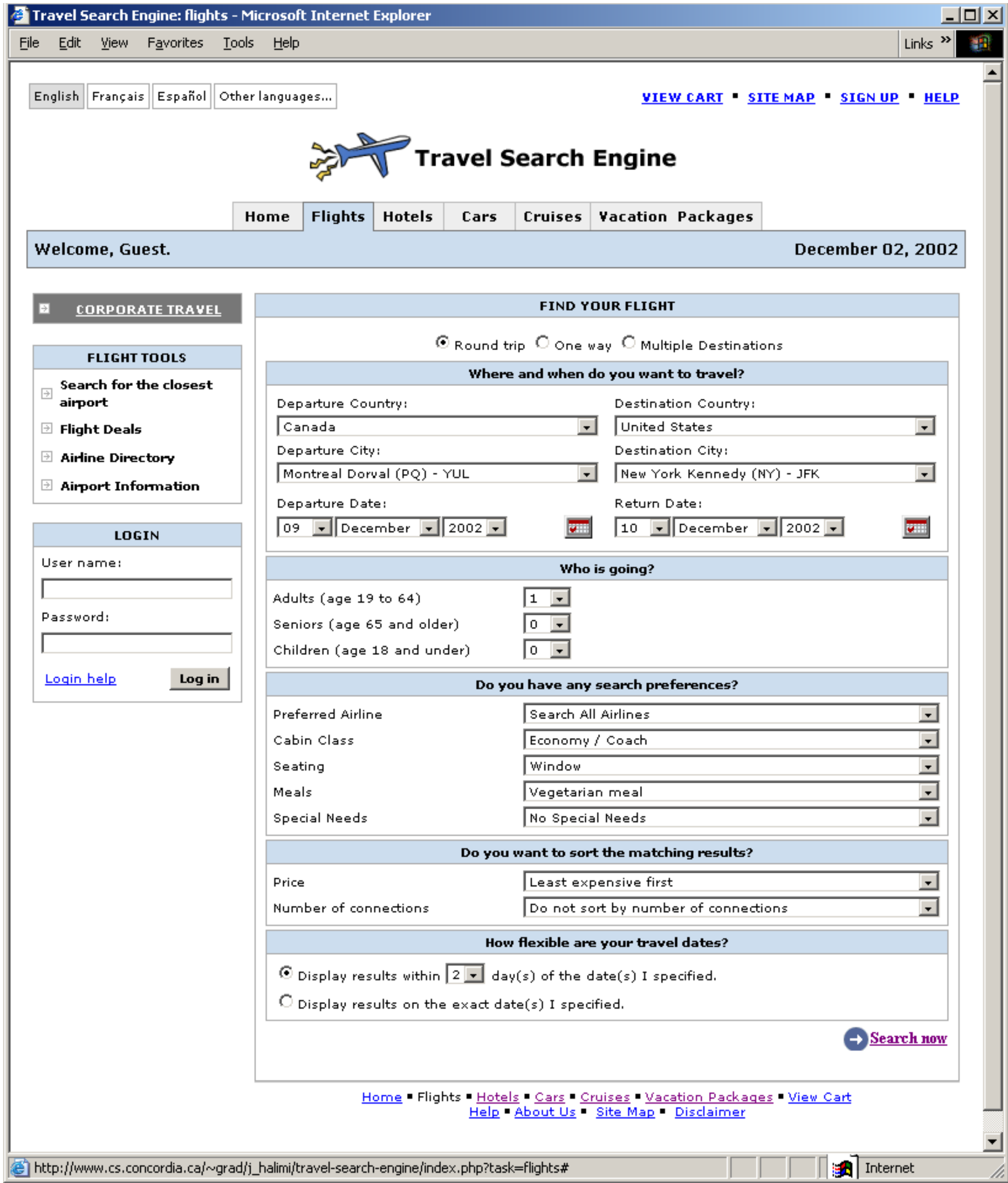

## Specifying a Multi City Trip

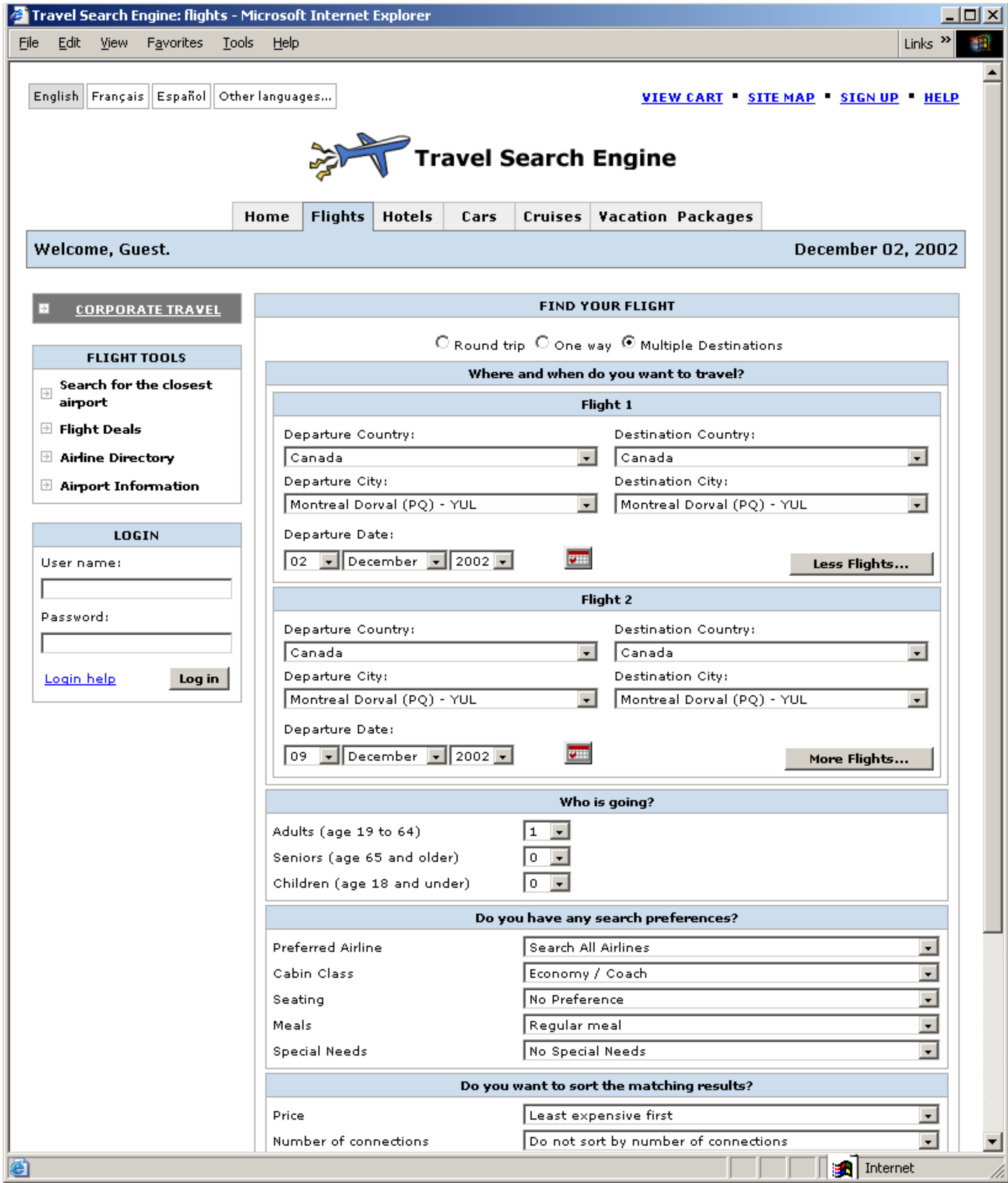

# Browsing & Selecting the Search Results

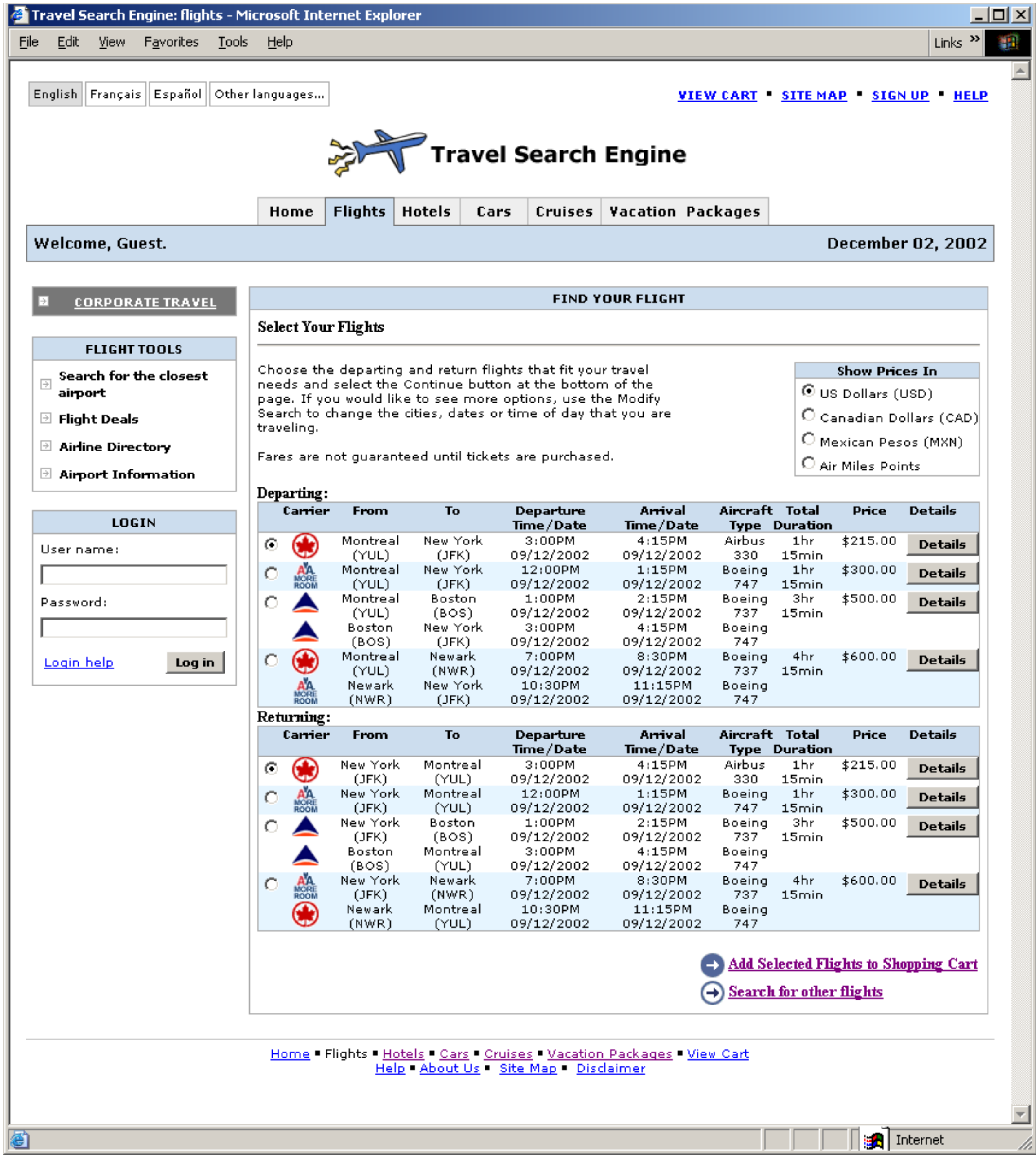

## Viewing the Details About a Particular Non-stop / Connection Flight

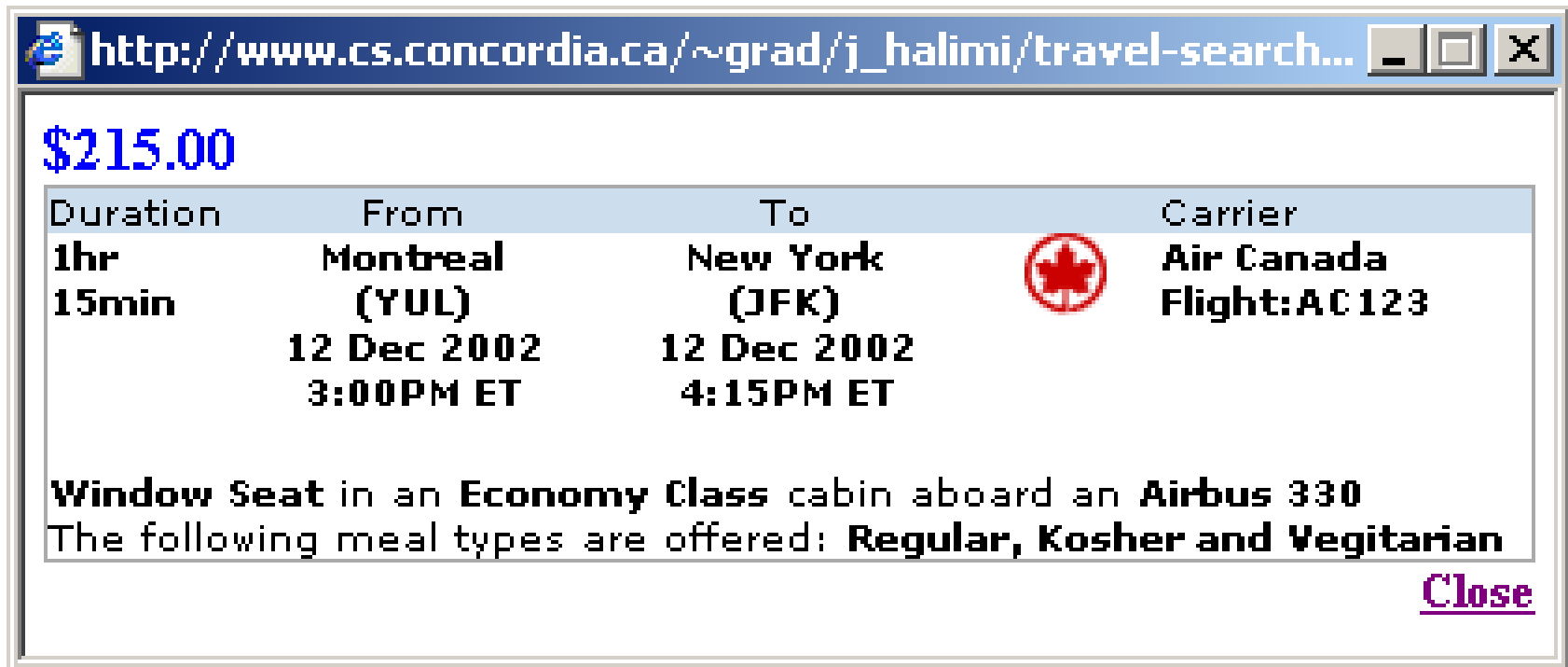

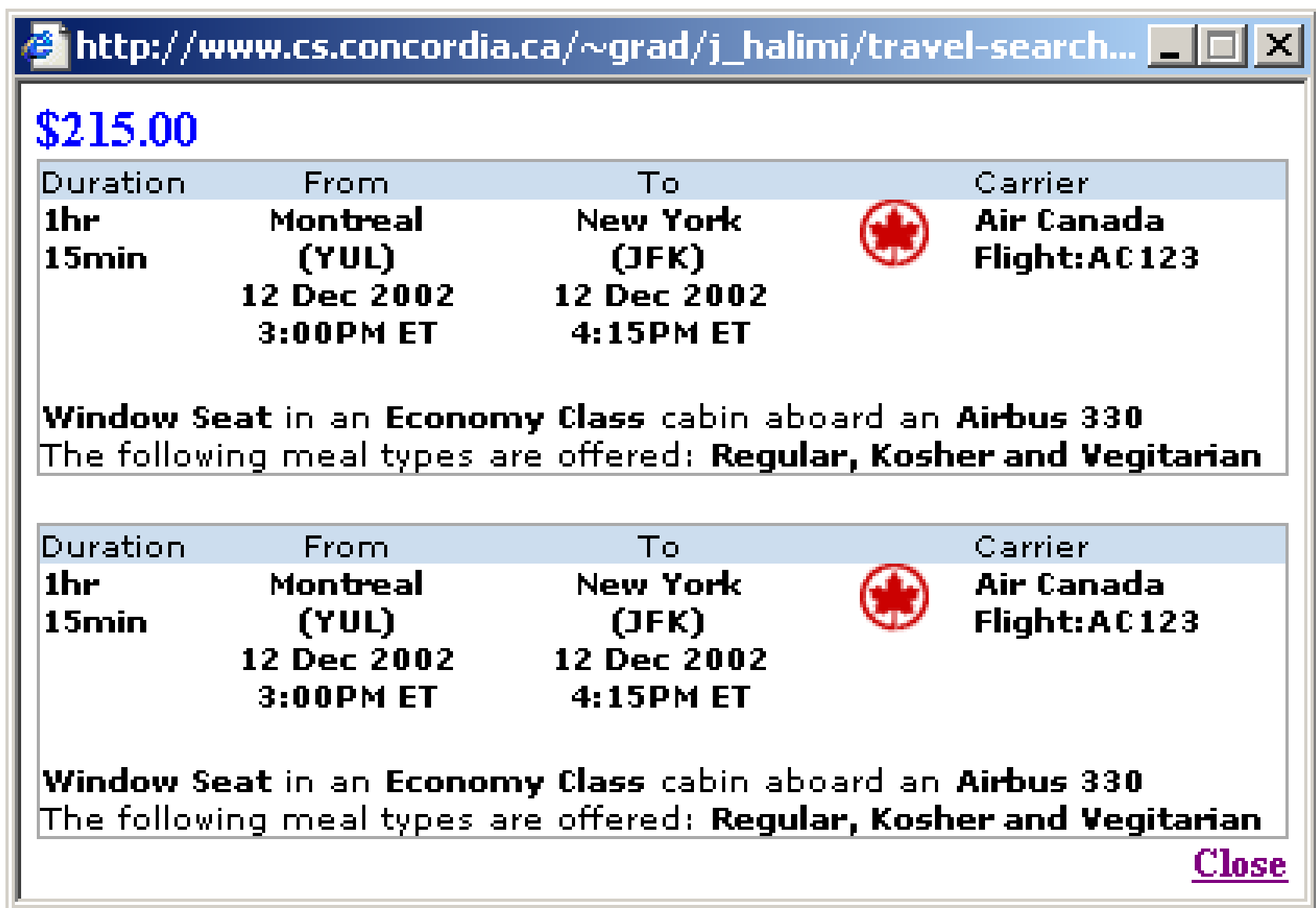

## Adding the Selected Flights to the Shopping Cart

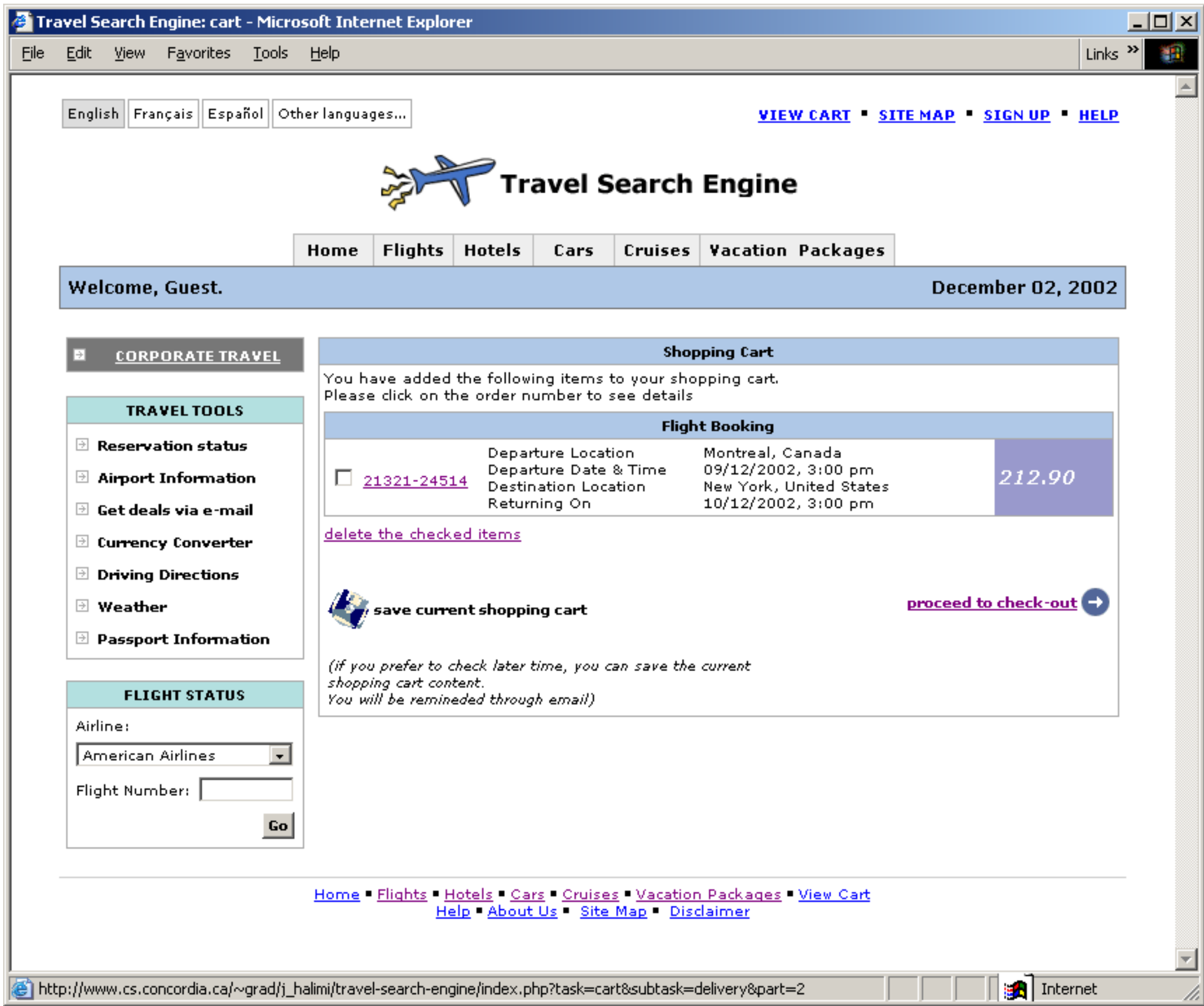

## Checking Out the Newly Added Item from the Cart

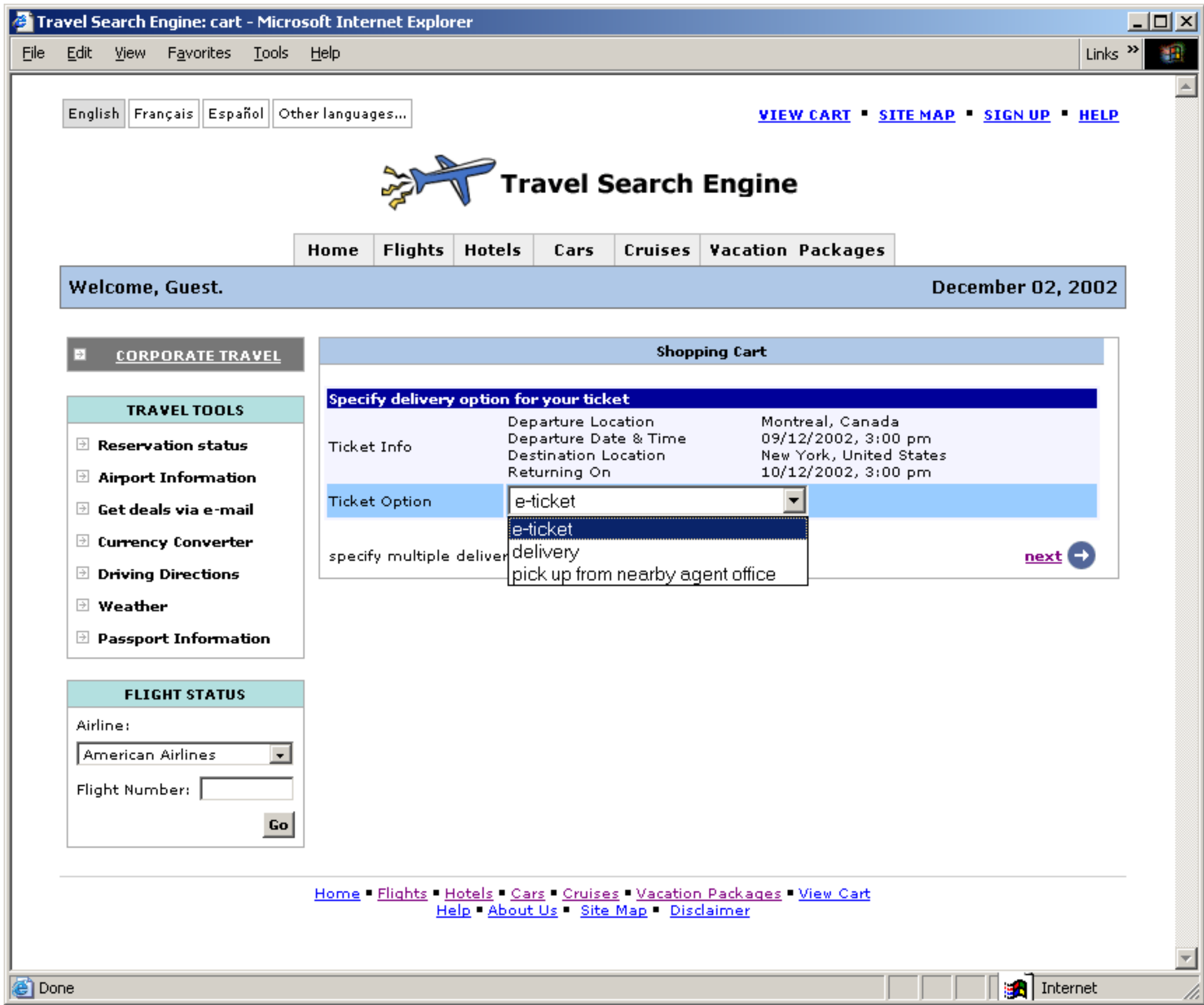

# Viewing the Invoice via the Shopping Cart

.

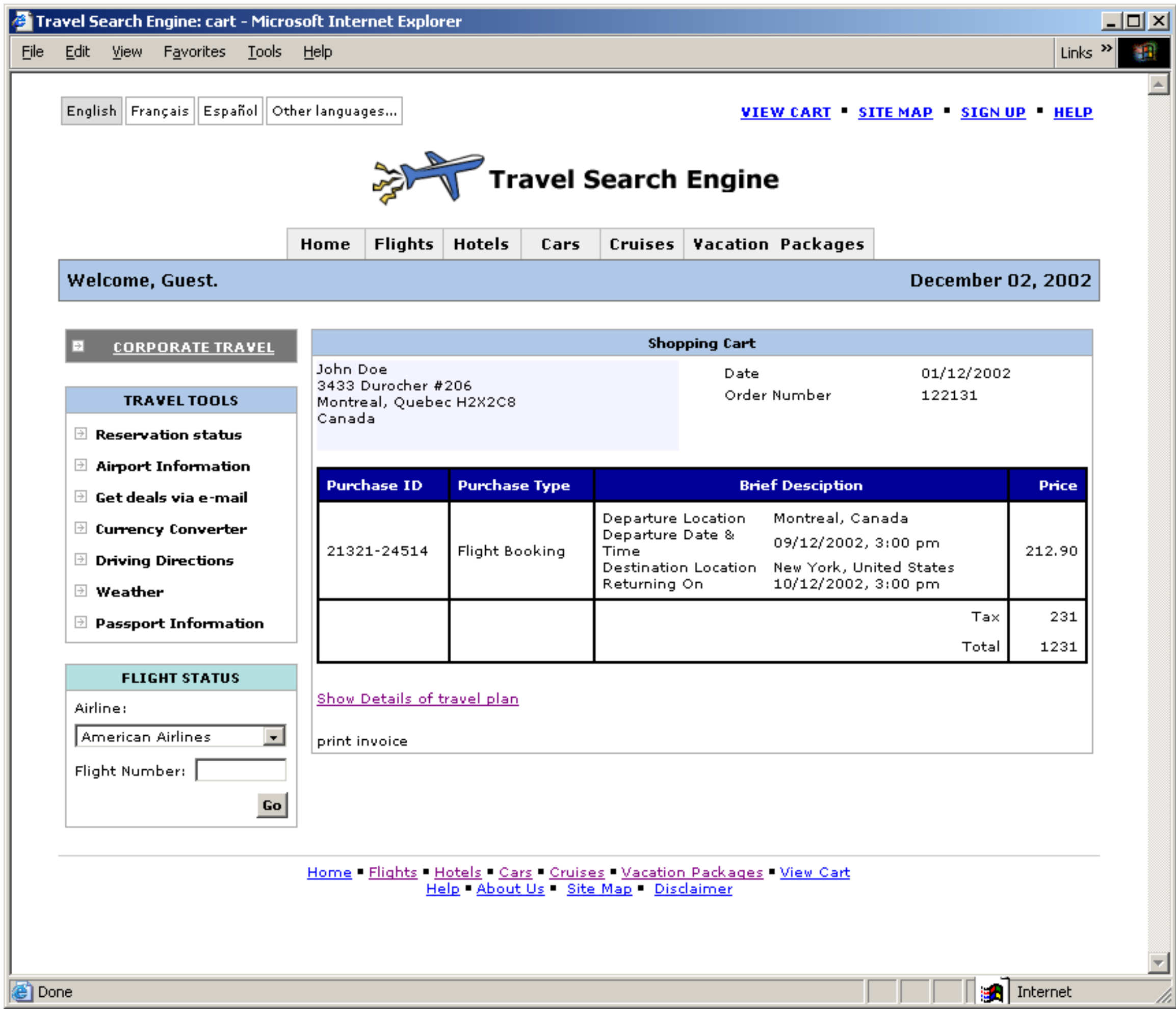

# Viewing the Generated Itinerary

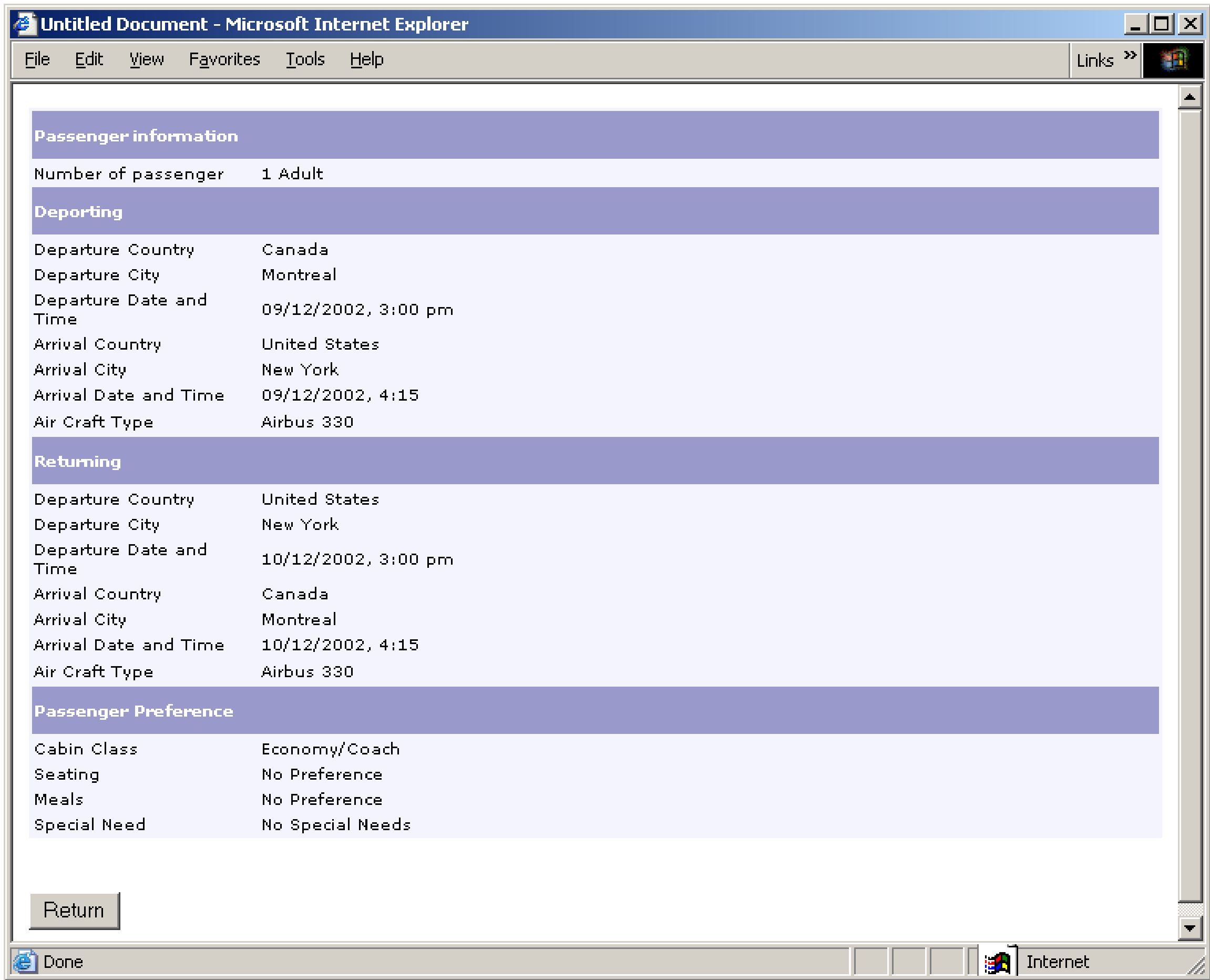

# Scenario2: Bulk Corporate Reservation

The objective is to book a set of flight seats for a group of corporate users having different travel preferences. The users will travel from Montreal to New York.

### The Corporate Main Page

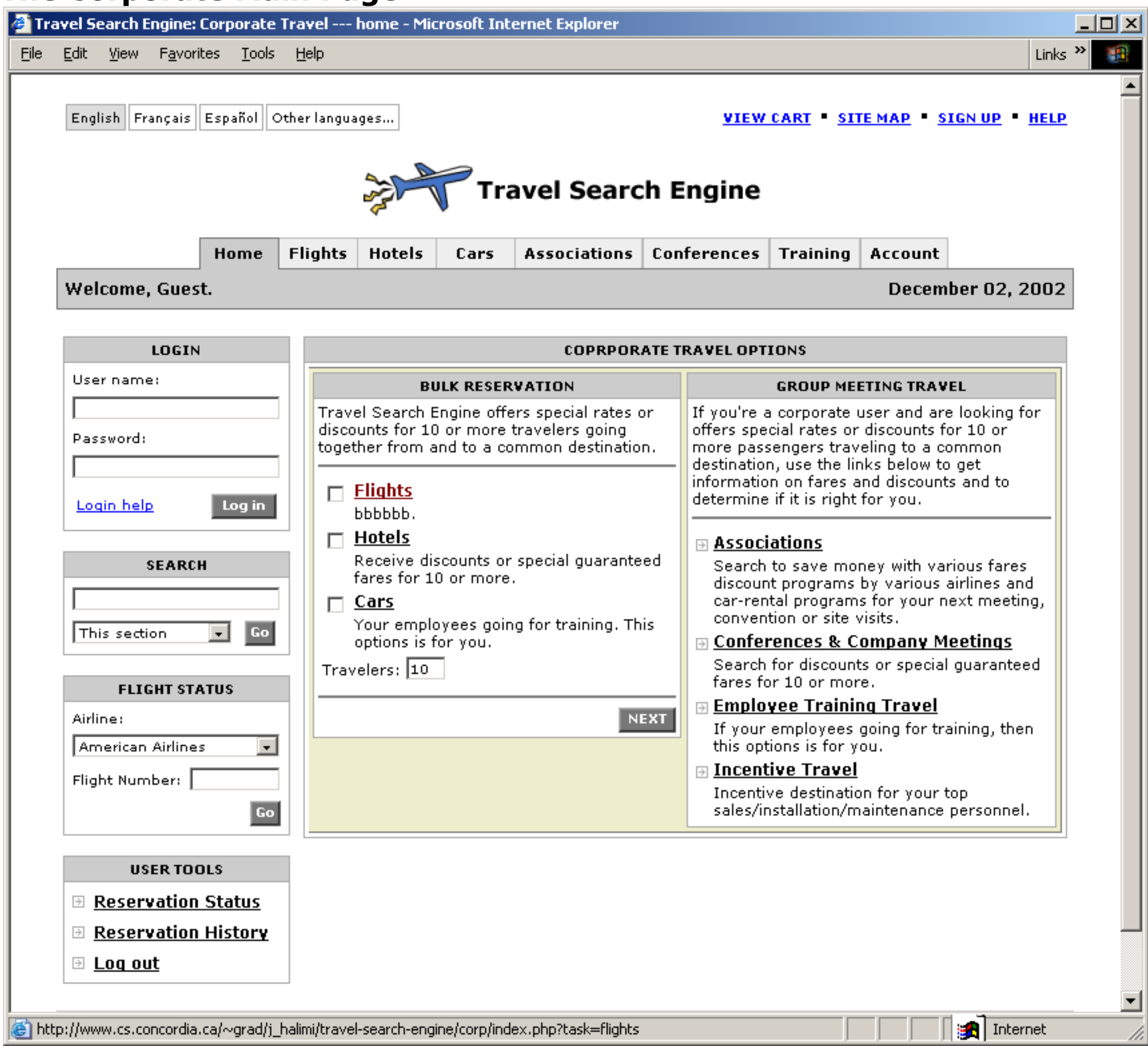

# The Corporate Bulk Flight Reservation

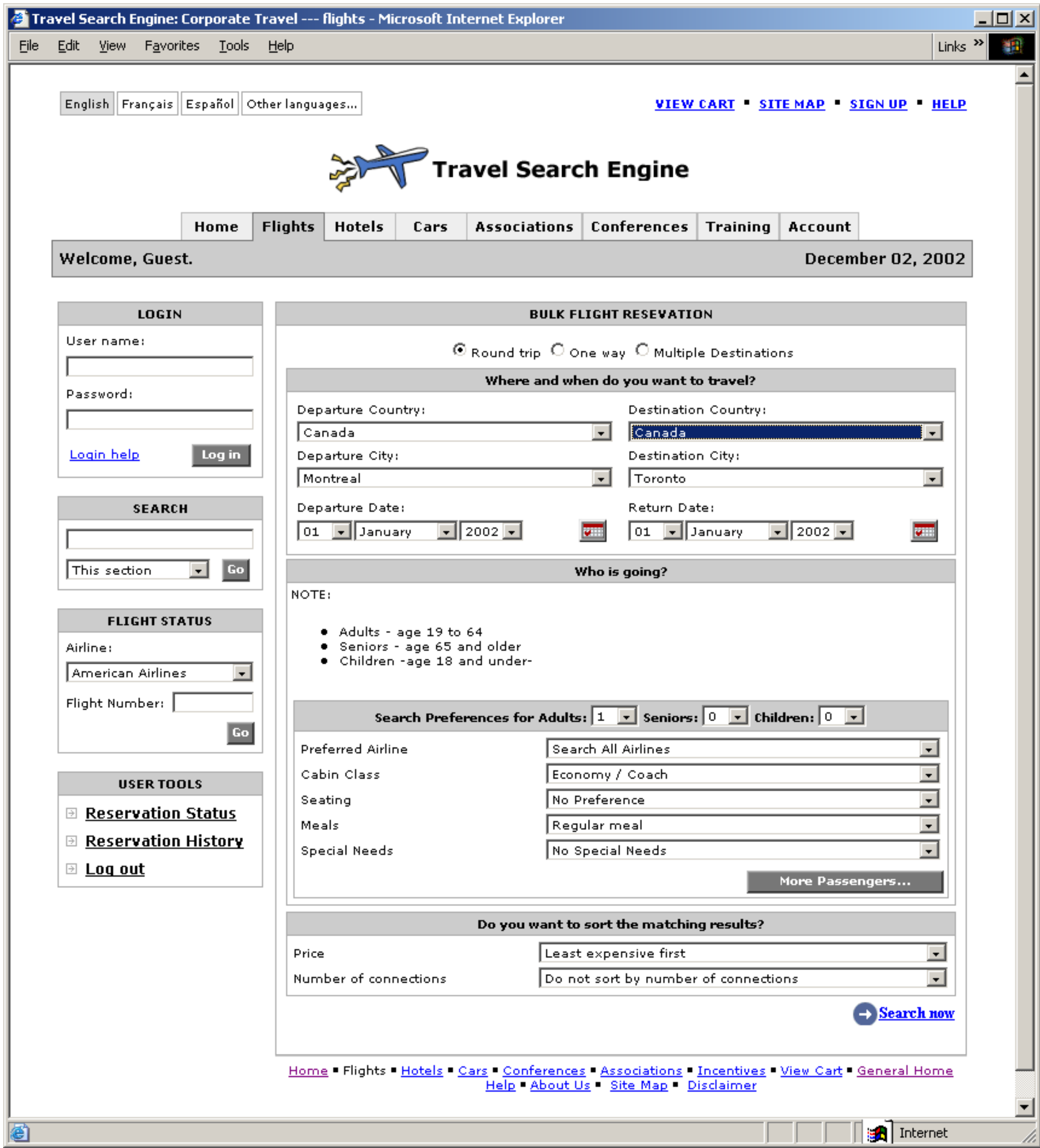

## Specifying Different Preferences for Different Travelers

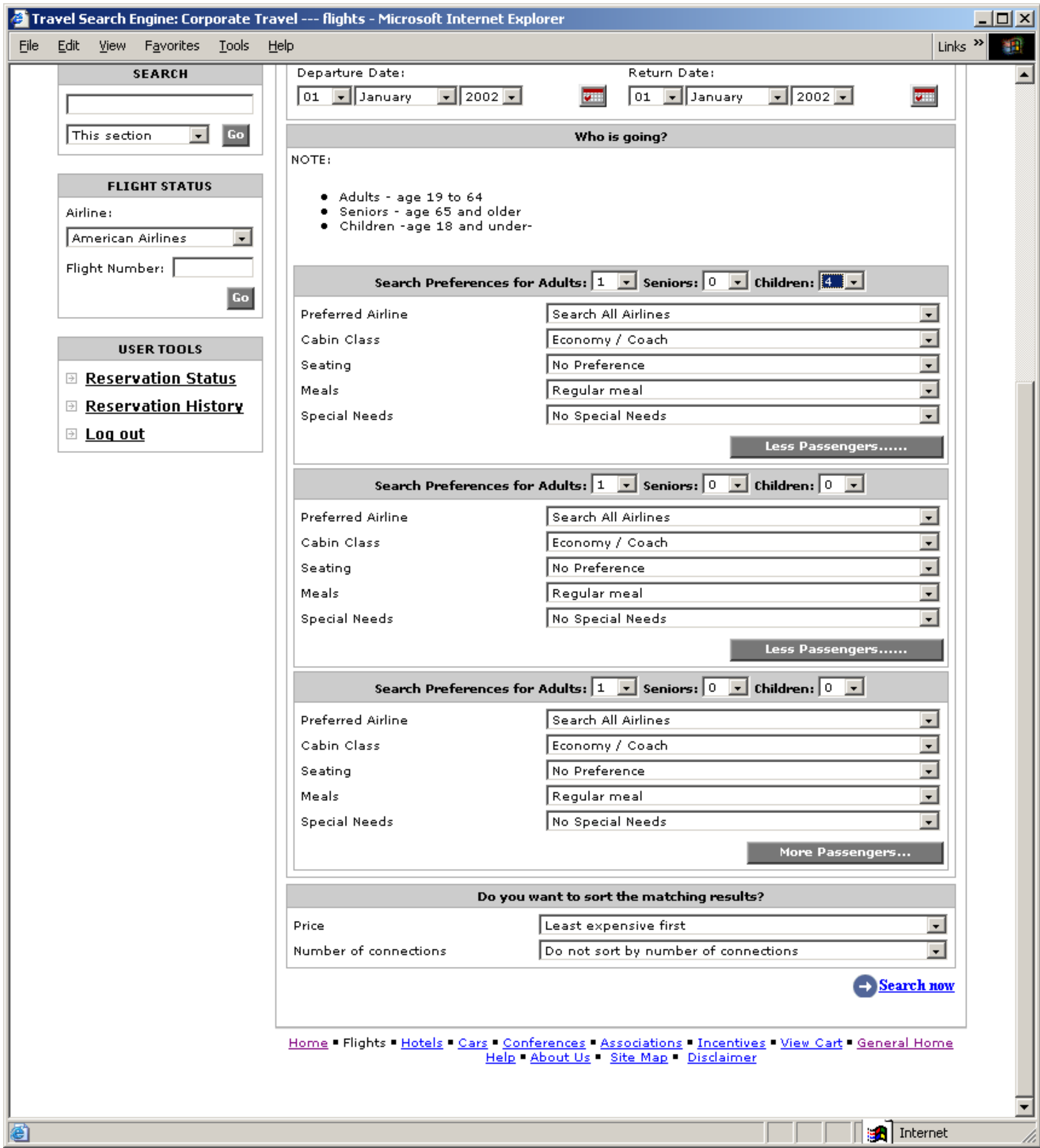

Comments about the screen:

1. Item.

2. Item.

## Browsing and Selecting from Search Results

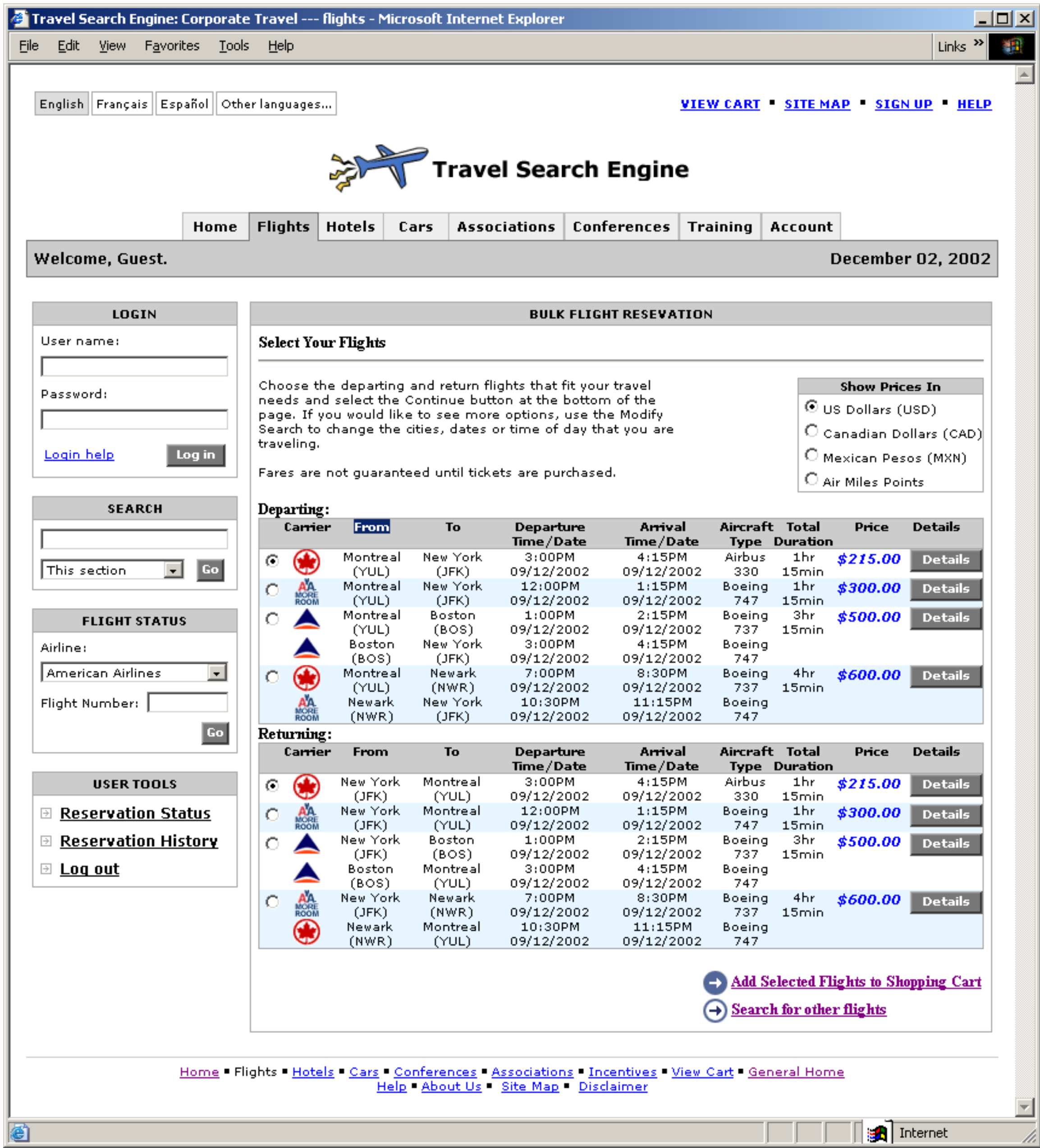

## Detailed Information About a Matching Flight

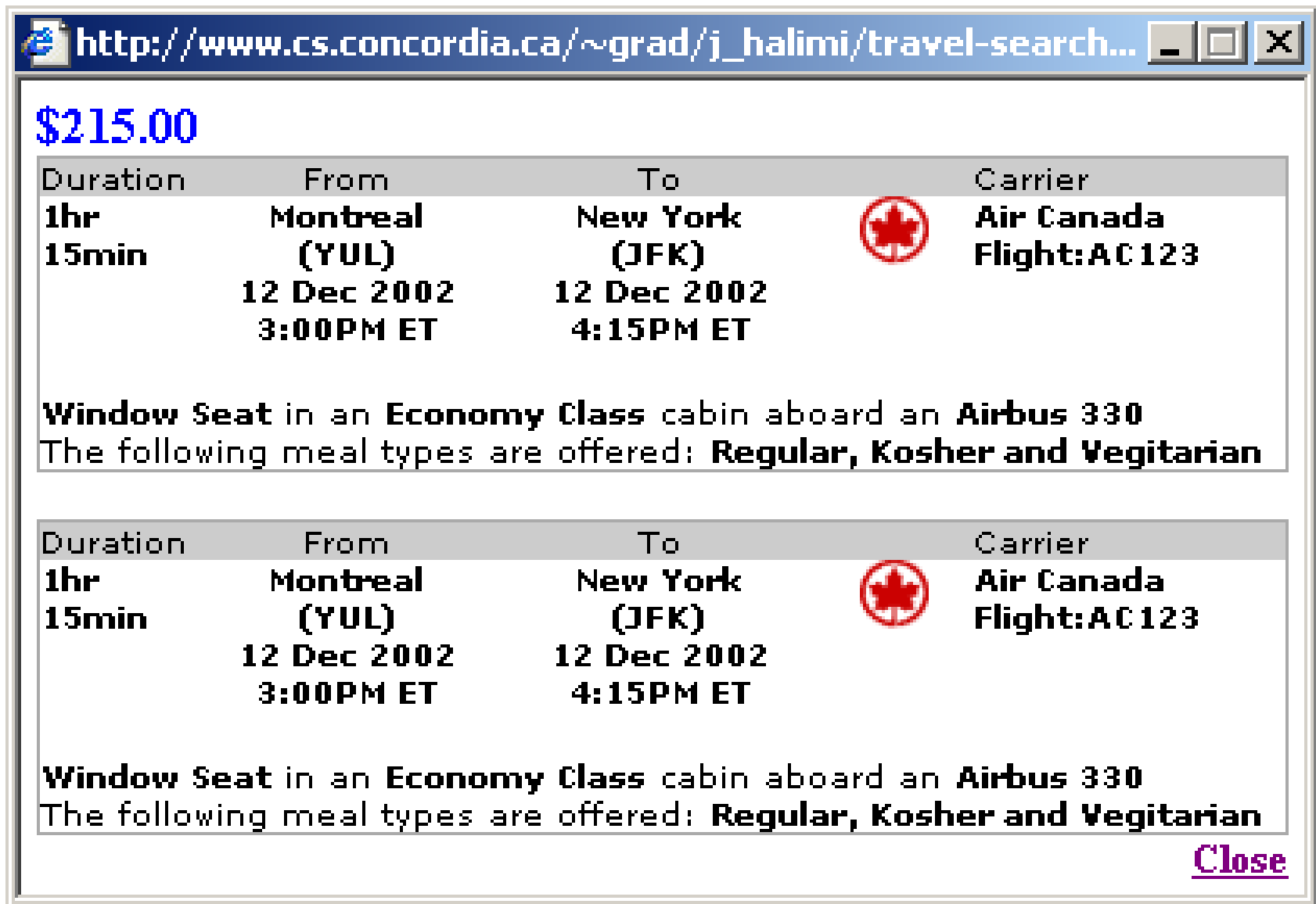
http://www.cs.concordia.ca/~grad/j_halimi/travel-search...

$215.00

| Duration | From | To | | Carrier |
|---|---|---|---|---|
| **1hr 15min** | **Montreal (YUL) 12 Dec 2002 3:00PM ET** | **New York (JFK) 12 Dec 2002 4:15PM ET** | | **Air Canada Flight:AC123** |

**Window Seat** in an **Economy Class** cabin aboard an **Airbus 330**
The following meal types are offered: **Regular, Kosher and Vegitarian**

| Duration | From | To | | Carrier |
|---|---|---|---|---|
| **1hr 15min** | **Montreal (YUL) 12 Dec 2002 3:00PM ET** | **New York (JFK) 12 Dec 2002 4:15PM ET** | | **Air Canada Flight:AC123** |

**Window Seat** in an **Economy Class** cabin aboard an **Airbus 330**
The following meal types are offered: **Regular, Kosher and Vegitarian**

Close

# Searching for a Hotel with the Following Criteria:

- in downtown Montreal
- for a group of 6 adults and 1 child
- within 1 mile of downtown Montreal
- check in date: 8th November, 2002
- check out date: 10th November, 2002
- approved by CAA/AAA
- three rooms with two double beds in each room
- 2 smoking room and 1 non-smoking room
- with indoor pool
- small pets allowed
- shuttle service to and from airport is required
- the rooms MUST be adjacent to each other
- must have a view
- display the result sorted by price

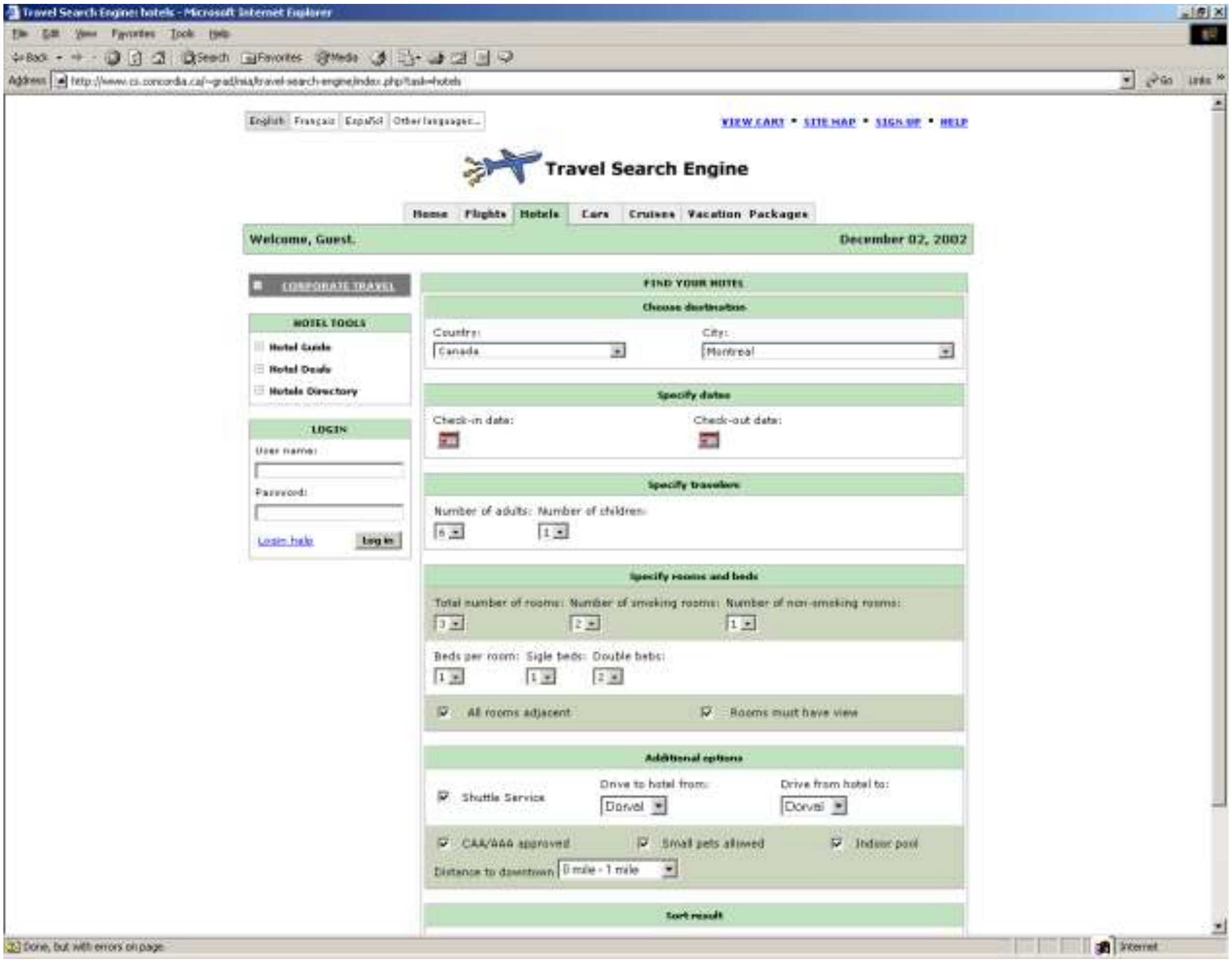
Travel Search Engine: hotels - Microsoft Internet Explorer
English Français Español Other languages...
VIEW CART • SITE MAP • SIGN UP • HELP
Travel Search Engine
Home Flights Hotels Cars Cruises Vacation Packages
Welcome, Guest.
December 02, 2002
CORPORATE TRAVEL
HOTEL TOOLS
Hotel Guide
Hotel Deals
Hotels Directory
LOGIN
User name:
Password:
Login help
Log in
FIND YOUR HOTEL
Choose destination
Country:
Canada
City:
Montreal
Specify dates
Check-in date:
Check-out date:
Specify travelers
Number of adults: Number of children:
Specify rooms and beds
Total number of rooms: Number of smoking rooms: Number of non-smoking rooms:
Beds per room: Sigle beds: Double bebs:
All rooms adjacent
Rooms must have view
Additional options
Shuttle Service
Drive to hotel from:
Dorval
Drive from hotel to:
Dorval
CAA/AAA approved
Small pets allowed
Indoor pool
Distance to downtown
0 mile - 1 mile
Sort result
Done, but with errors on page
Internet

## By mistake John Smith entered the following invalid data

- The check-out date was before the check-in date
- The number of adults was selected as 0.
- The sum of smoking rooms and non-smoking rooms are not equal to the total rooms requested.
- The sum of single bed and double bed in each room are not equal to the No. of beds in each room requested.
- The system prompt John Smith with the error message page to ask him re-enter the invalid fields.
- John Smith enters the correct data on the page.
- Performs the search.

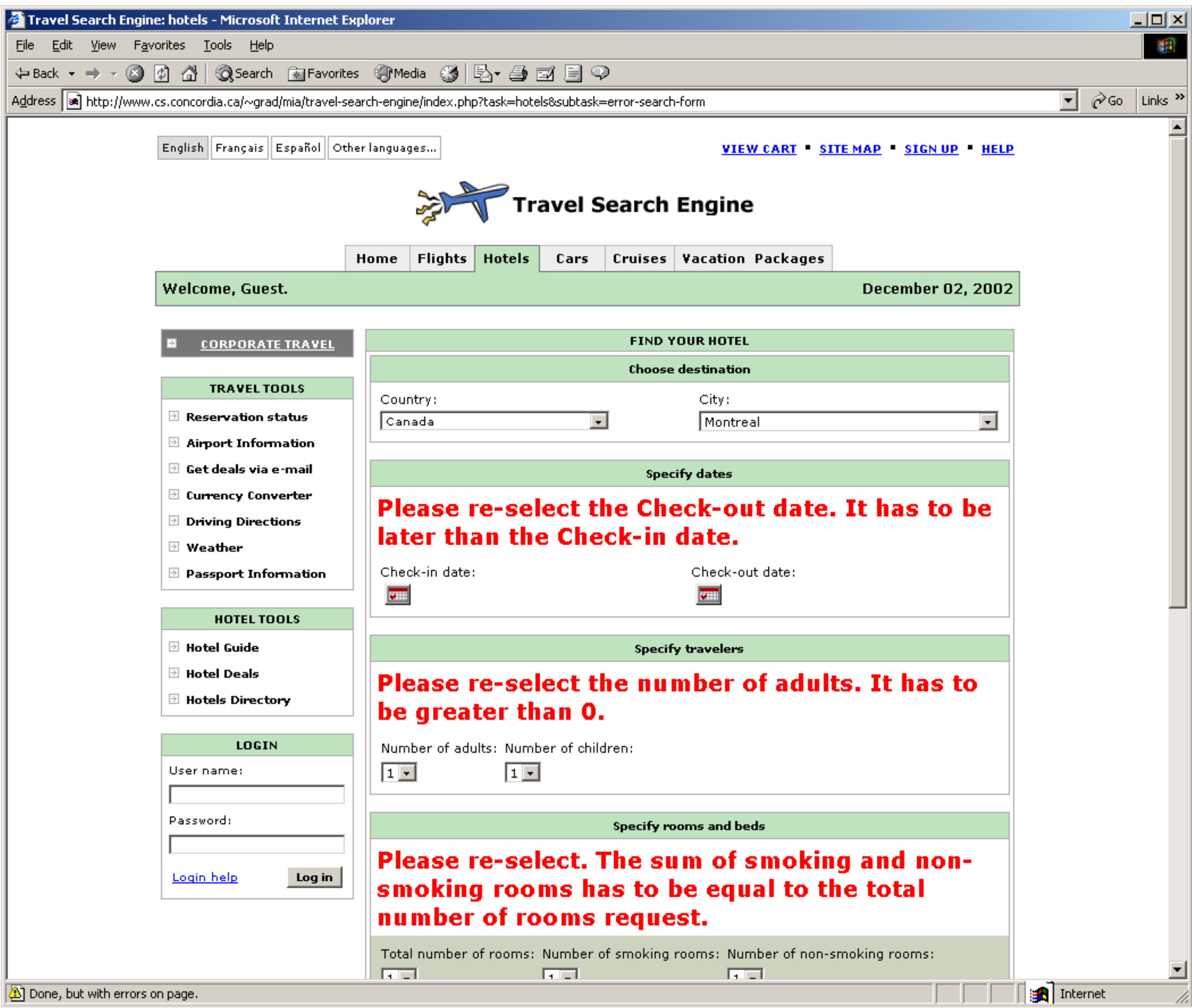

The system displays the search result.

- John Smith books the Holiday Inn city view room.

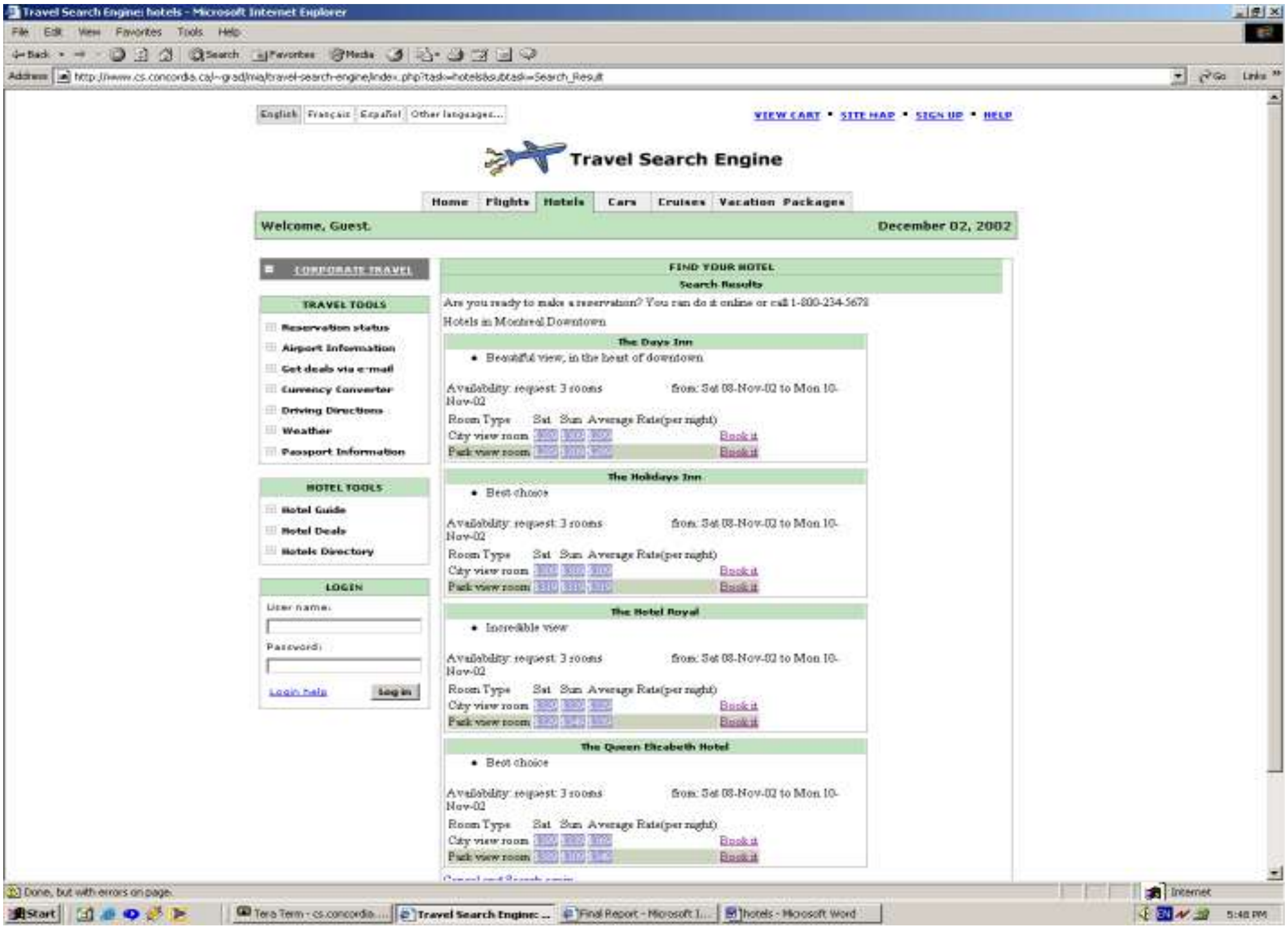

The system displays all the reservation details and policy.

- John Smith reviews the details and policy.
- John Smith accepts the term and continues the reservation

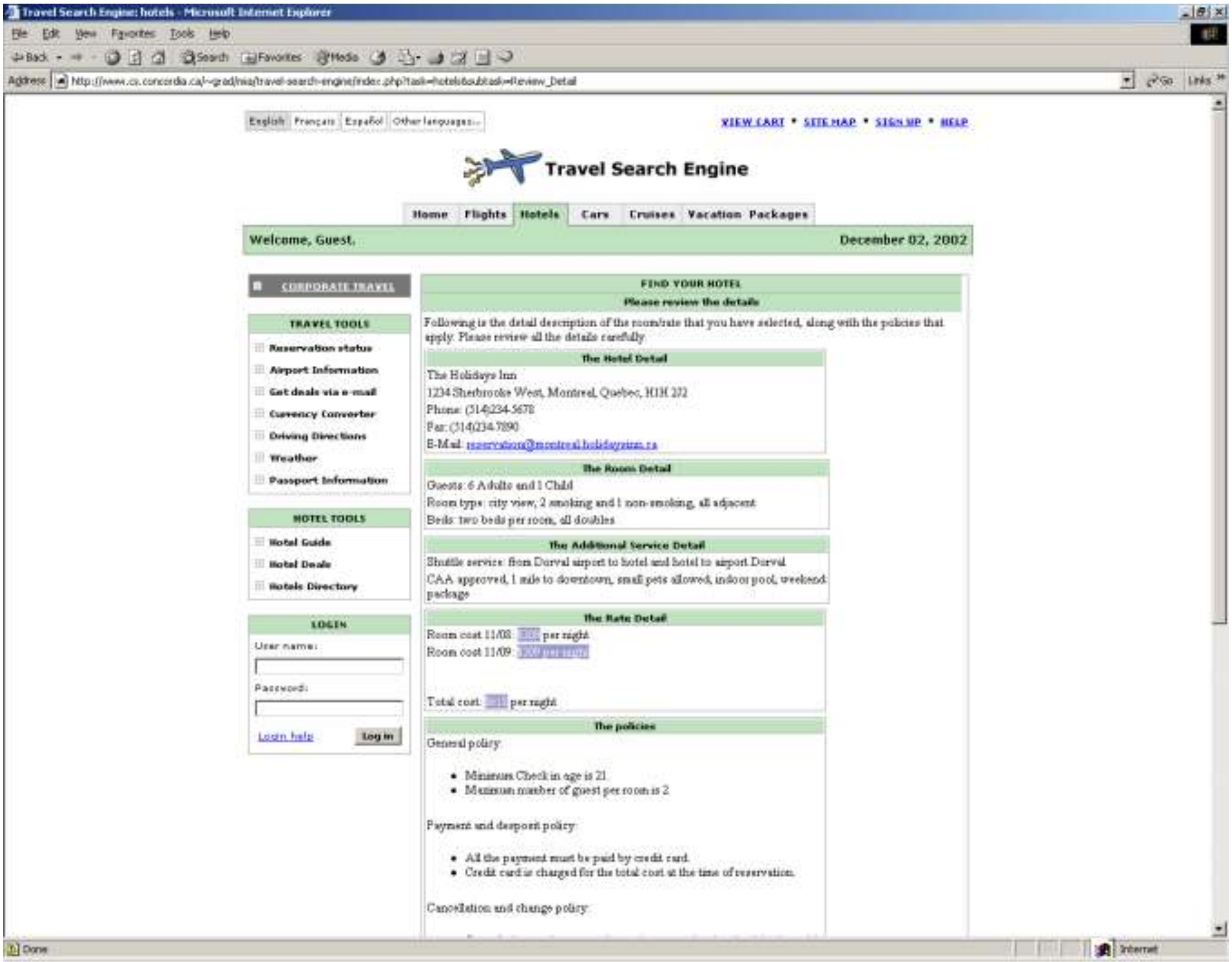

# John Smith enters the personal information

To create a new profile, john smith, enters personal information and moves to shopping cart for checking out.

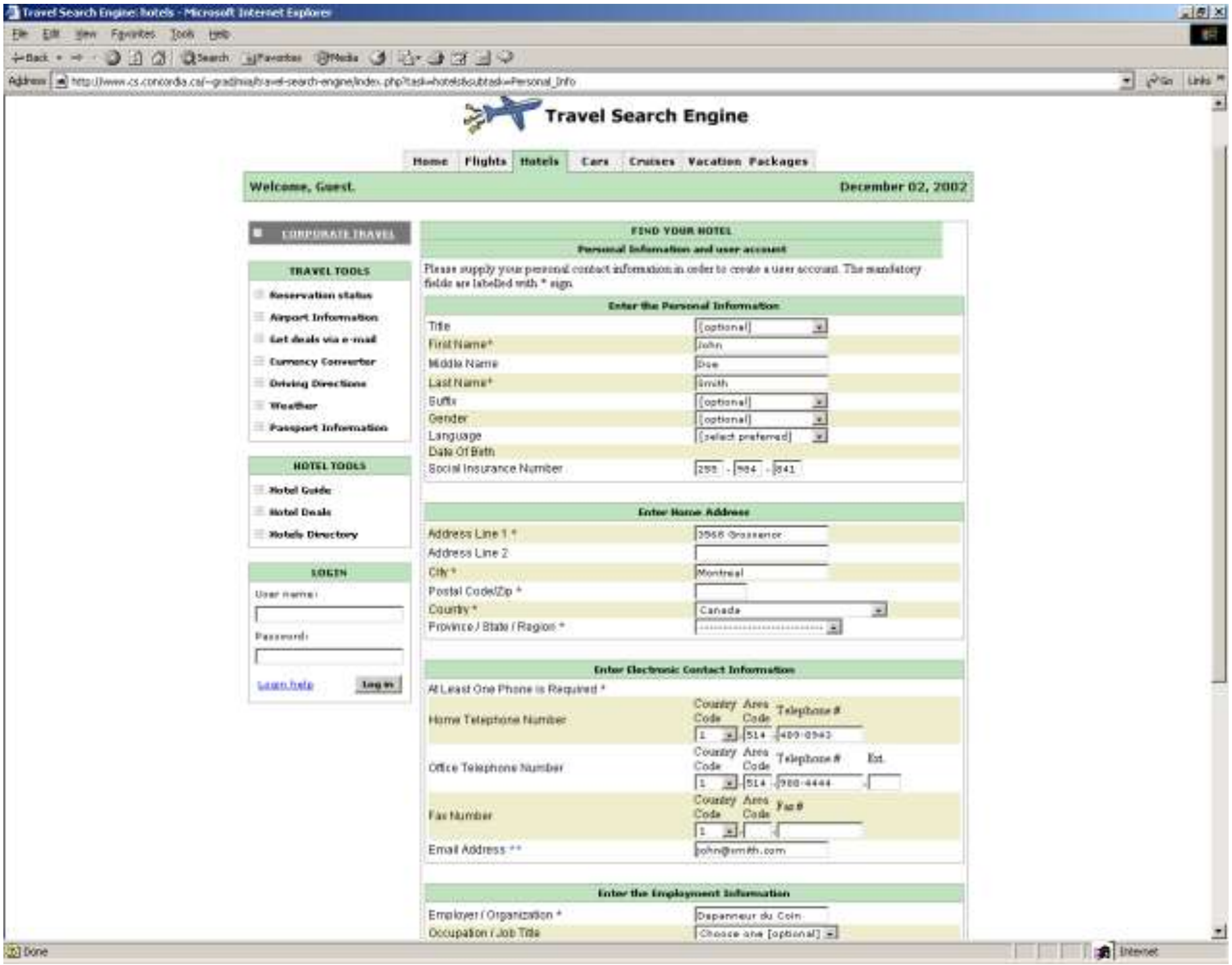

## The System Displays the Shopping Cart

Here john smith can review his reservation in cart.

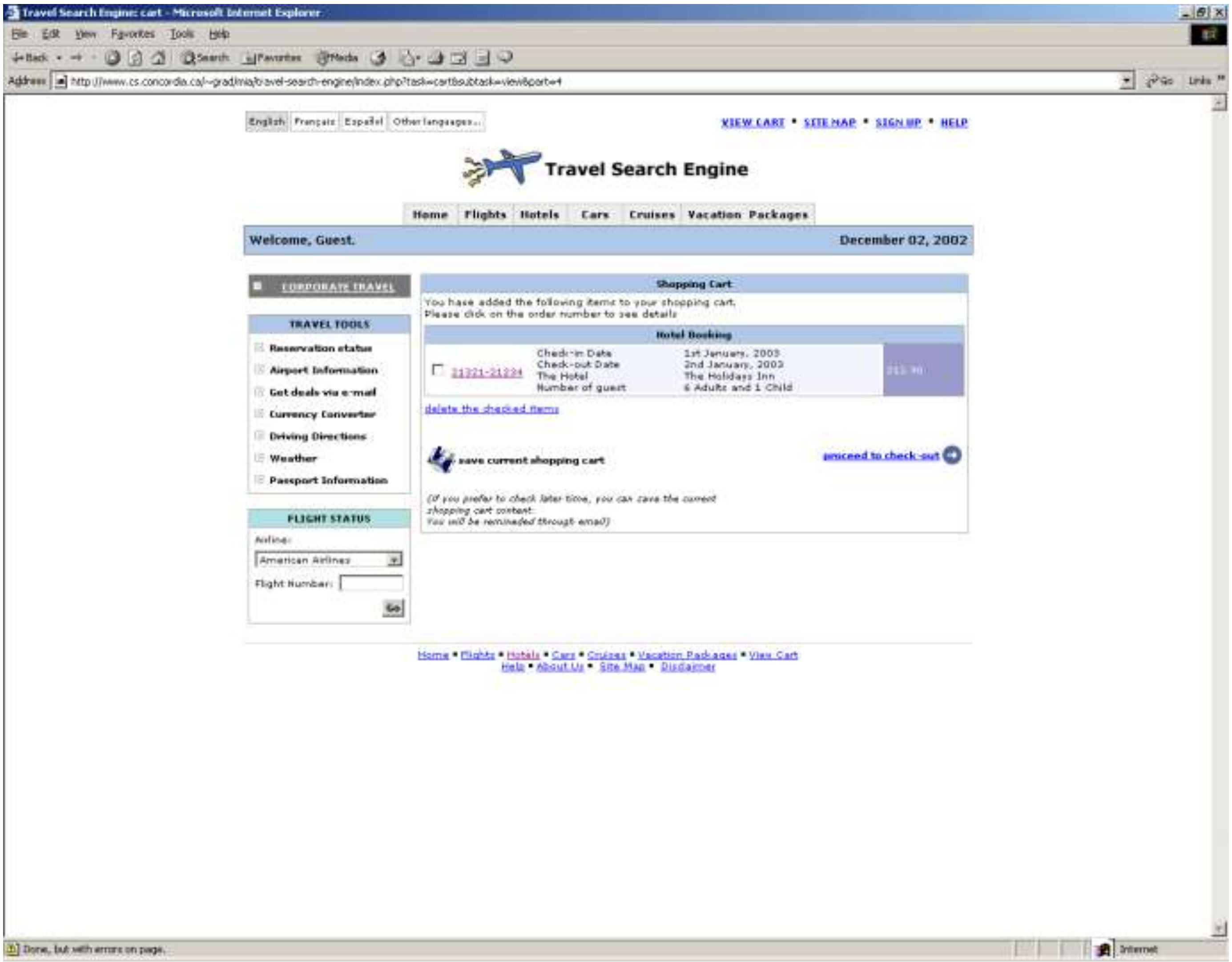

## Billing and Information Shipping

John Smith’s payment information is carried out here in the checkout page. John can specify a different payment method.

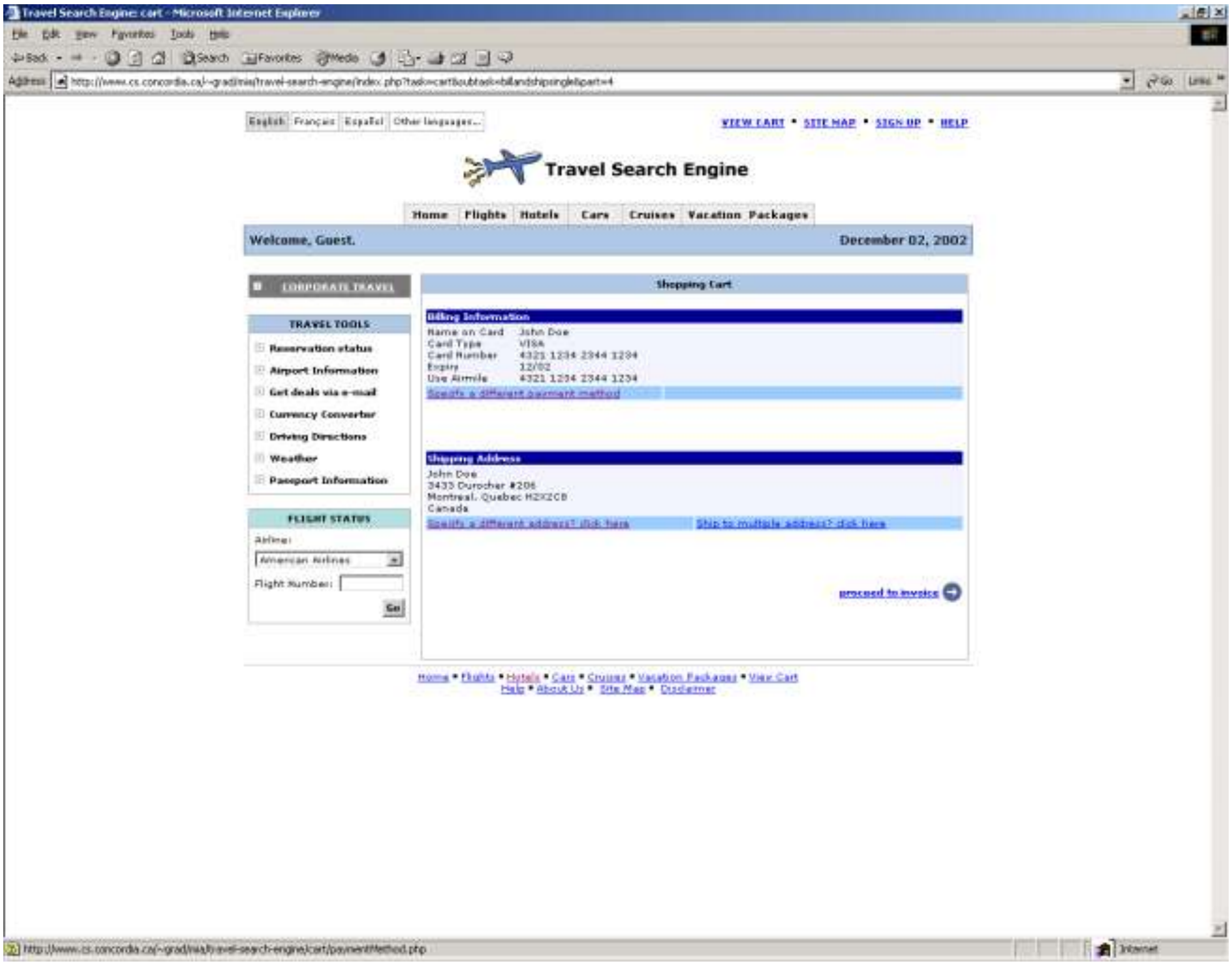

## Invoice

Finally, John can take a printout of the invoice for his purchase.

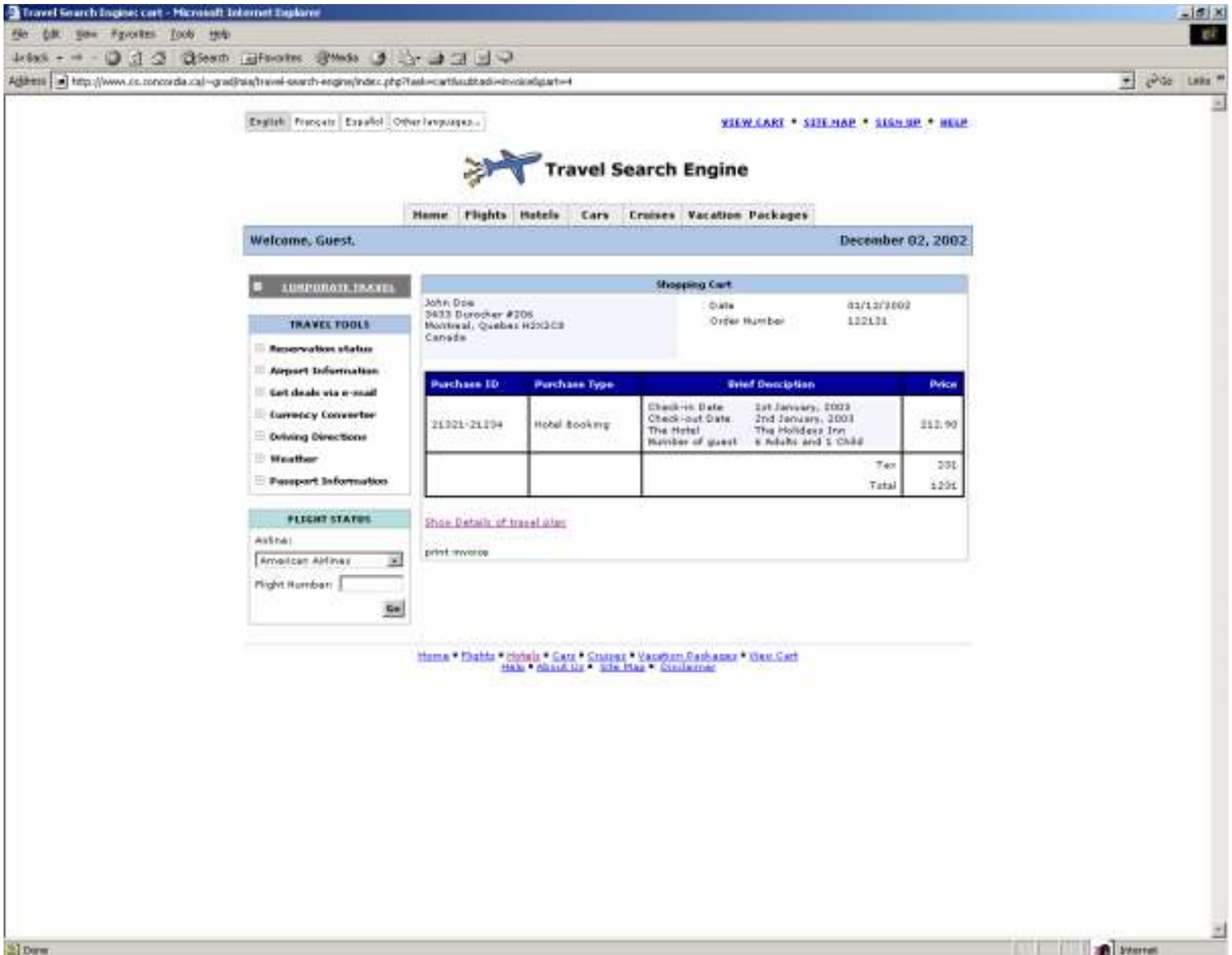

## Typical Car Scenario

Details concerning rental fees and a vehicle's available options will be indicated to facilitate the client's rental selection. Once a client has identified the vehicle best suited to his/her needs, the client will fill an order form with all required personal and/or login information, along with the vehicle selected and duration of the rental request, etc. The client now submits the order by clicking on the submit button at the bottom of the screen. The client confirmation and details concerning the dispatch of the vehicle get emailed to the customer in addition to displaying this info on the confirmation page.

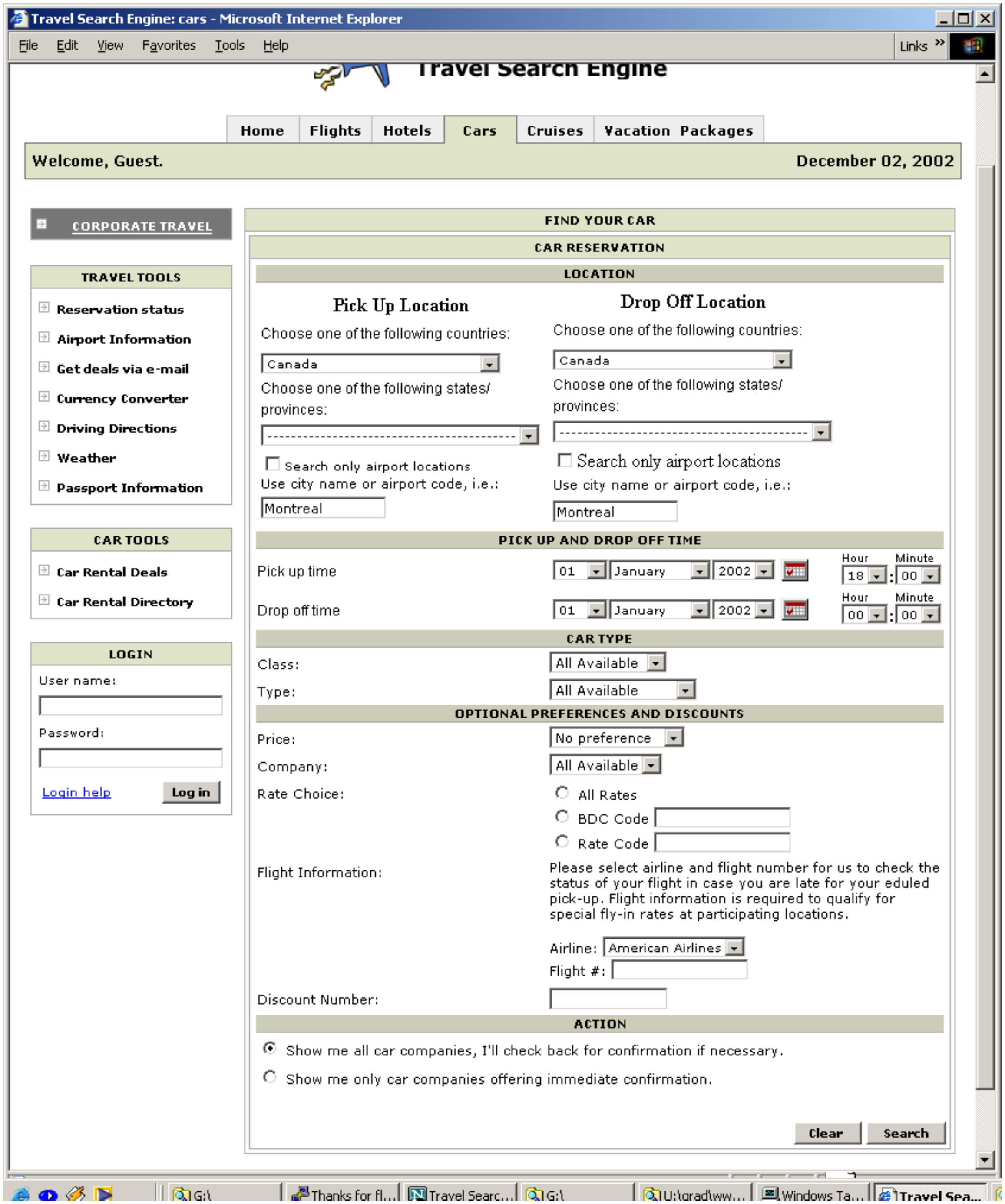
Travel Search Engine: cars - Microsoft Internet Explorer
File Edit View Favorites Tools Help
Links
Travel Search Engine
Home Flights Hotels Cars Cruises Vacation Packages
Welcome, Guest.
December 02, 2002
CORPORATE TRAVEL
TRAVEL TOOLS
Reservation status
Airport Information
Get deals via e-mail
Currency Converter
Driving Directions
Weather
Passport Information
CAR TOOLS
Car Rental Deals
Car Rental Directory
LOGIN
User name:
Password:
Login help
Log in
FIND YOUR CAR
CAR RESERVATION
LOCATION
Pick Up Location
Choose one of the following countries:
Canada
Choose one of the following states/ provinces:
Search only airport locations
Use city name or airport code, i.e.:
Montreal
Drop Off Location
Choose one of the following countries:
Canada
Choose one of the following states/ provinces:
Search only airport locations
Use city name or airport code, i.e.:
Montreal
PICK UP AND DROP OFF TIME
Pick up time
01 January 2002
Hour Minute
18 : 00
Drop off time
01 January 2002
Hour Minute
00 : 00
CAR TYPE
Class:
All Available
Type:
All Available
OPTIONAL PREFERENCES AND DISCOUNTS
Price:
No preference
Company:
All Available
Rate Choice:
All Rates
BDC Code
Rate Code
Flight Information:
Please select airline and flight number for us to check the status of your flight in case you are late for your eduled pick-up. Flight information is required to qualify for special fly-in rates at participating locations.
Airline: American Airlines
Flight #:
Discount Number:
ACTION
Show me all car companies, I'll check back for confirmation if necessary.
Show me only car companies offering immediate confirmation.
Clear
Search
G:\
Thanks for fl...
Travel Searc...
G:\
U:\grad\ww...
Windows Ta...
Travel Sea...

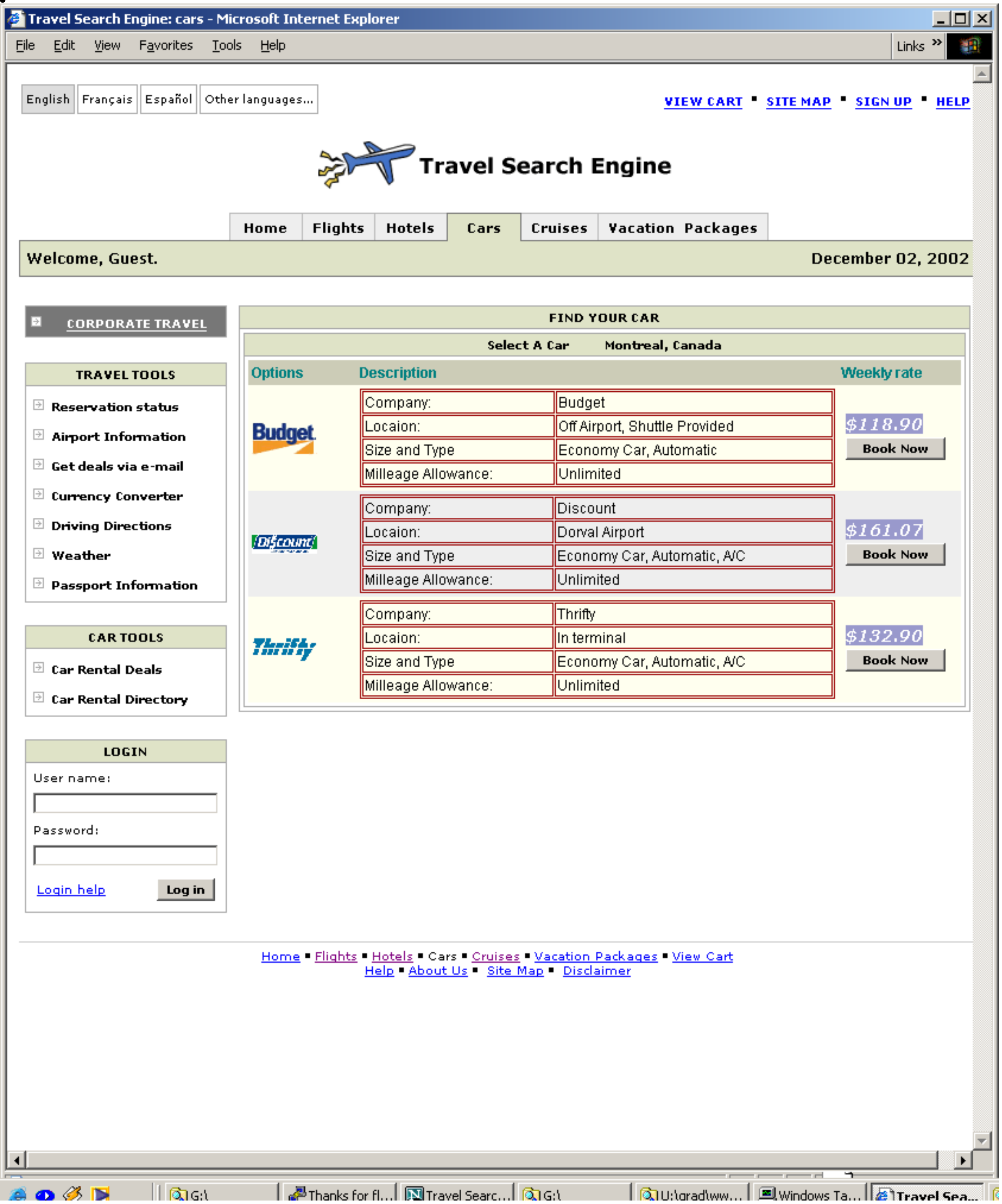
Travel Search Engine: cars - Microsoft Internet Explorer
File Edit View Favorites Tools Help
Links
English
Français
Español
Other languages...
VIEW CART ▪ SITE MAP ▪ SIGN UP ▪ HELP
Travel Search Engine
Home
Flights
Hotels
Cars
Cruises
Vacation Packages
Welcome, Guest.
December 02, 2002
CORPORATE TRAVEL
TRAVEL TOOLS
Reservation status
Airport Information
Get deals via e-mail
Currency Converter
Driving Directions
Weather
Passport Information
CAR TOOLS
Car Rental Deals
Car Rental Directory
LOGIN
User name:
Password:
Login help
Log in
FIND YOUR CAR
Select A Car Montreal, Canada
Options
Description
Weekly rate
Budget
Company: Budget
Locaion: Off Airport, Shuttle Provided
Size and Type: Economy Car, Automatic
Milleage Allowance: Unlimited
$118.90
Book Now
Discount
Company: Discount
Locaion: Dorval Airport
Size and Type: Economy Car, Automatic, A/C
Milleage Allowance: Unlimited
$161.07
Book Now
Thrifty
Company: Thrifty
Locaion: In terminal
Size and Type: Economy Car, Automatic, A/C
Milleage Allowance: Unlimited
$132.90
Book Now
Home ▪ Flights ▪ Hotels ▪ Cars ▪ Cruises ▪ Vacation Packages ▪ View Cart
Help ▪ About Us ▪ Site Map ▪ Disclaimer
G:\
Thanks for fl...
Travel Searc...
G:\
U:\grad\ww...
Windows Ta...
Travel Sea...

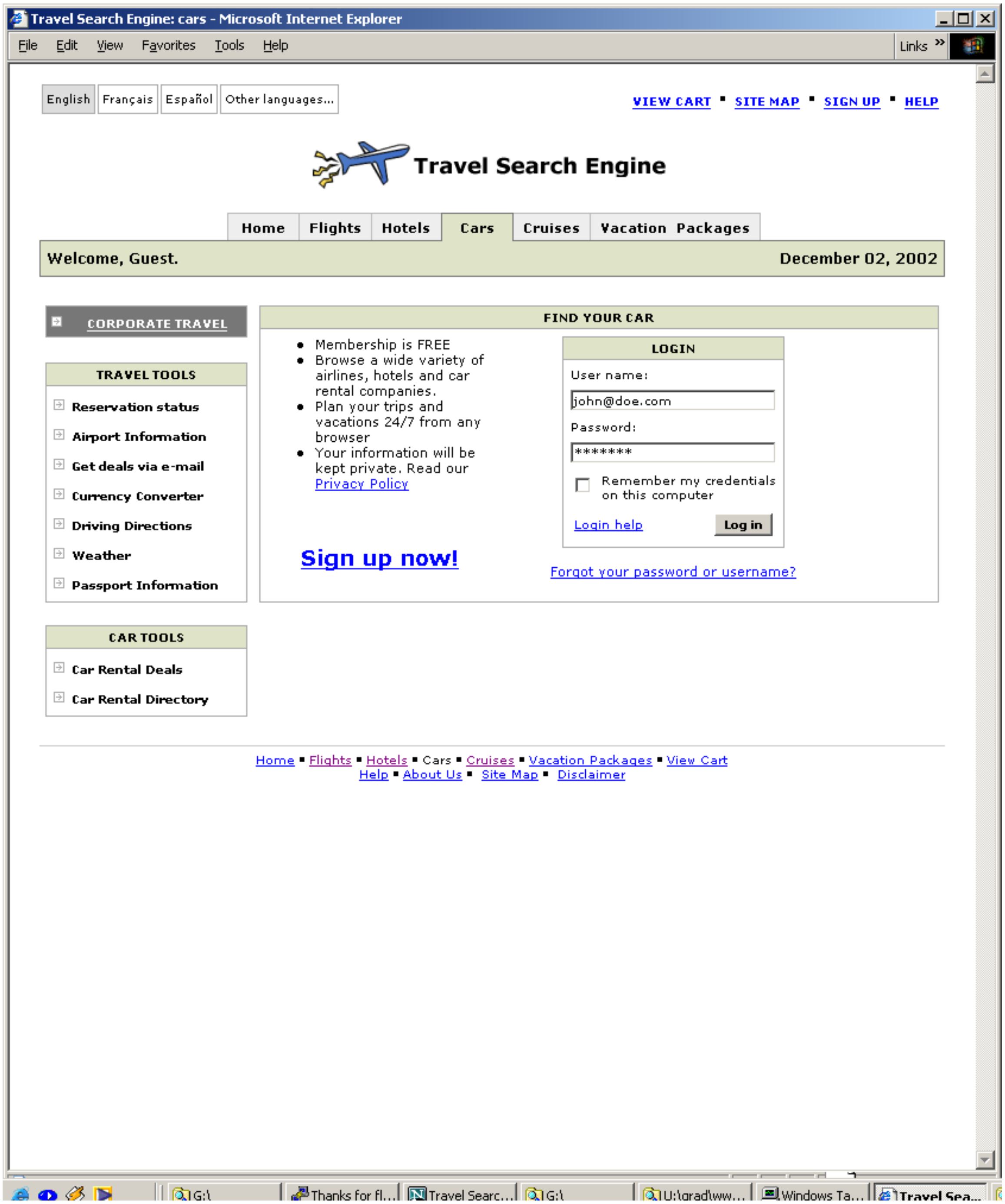
Travel Search Engine: cars - Microsoft Internet Explorer
File Edit View Favorites Tools Help
Links
English Français Español Other languages...
VIEW CART ▪ SITE MAP ▪ SIGN UP ▪ HELP
Travel Search Engine
Home Flights Hotels Cars Cruises Vacation Packages
Welcome, Guest.
December 02, 2002
CORPORATE TRAVEL
TRAVEL TOOLS
Reservation status
Airport Information
Get deals via e-mail
Currency Converter
Driving Directions
Weather
Passport Information
CAR TOOLS
Car Rental Deals
Car Rental Directory
FIND YOUR CAR
Membership is FREE
Browse a wide variety of airlines, hotels and car rental companies.
Plan your trips and vacations 24/7 from any browser
Your information will be kept private. Read our Privacy Policy
LOGIN
User name:
john@doe.com
Password:
*******
Remember my credentials on this computer
Login help
Log in
Sign up now!
Forgot your password or username?
Home ▪ Flights ▪ Hotels ▪ Cars ▪ Cruises ▪ Vacation Packages ▪ View Cart
Help ▪ About Us ▪ Site Map ▪ Disclaimer
G:\
Thanks for fl...
Travel Searc...
G:\
U:\grad\ww...
Windows Ta...
Travel Sea...

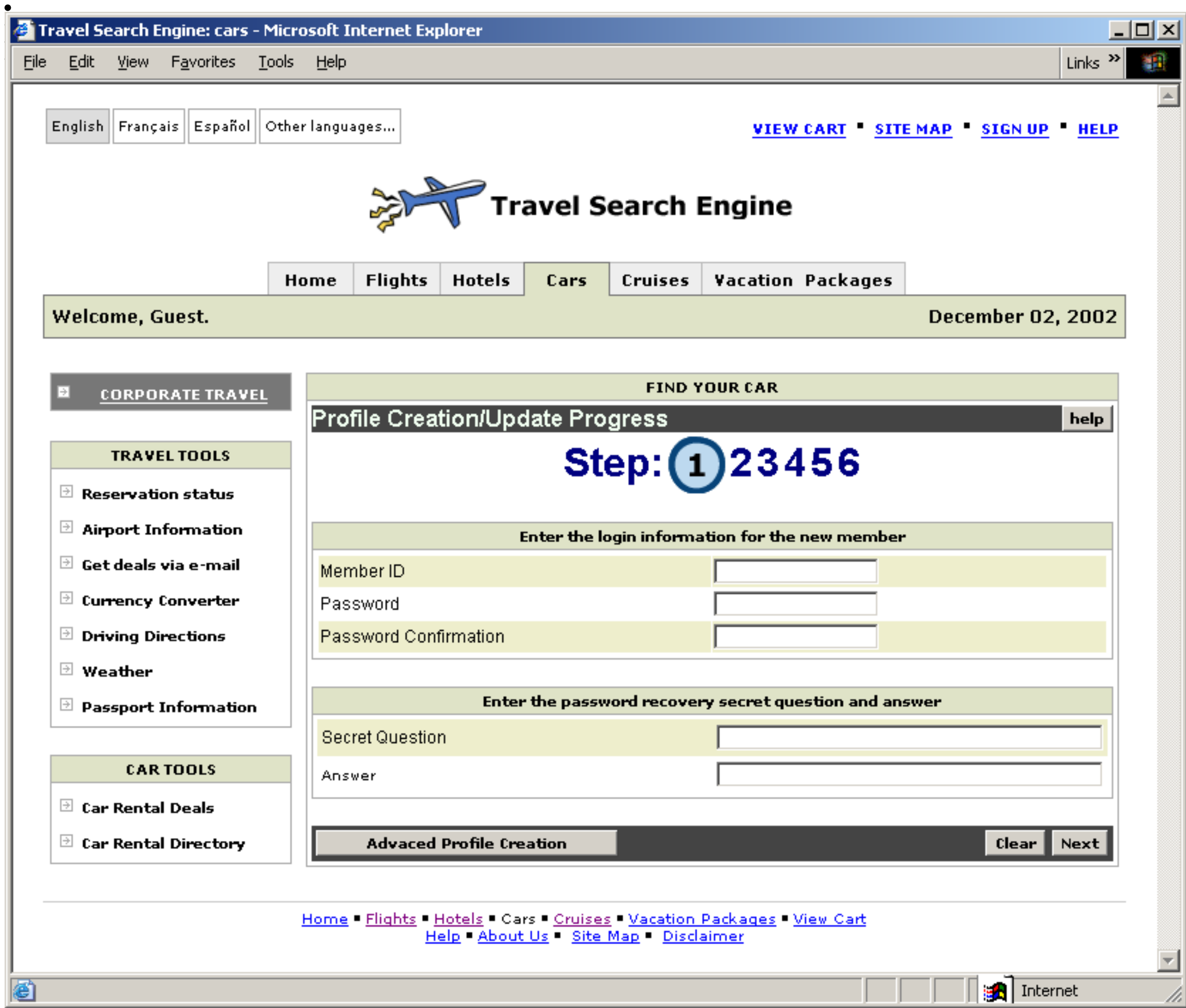
Travel Search Engine: cars - Microsoft Internet Explorer
File Edit View Favorites Tools Help
Links
English
Français
Español
Other languages...
VIEW CART
SITE MAP
SIGN UP
HELP
Travel Search Engine
Home
Flights
Hotels
Cars
Cruises
Vacation Packages
Welcome, Guest.
December 02, 2002
CORPORATE TRAVEL
TRAVEL TOOLS
Reservation status
Airport Information
Get deals via e-mail
Currency Converter
Driving Directions
Weather
Passport Information
CAR TOOLS
Car Rental Deals
Car Rental Directory
FIND YOUR CAR
Profile Creation/Update Progress
help
Step: 1 2 3 4 5 6
Enter the login information for the new member
Member ID
Password
Password Confirmation
Enter the password recovery secret question and answer
Secret Question
Answer
Advaced Profile Creation
Clear
Next
Home ▪ Flights ▪ Hotels ▪ Cars ▪ Cruises ▪ Vacation Packages ▪ View Cart
Help ▪ About Us ▪ Site Map ▪ Disclaimer
Internet

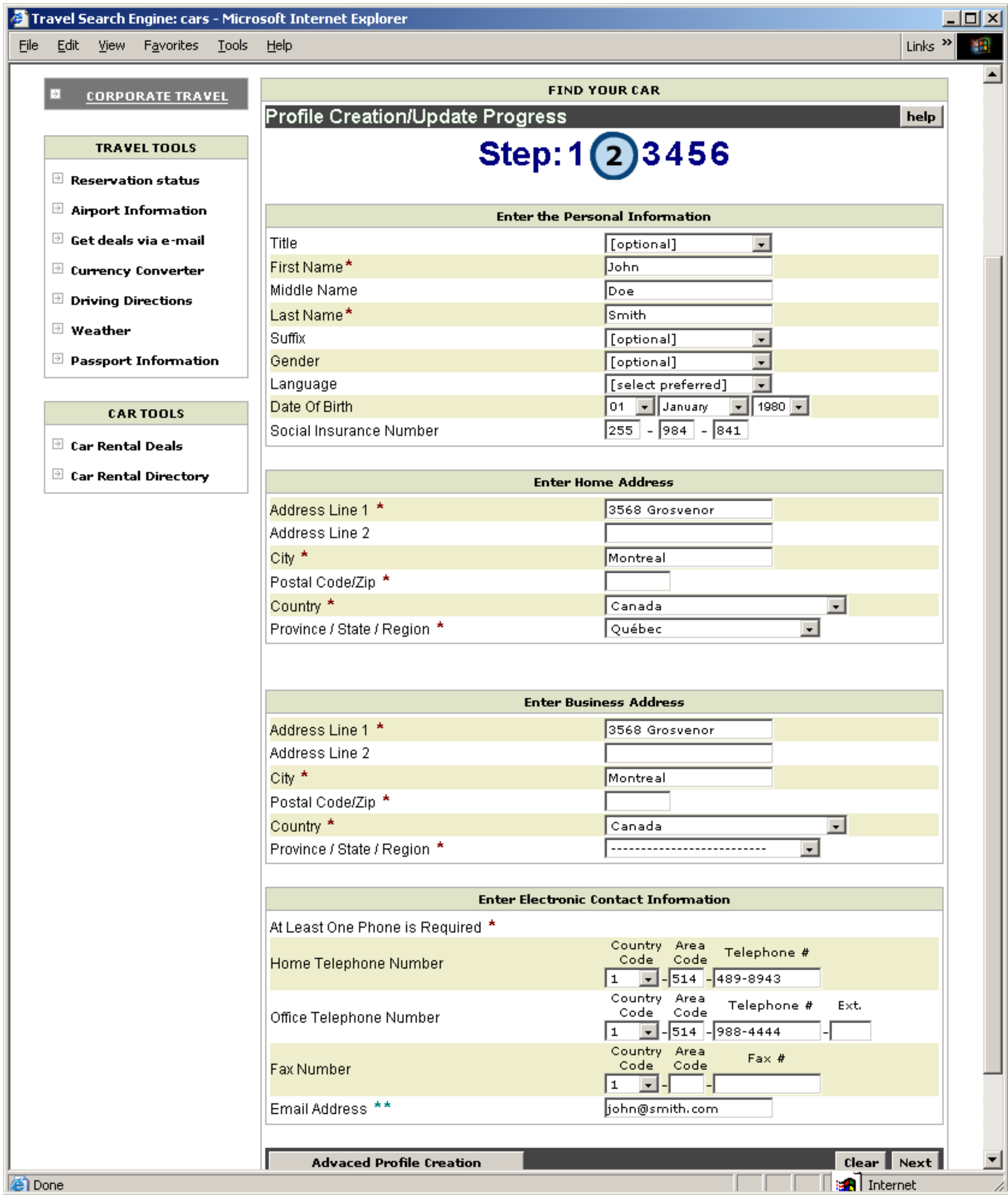
Travel Search Engine: cars - Microsoft Internet Explorer
File Edit View Favorites Tools Help
Links
CORPORATE TRAVEL
TRAVEL TOOLS
Reservation status
Airport Information
Get deals via e-mail
Currency Converter
Driving Directions
Weather
Passport Information
CAR TOOLS
Car Rental Deals
Car Rental Directory
FIND YOUR CAR
Profile Creation/Update Progress
help
Step: 1 2 3 4 5 6
Enter the Personal Information
Title [optional]
First Name* John
Middle Name Doe
Last Name* Smith
Suffix [optional]
Gender [optional]
Language [select preferred]
Date Of Birth 01 January 1980
Social Insurance Number 255 - 984 - 841
Enter Home Address
Address Line 1 * 3568 Grosvenor
Address Line 2
City * Montreal
Postal Code/Zip *
Country * Canada
Province / State / Region * Québec
Enter Business Address
Address Line 1 * 3568 Grosvenor
Address Line 2
City * Montreal
Postal Code/Zip *
Country * Canada
Province / State / Region * -------------------------
Enter Electronic Contact Information
At Least One Phone is Required *
Country Code Area Code Telephone #
Home Telephone Number 1 - 514 - 489-8943
Country Code Area Code Telephone # Ext.
Office Telephone Number 1 - 514 - 988-4444 -
Country Code Area Code Fax #
Fax Number 1 - -
Email Address ** john@smith.com
Advaced Profile Creation
Clear Next
Done
Internet

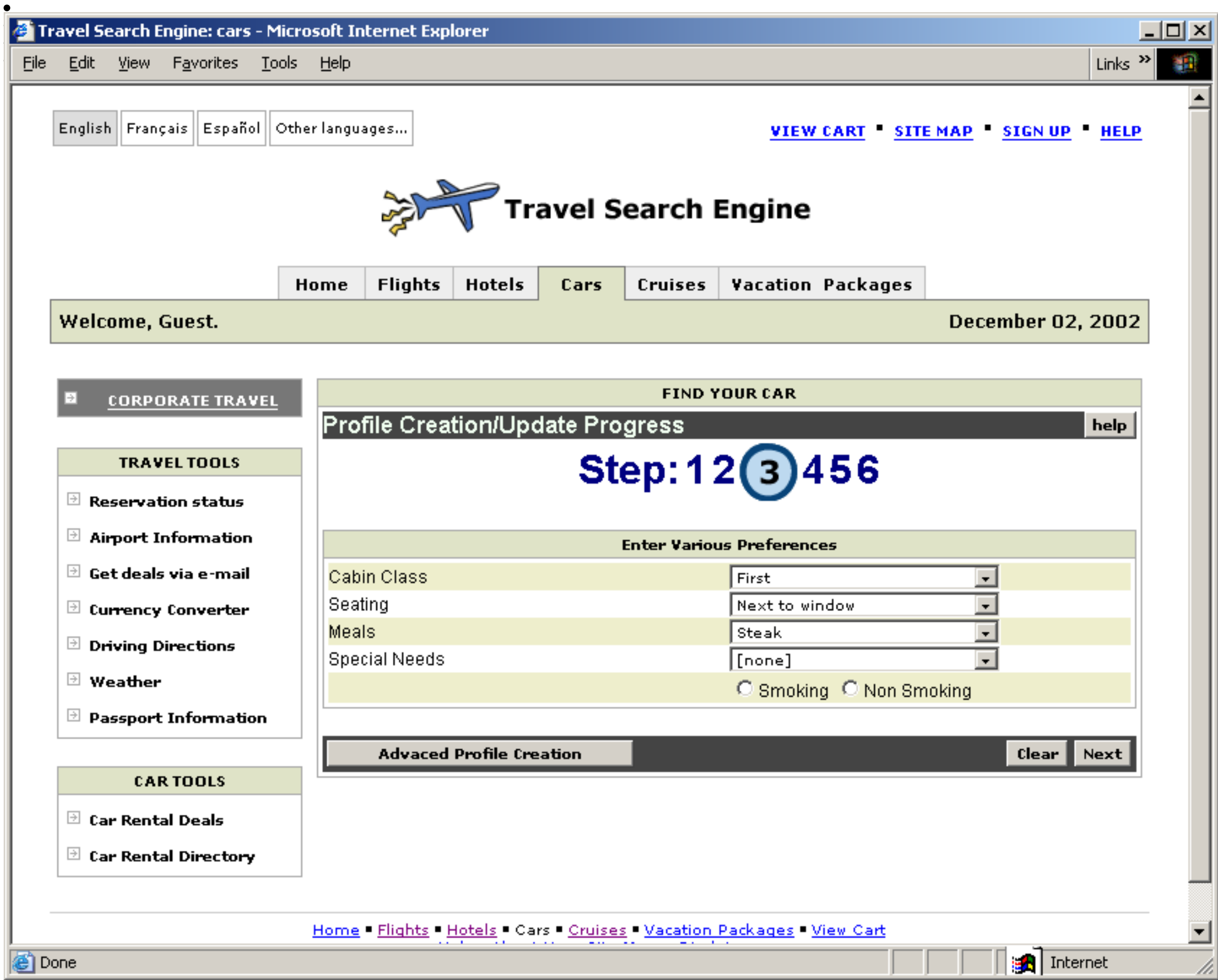
Travel Search Engine: cars - Microsoft Internet Explorer
File Edit View Favorites Tools Help
Links
English
Français
Español
Other languages...
VIEW CART
SITE MAP
SIGN UP
HELP
Travel Search Engine
Home
Flights
Hotels
Cars
Cruises
Vacation Packages
Welcome, Guest.
December 02, 2002
CORPORATE TRAVEL
TRAVEL TOOLS
Reservation status
Airport Information
Get deals via e-mail
Currency Converter
Driving Directions
Weather
Passport Information
CAR TOOLS
Car Rental Deals
Car Rental Directory
FIND YOUR CAR
Profile Creation/Update Progress
help
Step: 1 2 3 4 5 6
Enter Various Preferences
Cabin Class
First
Seating
Next to window
Meals
Steak
Special Needs
[none]
Smoking
Non Smoking
Advaced Profile Creation
Clear
Next
Home ▪ Flights ▪ Hotels ▪ Cars ▪ Cruises ▪ Vacation Packages ▪ View Cart
Done
Internet

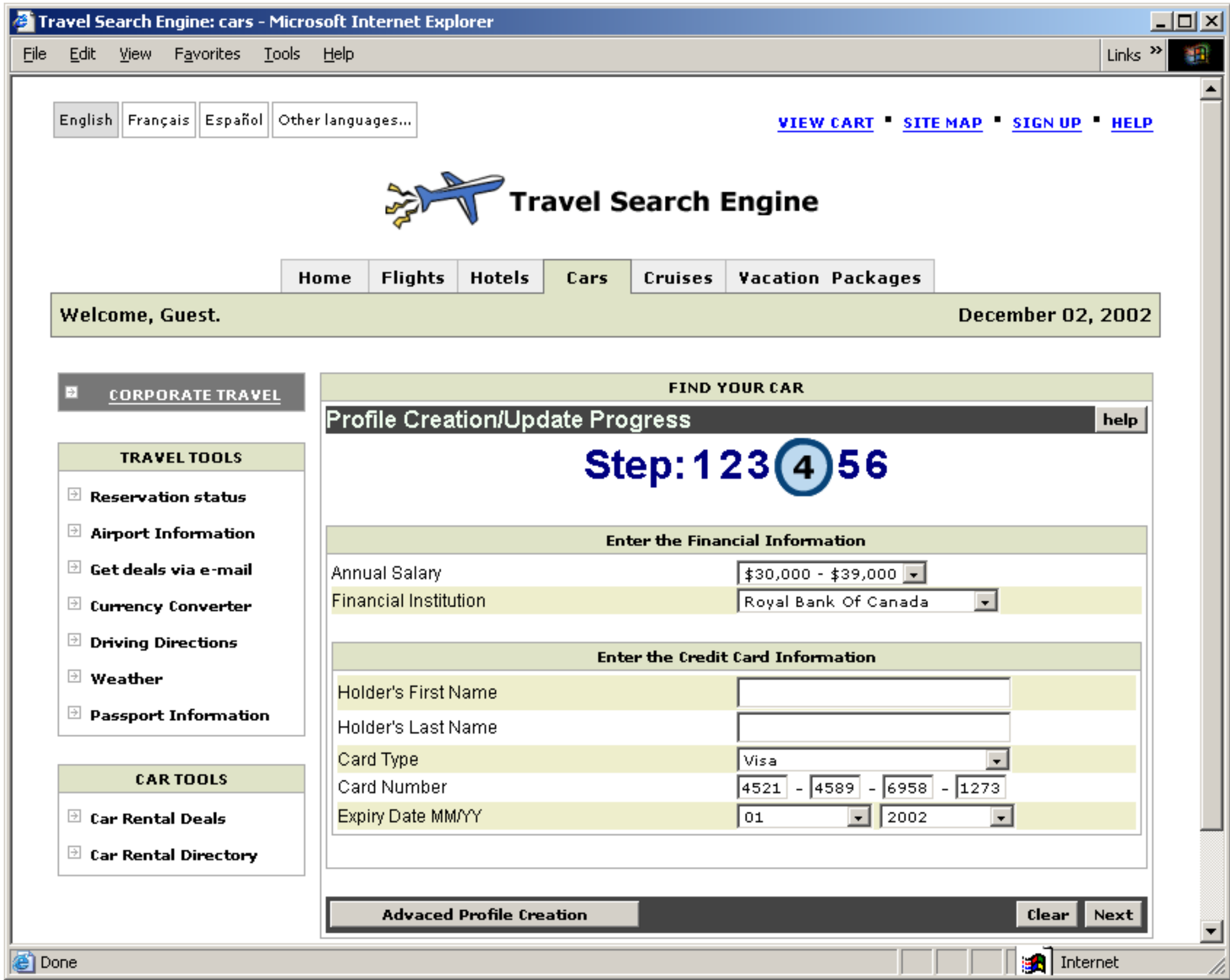

Travel Search Engine: cars - Microsoft Internet Explorer
File Edit View Favorites Tools Help
Links
English Français Español Other languages...
VIEW CART ▪ SITE MAP ▪ SIGN UP ▪ HELP
Travel Search Engine
Home Flights Hotels Cars Cruises Vacation Packages
Welcome, Guest.
December 02, 2002
CORPORATE TRAVEL
TRAVEL TOOLS
Reservation status
Airport Information
Get deals via e-mail
Currency Converter
Driving Directions
Weather
Passport Information
CAR TOOLS
Car Rental Deals
Car Rental Directory
FIND YOUR CAR
Profile Creation/Update Progress
help
Step: 1 2 3 4 5 6
Enter the Financial Information
Annual Salary
$30,000 - $39,000
Financial Institution
Royal Bank Of Canada
Enter the Credit Card Information
Holder's First Name
Holder's Last Name
Card Type
Visa
Card Number
4521 - 4589 - 6958 - 1273
Expiry Date MM/YY
01
2002
Advaced Profile Creation
Clear
Next
Done
Internet

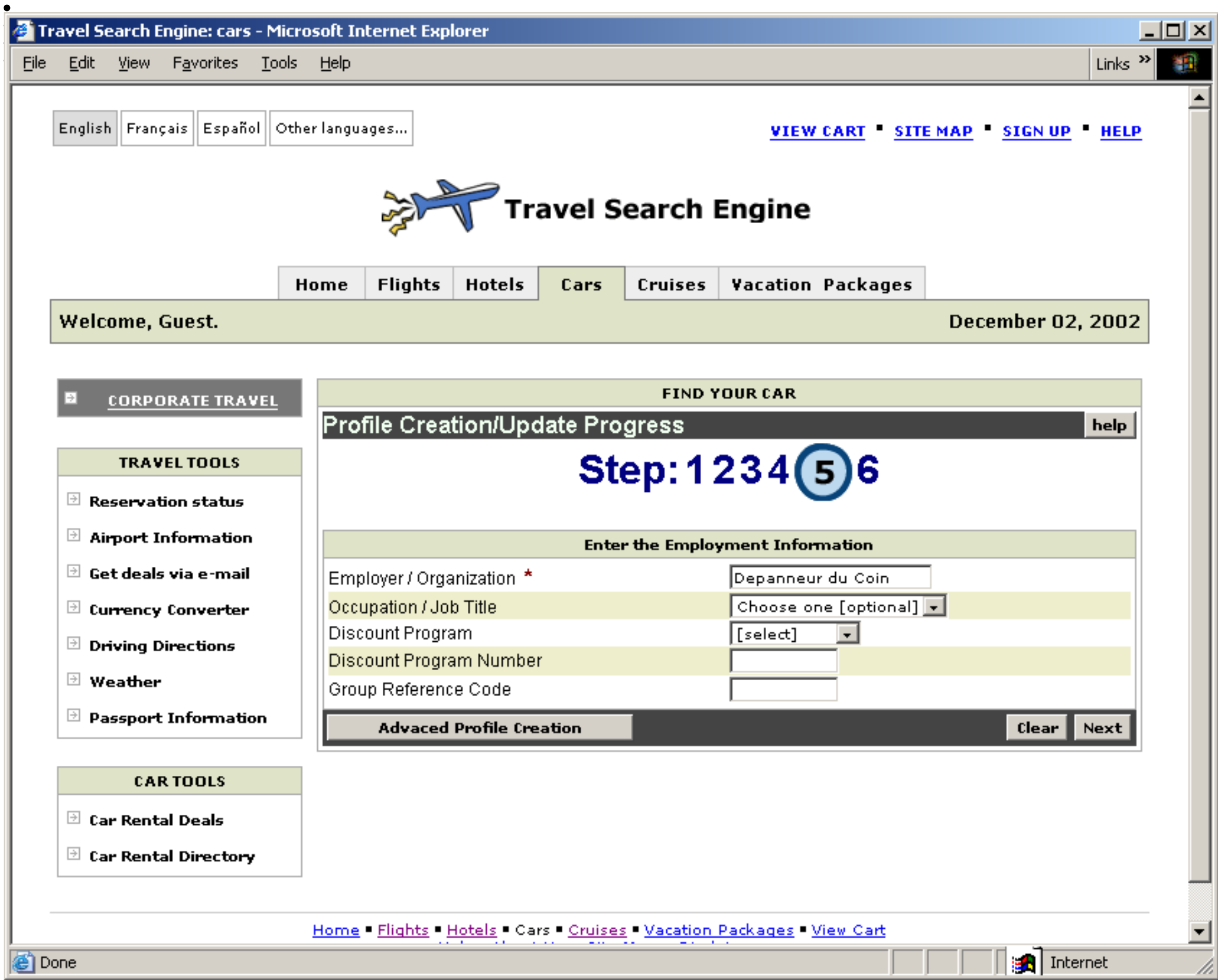
Travel Search Engine: cars - Microsoft Internet Explorer
File Edit View Favorites Tools Help
Links
English
Français
Español
Other languages...
VIEW CART
SITE MAP
SIGN UP
HELP
Travel Search Engine
Home
Flights
Hotels
Cars
Cruises
Vacation Packages
Welcome, Guest.
December 02, 2002
CORPORATE TRAVEL
TRAVEL TOOLS
Reservation status
Airport Information
Get deals via e-mail
Currency Converter
Driving Directions
Weather
Passport Information
CAR TOOLS
Car Rental Deals
Car Rental Directory
FIND YOUR CAR
Profile Creation/Update Progress
help
Step: 1 2 3 4 5 6
Enter the Employment Information
Employer / Organization *
Depanneur du Coin
Occupation / Job Title
Choose one [optional]
Discount Program
[select]
Discount Program Number
Group Reference Code
Advaced Profile Creation
Clear
Next
Home ▪ Flights ▪ Hotels ▪ Cars ▪ Cruises ▪ Vacation Packages ▪ View Cart
Done
Internet

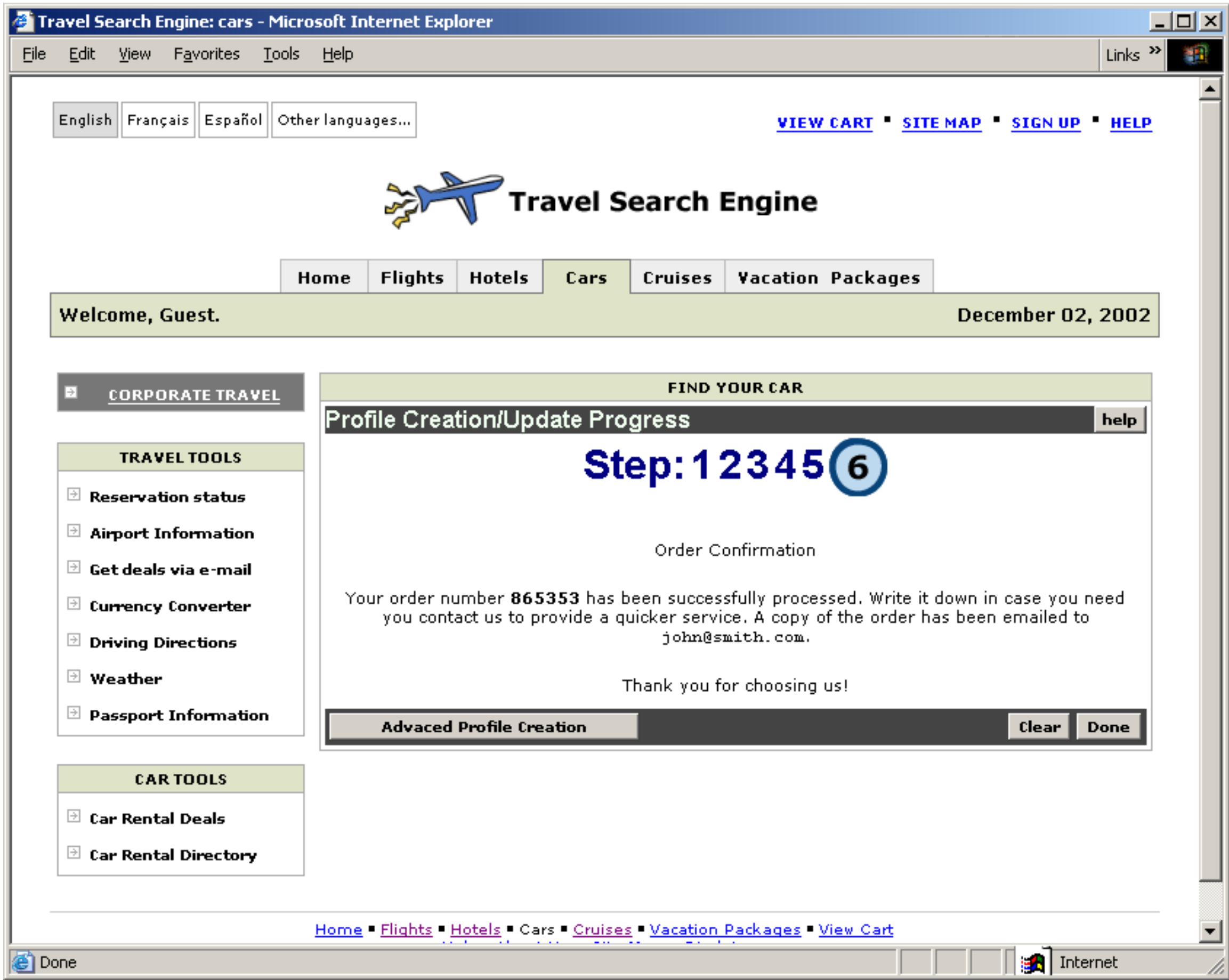
Travel Search Engine: cars - Microsoft Internet Explorer
File Edit View Favorites Tools Help
Links
English
Français
Español
Other languages...
VIEW CART
SITE MAP
SIGN UP
HELP
Travel Search Engine
Home
Flights
Hotels
Cars
Cruises
Vacation Packages
Welcome, Guest.
December 02, 2002
CORPORATE TRAVEL
TRAVEL TOOLS
Reservation status
Airport Information
Get deals via e-mail
Currency Converter
Driving Directions
Weather
Passport Information
CAR TOOLS
Car Rental Deals
Car Rental Directory
FIND YOUR CAR
Profile Creation/Update Progress
help
Step: 1 2 3 4 5 6
Order Confirmation
Your order number 865353 has been successfully processed. Write it down in case you need you contact us to provide a quicker service. A copy of the order has been emailed to john@smith.com.
Thank you for choosing us!
Advaced Profile Creation
Clear
Done
Home ▪ Flights ▪ Hotels ▪ Cars ▪ Cruises ▪ Vacation Packages ▪ View Cart
Done
Internet

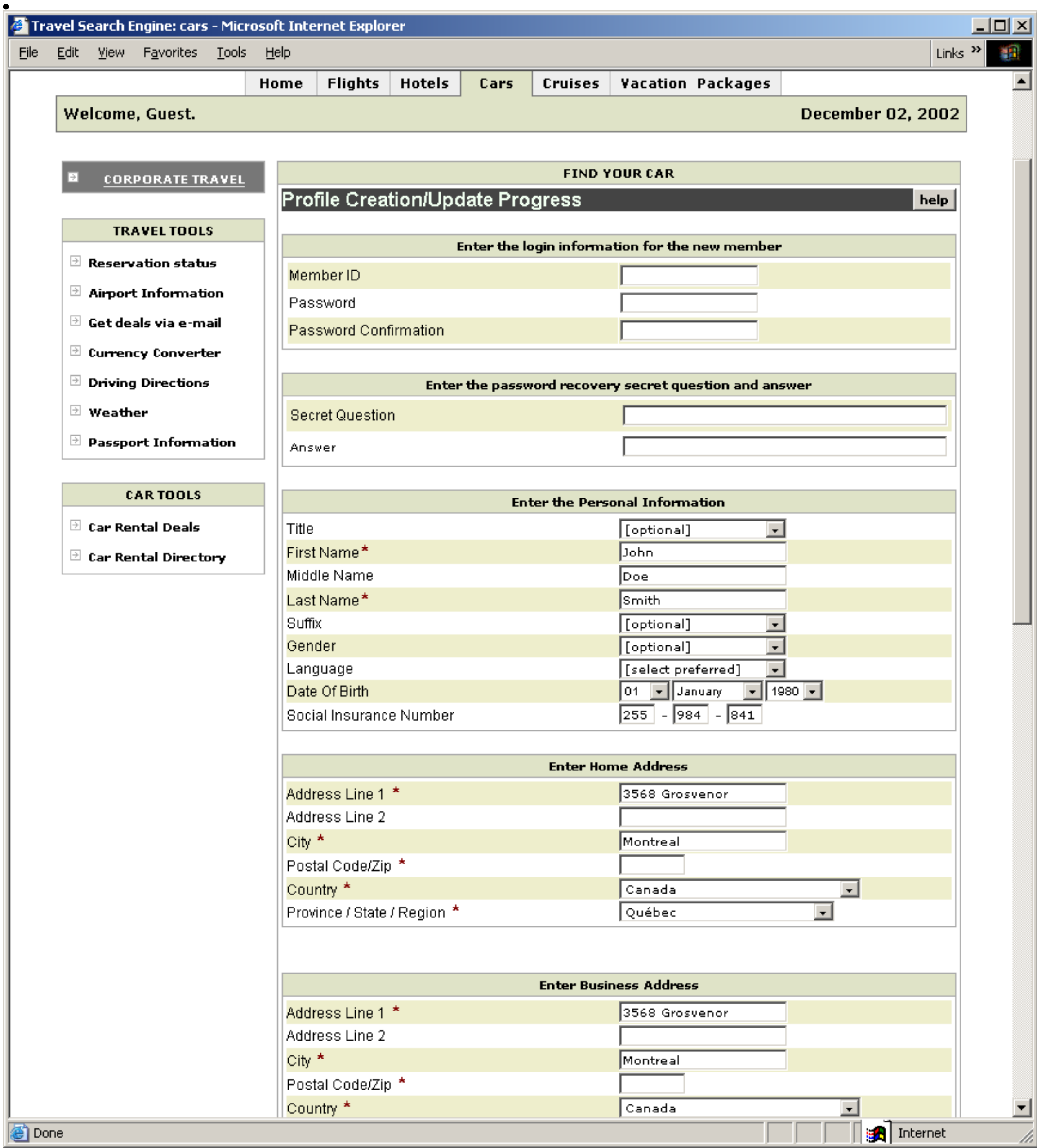
Travel Search Engine: cars - Microsoft Internet Explorer
File Edit View Favorites Tools Help
Links
Home Flights Hotels Cars Cruises Vacation Packages
Welcome, Guest.
December 02, 2002
CORPORATE TRAVEL
TRAVEL TOOLS
Reservation status
Airport Information
Get deals via e-mail
Currency Converter
Driving Directions
Weather
Passport Information
CAR TOOLS
Car Rental Deals
Car Rental Directory
FIND YOUR CAR
Profile Creation/Update Progress
help
Enter the login information for the new member
Member ID
Password
Password Confirmation
Enter the password recovery secret question and answer
Secret Question
Answer
Enter the Personal Information
Title
[optional]
First Name*
John
Middle Name
Doe
Last Name*
Smith
Suffix
[optional]
Gender
[optional]
Language
[select preferred]
Date Of Birth
01
January
1980
Social Insurance Number
255 - 984 - 841
Enter Home Address
Address Line 1 *
3568 Grosvenor
Address Line 2
City *
Montreal
Postal Code/Zip *
Country *
Canada
Province / State / Region *
Québec
Enter Business Address
Address Line 1 *
3568 Grosvenor
Address Line 2
City *
Montreal
Postal Code/Zip *
Country *
Canada
Done
Internet

**Corporate Travel: Associations**

This type of traveling is for corporation or an association meetings and conventions. This in the application of a person on behalf of the person's company for 10 or more people traveling from a common destination to a common destination or from different parts of North America or otherwise to a common destination. A user is usually asked to filling a detailed form at the end after having chosen a package.

The meeting attendees will receive special savings and bonus discounts on airfares and cars and other ways for your association to save money.

The user begins with a preliminary form with various topics, discount programs they might know of and special requirements. Then there is a search performed for travel packages. The user chooses one and applies. Then there is that long on-line form with various pieced of information, including personal info of the person making application on behalf of the company, travel agency (if any) the company is associated with and the ways of delivery of contract. This form also presents ways to schedule several meetings at once on different dates and places. The user can add more if required.

Next three screens present the Preliminary Form for Association, where user has to make their choice and click on "Search", followed by the list of search results of packages, and finally the application form.

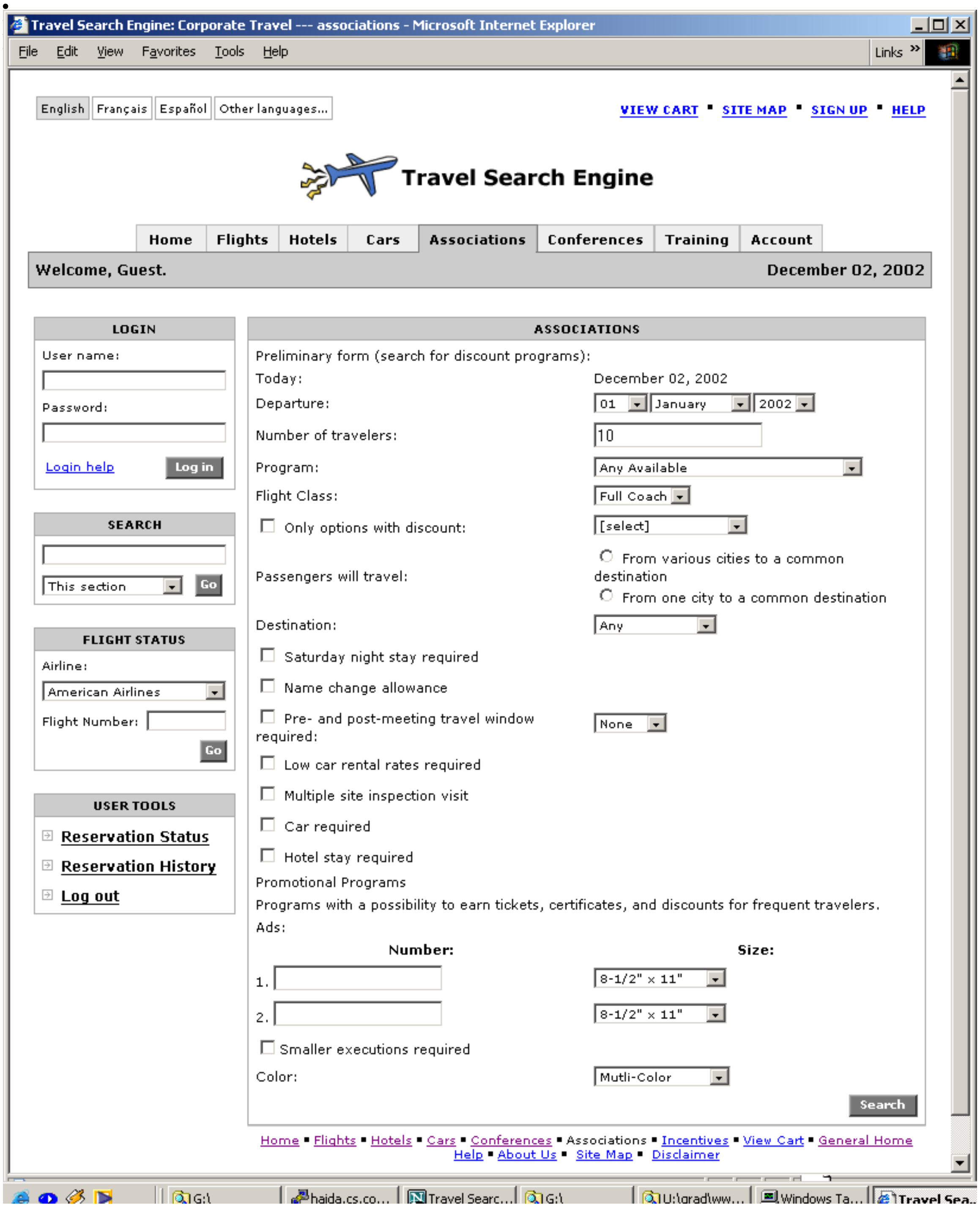
Travel Search Engine: Corporate Travel --- associations - Microsoft Internet Explorer
File Edit View Favorites Tools Help
Links
English Français Español Other languages...
VIEW CART ▪ SITE MAP ▪ SIGN UP ▪ HELP
Travel Search Engine
Home Flights Hotels Cars Associations Conferences Training Account
Welcome, Guest.
December 02, 2002
LOGIN
User name:
Password:
Login help
Log in
SEARCH
This section
Go
FLIGHT STATUS
Airline:
American Airlines
Flight Number:
Go
USER TOOLS
Reservation Status
Reservation History
Log out
ASSOCIATIONS
Preliminary form (search for discount programs):
Today: December 02, 2002
Departure: 01 January 2002
Number of travelers: 10
Program: Any Available
Flight Class: Full Coach
Only options with discount: [select]
Passengers will travel:
From various cities to a common destination
From one city to a common destination
Destination: Any
Saturday night stay required
Name change allowance
Pre- and post-meeting travel window required: None
Low car rental rates required
Multiple site inspection visit
Car required
Hotel stay required
Promotional Programs
Programs with a possibility to earn tickets, certificates, and discounts for frequent travelers.
Ads:
Number:
Size:
1. 8-1/2" x 11"
2. 8-1/2" x 11"
Smaller executions required
Color: Mutli-Color
Search
Home ▪ Flights ▪ Hotels ▪ Cars ▪ Conferences ▪ Associations ▪ Incentives ▪ View Cart ▪ General Home
Help ▪ About Us ▪ Site Map ▪ Disclaimer
G:\
haida.cs.co...
Travel Searc...
G:\
U:\grad\ww...
Windows Ta...
Travel Sea...

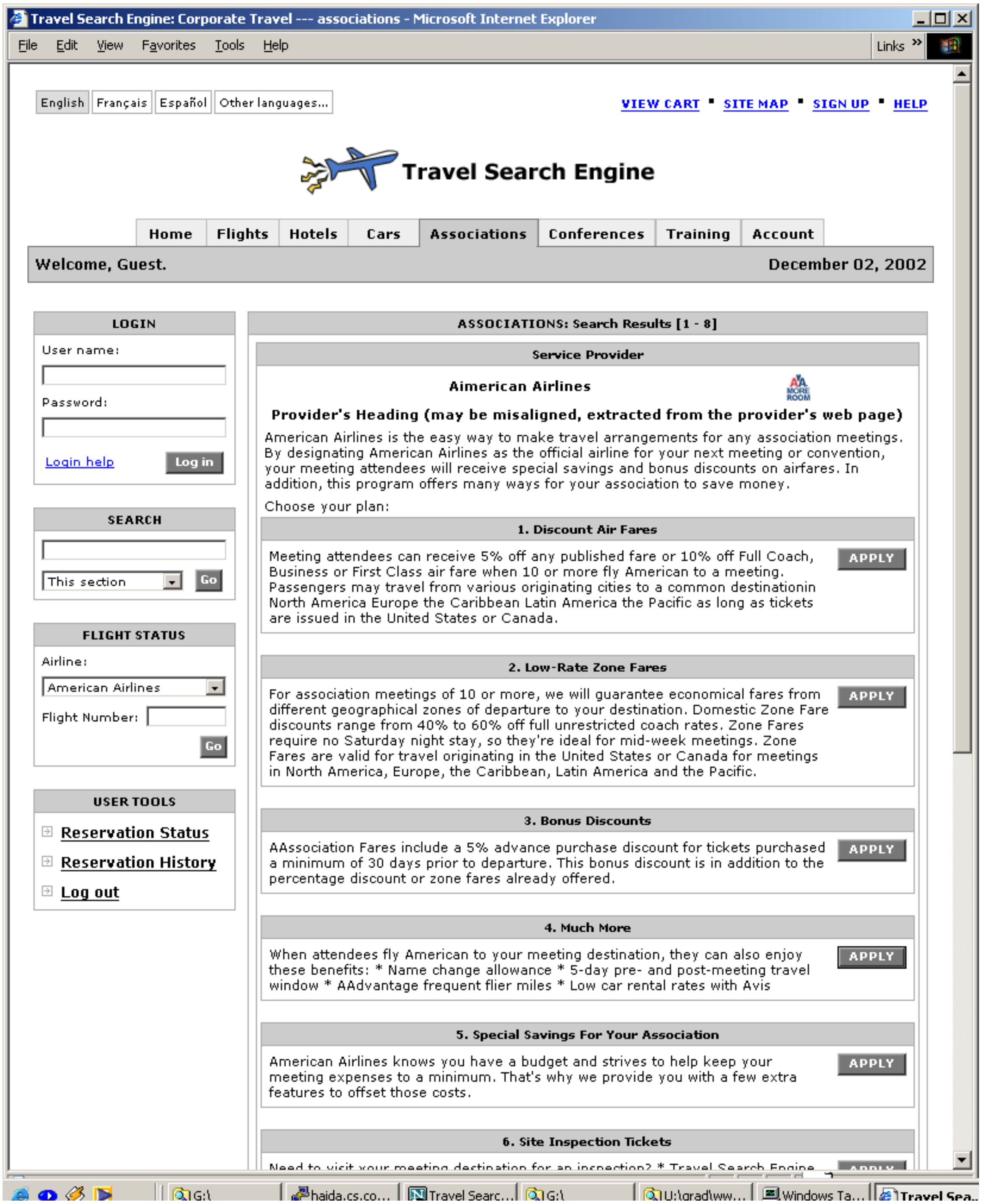
Travel Search Engine: Corporate Travel --- associations - Microsoft Internet Explorer
File Edit View Favorites Tools Help
Links
English Français Español Other languages...
VIEW CART ▪ SITE MAP ▪ SIGN UP ▪ HELP
Travel Search Engine
Home Flights Hotels Cars Associations Conferences Training Account
Welcome, Guest.
December 02, 2002
LOGIN
User name:
Password:
Login help
Log in
SEARCH
This section
Go
FLIGHT STATUS
Airline:
American Airlines
Flight Number:
Go
USER TOOLS
Reservation Status
Reservation History
Log out
ASSOCIATIONS: Search Results [1 - 8]
Service Provider
Aimerican Airlines
Provider's Heading (may be misaligned, extracted from the provider's web page)
American Airlines is the easy way to make travel arrangements for any association meetings. By designating American Airlines as the official airline for your next meeting or convention, your meeting attendees will receive special savings and bonus discounts on airfares. In addition, this program offers many ways for your association to save money.
Choose your plan:
1. Discount Air Fares
Meeting attendees can receive 5% off any published fare or 10% off Full Coach, Business or First Class air fare when 10 or more fly American to a meeting. Passengers may travel from various originating cities to a common destinationin North America Europe the Caribbean Latin America the Pacific as long as tickets are issued in the United States or Canada.
APPLY
2. Low-Rate Zone Fares
For association meetings of 10 or more, we will guarantee economical fares from different geographical zones of departure to your destination. Domestic Zone Fare discounts range from 40% to 60% off full unrestricted coach rates. Zone Fares require no Saturday night stay, so they're ideal for mid-week meetings. Zone Fares are valid for travel originating in the United States or Canada for meetings in North America, Europe, the Caribbean, Latin America and the Pacific.
APPLY
3. Bonus Discounts
AAssociation Fares include a 5% advance purchase discount for tickets purchased a minimum of 30 days prior to departure. This bonus discount is in addition to the percentage discount or zone fares already offered.
APPLY
4. Much More
When attendees fly American to your meeting destination, they can also enjoy these benefits: * Name change allowance * 5-day pre- and post-meeting travel window * AAdvantage frequent flier miles * Low car rental rates with Avis
APPLY
5. Special Savings For Your Association
American Airlines knows you have a budget and strives to help keep your meeting expenses to a minimum. That's why we provide you with a few extra features to offset those costs.
APPLY
6. Site Inspection Tickets
G:\ haida.cs.co... Travel Searc... G:\ U:\grad\ww... Windows Ta... Travel Sea...

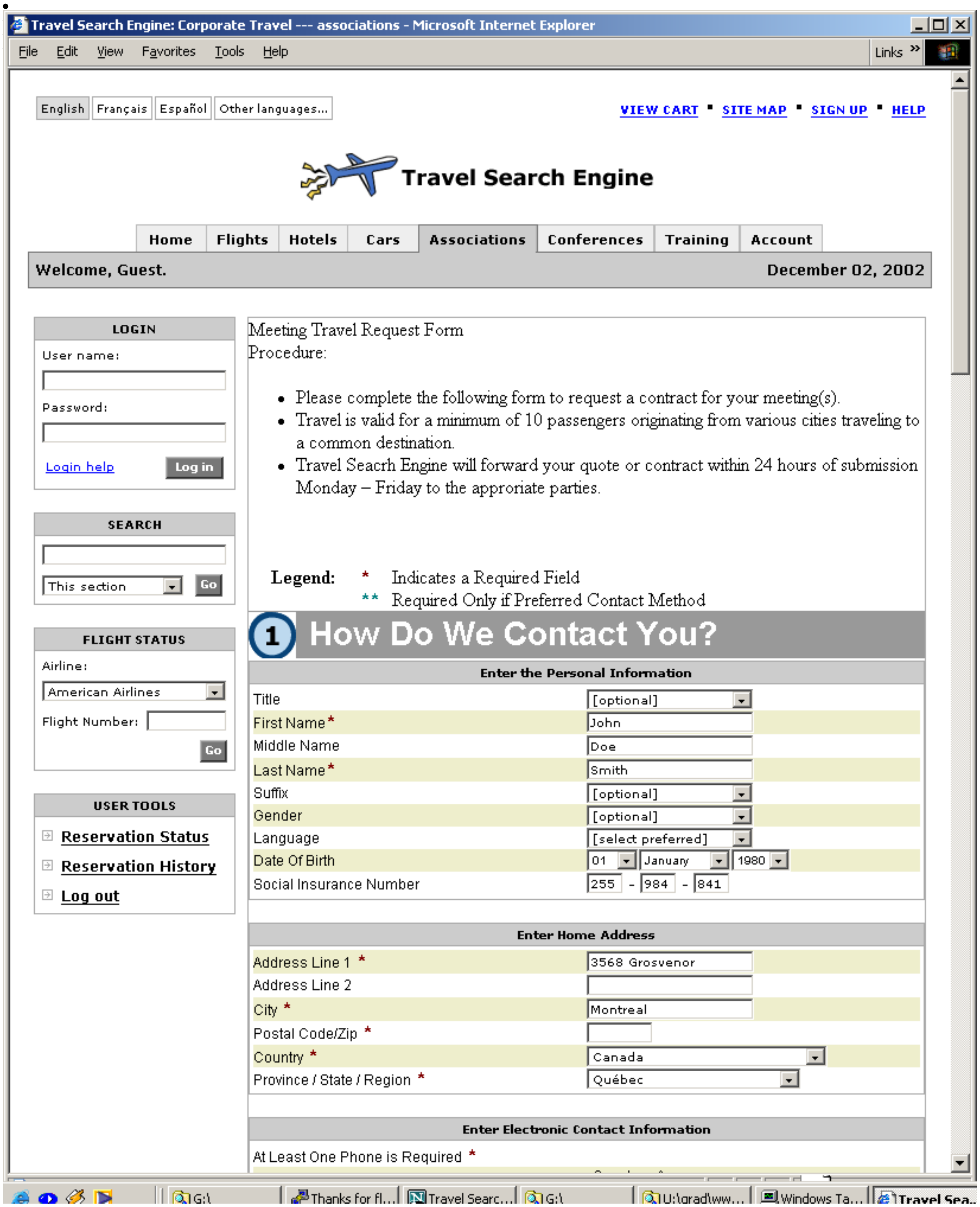
Travel Search Engine: Corporate Travel --- associations - Microsoft Internet Explorer
File Edit View Favorites Tools Help
Links
English Français Español Other languages...
VIEW CART ▪ SITE MAP ▪ SIGN UP ▪ HELP
Travel Search Engine
Home Flights Hotels Cars Associations Conferences Training Account
Welcome, Guest.
December 02, 2002
LOGIN
User name:
Password:
Login help
Log in
SEARCH
This section
Go
FLIGHT STATUS
Airline:
American Airlines
Flight Number:
Go
USER TOOLS
Reservation Status
Reservation History
Log out
Meeting Travel Request Form
Procedure:
Please complete the following form to request a contract for your meeting(s).
Travel is valid for a minimum of 10 passengers originating from various cities traveling to a common destination.
Travel Seacrh Engine will forward your quote or contract within 24 hours of submission Monday – Friday to the approriate parties.
Legend: * Indicates a Required Field
** Required Only if Preferred Contact Method
1 How Do We Contact You?
Enter the Personal Information
Title [optional]
First Name* John
Middle Name Doe
Last Name* Smith
Suffix [optional]
Gender [optional]
Language [select preferred]
Date Of Birth 01 January 1980
Social Insurance Number 255 - 984 - 841
Enter Home Address
Address Line 1 * 3568 Grosvenor
Address Line 2
City * Montreal
Postal Code/Zip *
Country * Canada
Province / State / Region * Québec
Enter Electronic Contact Information
At Least One Phone is Required *
G:\
Thanks for fl...
Travel Searc...
G:\
U:\grad\ww...
Windows Ta...
Travel Sea...

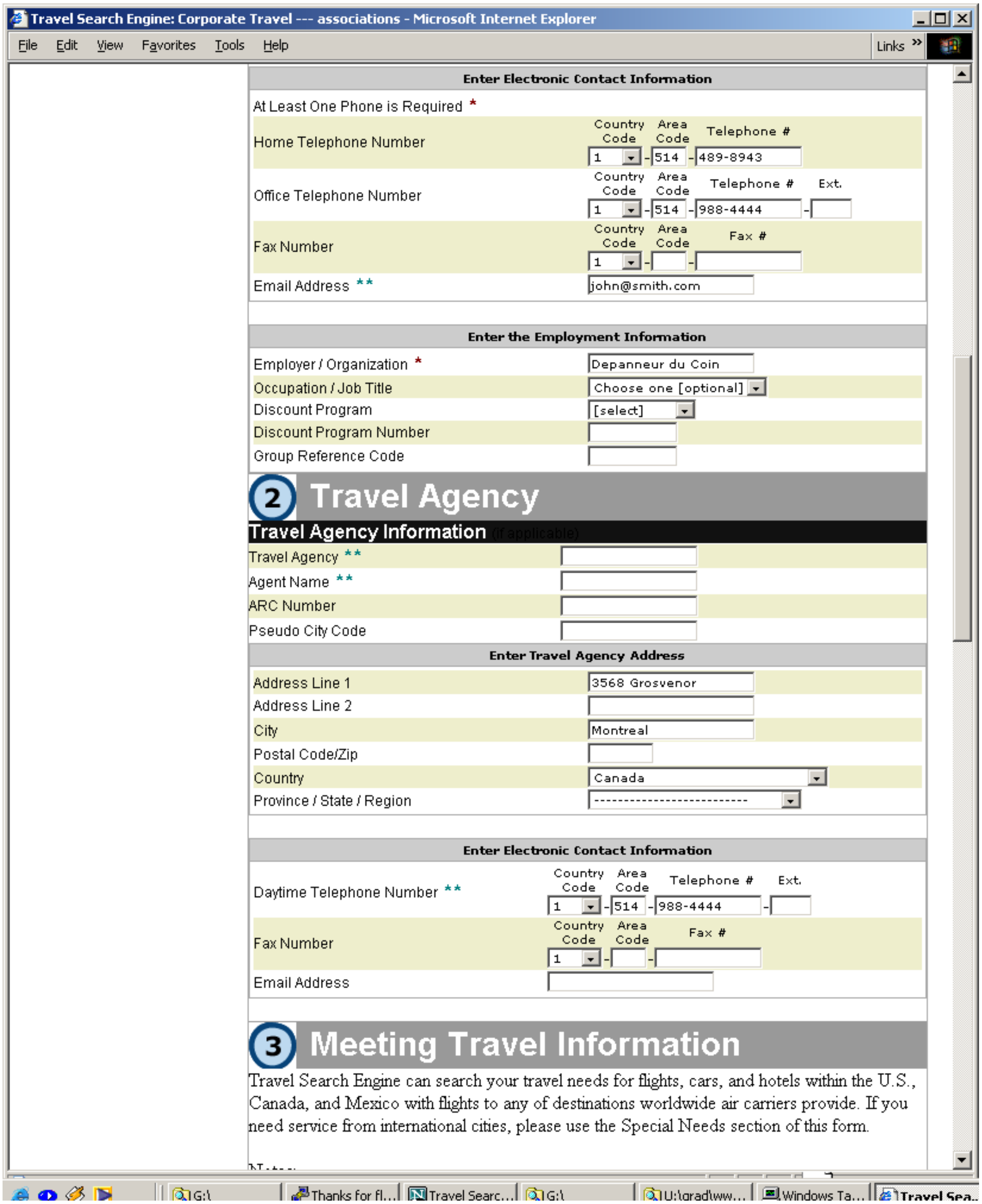

Travel Search Engine: Corporate Travel --- associations - Microsoft Internet Explorer

File Edit View Favorites Tools Help Links

**Enter Electronic Contact Information**

At Least One Phone is Required *

| | Country Code | Area Code | Telephone # | Ext. |
|---|---|---|---|---|
| Home Telephone Number | 1 | 514 | 489-8943 | |
| Office Telephone Number | 1 | 514 | 988-4444 | |
| Fax Number | 1 | | | |
| Email Address ** | john@smith.com | | | |

**Enter the Employment Information**

| | |
|---|---|
| Employer / Organization * | Depanneur du Coin |
| Occupation / Job Title | Choose one [optional] |
| Discount Program | [select] |
| Discount Program Number | |
| Group Reference Code | |

## 2 Travel Agency

**Travel Agency Information** (if applicable)

| | |
|---|---|
| Travel Agency ** | |
| Agent Name ** | |
| ARC Number | |
| Pseudo City Code | |

**Enter Travel Agency Address**

| | |
|---|---|
| Address Line 1 | 3568 Grosvenor |
| Address Line 2 | |
| City | Montreal |
| Postal Code/Zip | |
| Country | Canada |
| Province / State / Region | ------------------------- |

**Enter Electronic Contact Information**

| | Country Code | Area Code | Telephone # | Ext. |
|---|---|---|---|---|
| Daytime Telephone Number ** | 1 | 514 | 988-4444 | |
| Fax Number | 1 | | | |
| Email Address | | | | |

## 3 Meeting Travel Information

Travel Search Engine can search your travel needs for flights, cars, and hotels within the U.S., Canada, and Mexico with flights to any of destinations worldwide air carriers provide. If you need service from international cities, please use the Special Needs section of this form.

G:\ Thanks for fl... Travel Searc... G:\ U:\grad\ww... Windows Ta... Travel Sea...

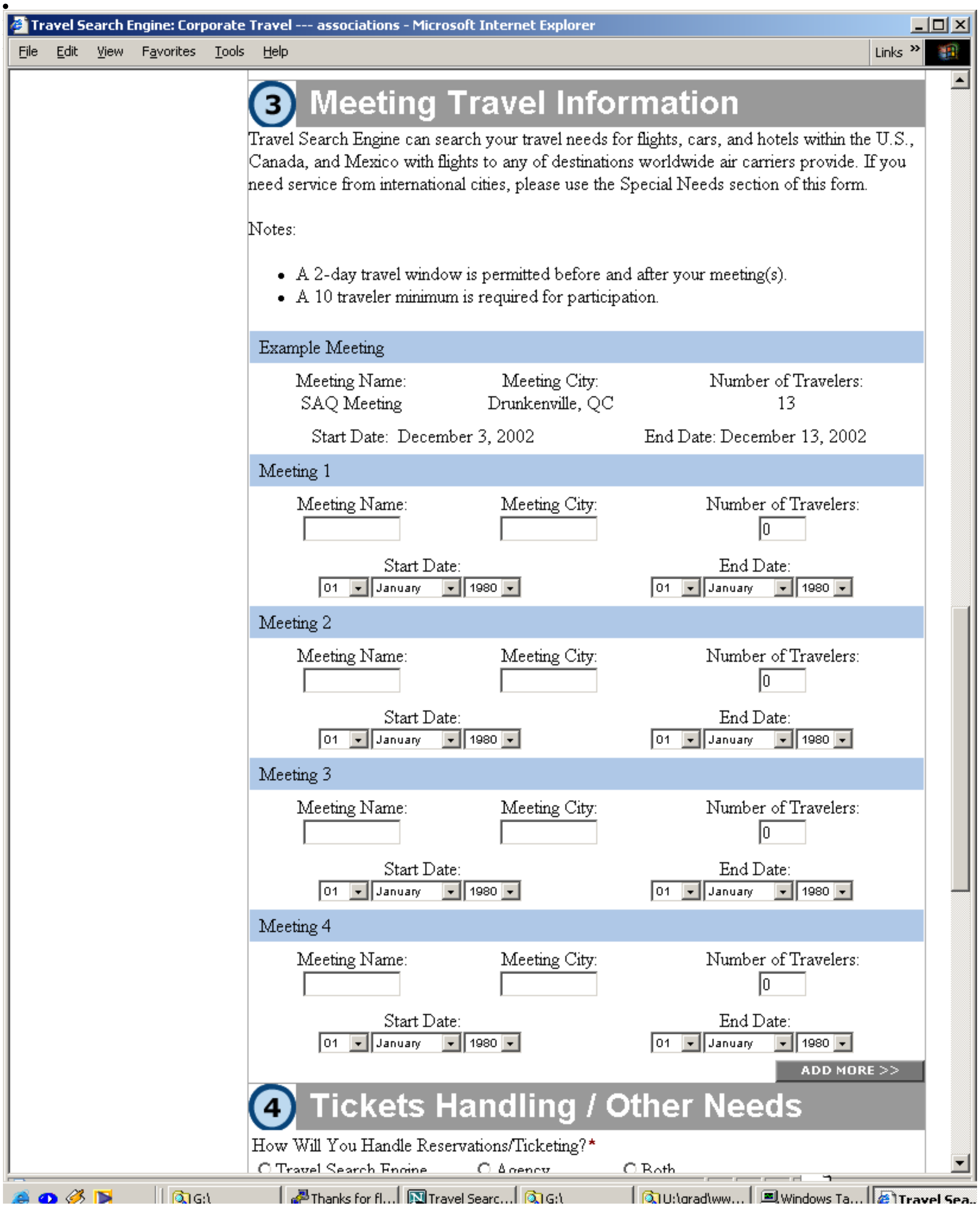
Travel Search Engine: Corporate Travel --- associations - Microsoft Internet Explorer

File Edit View Favorites Tools Help Links »

## 3 Meeting Travel Information

Travel Search Engine can search your travel needs for flights, cars, and hotels within the U.S., Canada, and Mexico with flights to any of destinations worldwide air carriers provide. If you need service from international cities, please use the Special Needs section of this form.

Notes:

- A 2-day travel window is permitted before and after your meeting(s).
- A 10 traveler minimum is required for participation.

Example Meeting

| Meeting Name: | Meeting City: | Number of Travelers: |
|---|---|---|
| SAQ Meeting | Drunkenville, QC | 13 |

Start Date: December 3, 2002 End Date: December 13, 2002

Meeting 1

| Meeting Name: | Meeting City: | Number of Travelers: |
|---|---|---|
| | | 0 |

Start Date: 01 January 1980 End Date: 01 January 1980

Meeting 2

| Meeting Name: | Meeting City: | Number of Travelers: |
|---|---|---|
| | | 0 |

Start Date: 01 January 1980 End Date: 01 January 1980

Meeting 3

| Meeting Name: | Meeting City: | Number of Travelers: |
|---|---|---|
| | | 0 |

Start Date: 01 January 1980 End Date: 01 January 1980

Meeting 4

| Meeting Name: | Meeting City: | Number of Travelers: |
|---|---|---|
| | | 0 |

Start Date: 01 January 1980 End Date: 01 January 1980

ADD MORE >>

## 4 Tickets Handling / Other Needs

How Will You Handle Reservations/Ticketing?*

○ Travel Search Engine ○ Agency ○ Both

G:\ | Thanks for fl... | Travel Searc... | G:\ | U:\grad\ww... | Windows Ta... | Travel Sea...

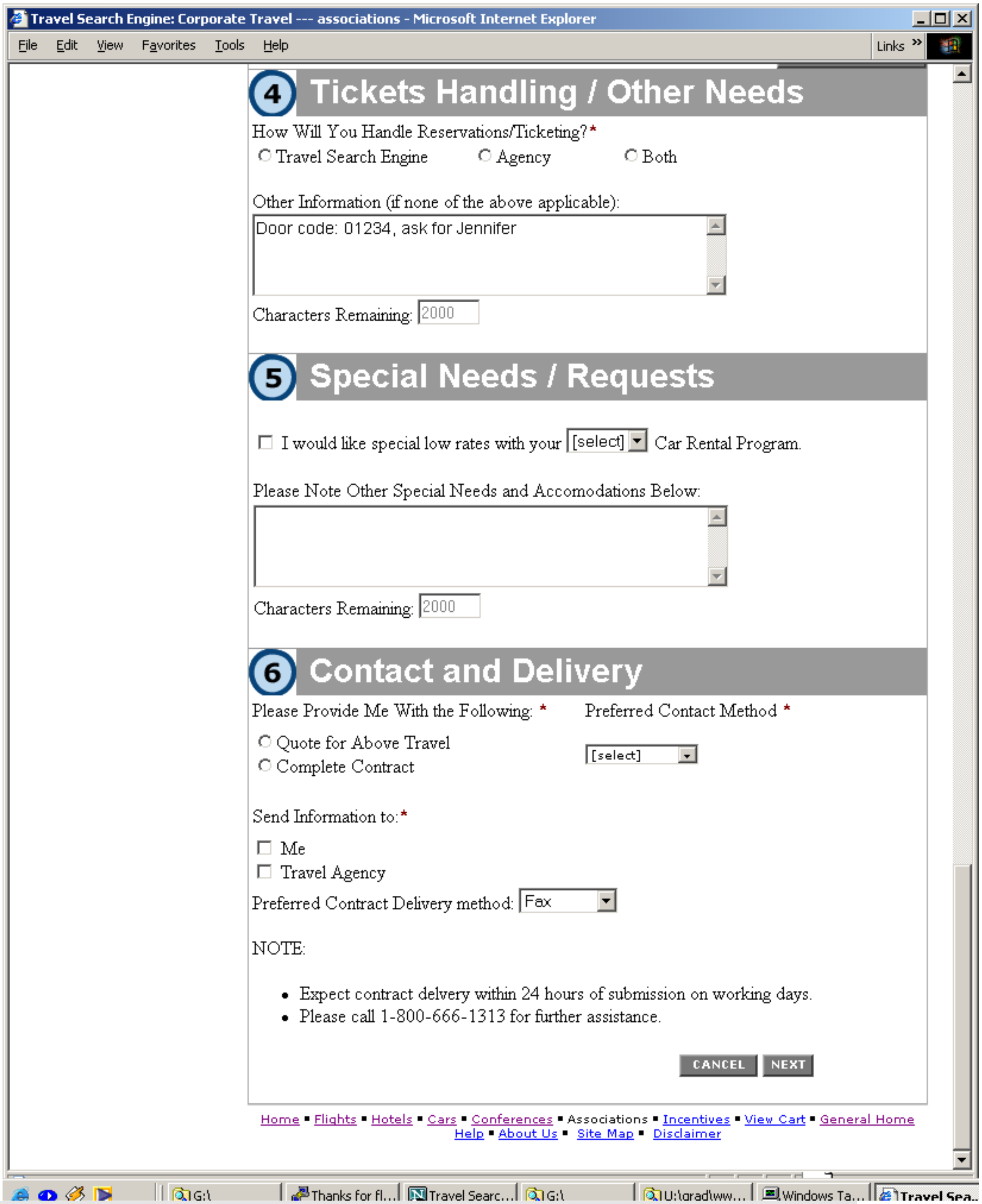

Travel Search Engine: Corporate Travel --- associations - Microsoft Internet Explorer
File Edit View Favorites Tools Help
Links
4 Tickets Handling / Other Needs
How Will You Handle Reservations/Ticketing?*
Travel Search Engine
Agency
Both
Other Information (if none of the above applicable):
Door code: 01234, ask for Jennifer
Characters Remaining: 2000
5 Special Needs / Requests
I would like special low rates with your [select] Car Rental Program.
Please Note Other Special Needs and Accomodations Below:
Characters Remaining: 2000
6 Contact and Delivery
Please Provide Me With the Following: *
Preferred Contact Method *
Quote for Above Travel
Complete Contract
[select]
Send Information to:*
Me
Travel Agency
Preferred Contract Delivery method: Fax
NOTE:
Expect contract delivery within 24 hours of submission on working days.
Please call 1-800-666-1313 for further assistance.
CANCEL
NEXT
Home ▪ Flights ▪ Hotels ▪ Cars ▪ Conferences ▪ Associations ▪ Incentives ▪ View Cart ▪ General Home
Help ▪ About Us ▪ Site Map ▪ Disclaimer
G:\
Thanks for fl...
Travel Searc...
G:\
U:\grad\ww...
Windows Ta...
Travel Sea...

## Scenarios

### **Reserve a cruise**:

- Reserve a Canada/New England
- 1 week
- $999.99 per person
- Carnival line
- Carnival Destiny ship
- Three stars rating
- Departs from New York on November 10
- Sort by price
- Two adults, one senior
- Also reserve flight tickets

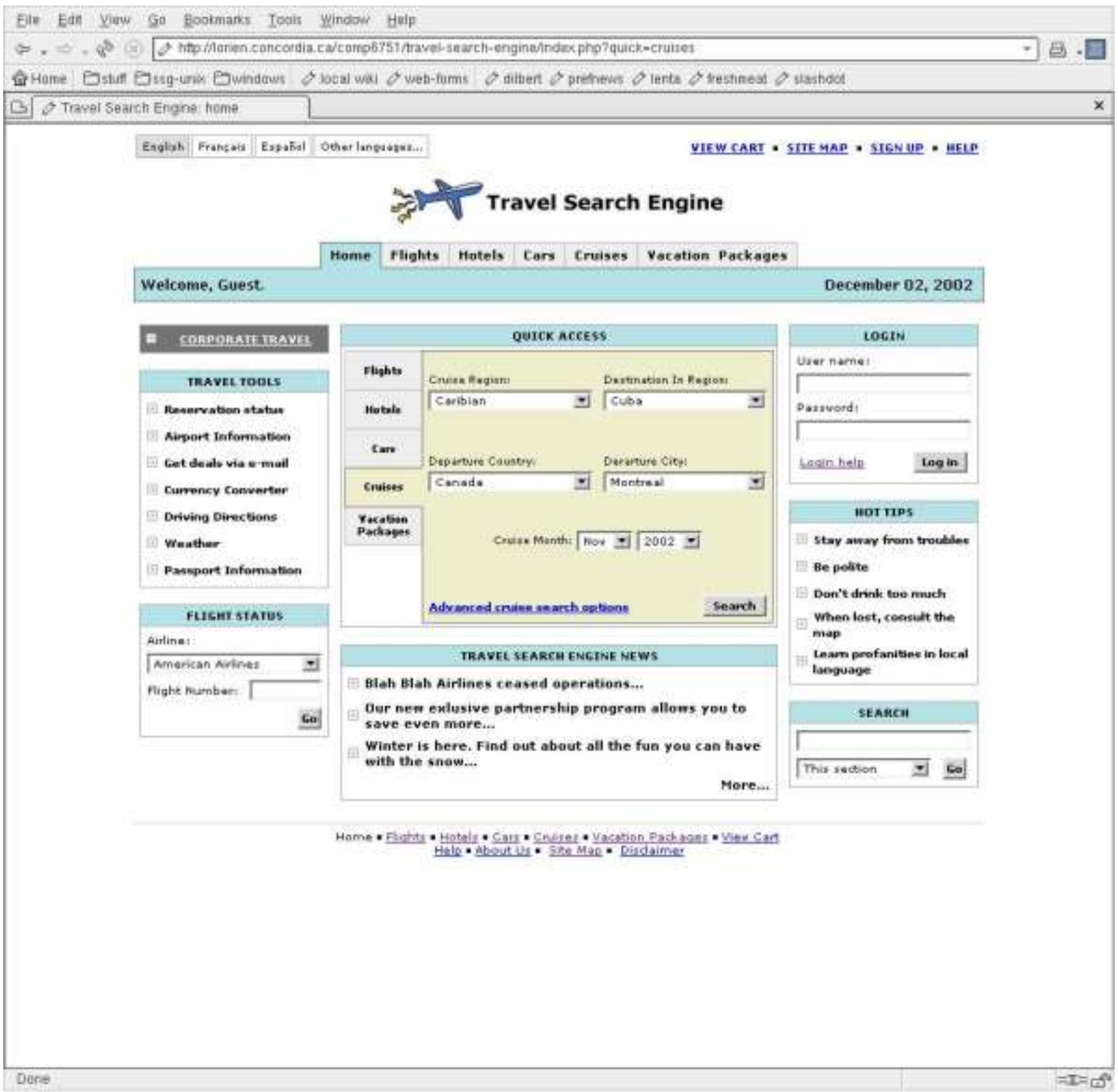

English | Français | Español | Other languages...

VIEW CART ▪ SITE MAP ▪ SIGN UP ▪ HELP

# Travel Search Engine

Home | Flights | Hotels | Cars | Cruises | Vacation Packages

Welcome, Guest.

December 02, 2002

CORPORATE TRAVEL

**TRAVEL TOOLS**

- Reservation status
- Airport Information
- Get deals via e-mail
- Currency Converter
- Driving Directions
- Weather
- Passport Information

**CRUISE TOOLS**

- Cruise Specials
- Cruise Lines
- Cruise Destinations
- Ship/Port Search

**LOGIN**

User name:

Password:

Login help | Log in

**FIND YOUR CRUISE**

**Where and when do you want to travel?**

Cruise region: Caribbean

Destination within region: Cuba

Cruise departure port: Any port

Departure country: Canada

Derarture city: Montreal

Cruise departure date: November 1 2002

Cruise duration: Any duration

**Who is going?**

Adults: 2

Children: 1

Seniors: 0

**Do you have ship preferences?**

Cruise line: Show all lines

Ship: Show all ships

Ship rating: Show all ratings

**Do you have any additional search preferences?**

Price range (per person): Show all prices

Sort results by: Price

Start over | Search

Home ▪ Flights ▪ Hotels ▪ Cars ▪ Cruises ▪ Vacation Packages ▪ View Cart
Help ▪ About Us ▪ Site Map ▪ Disclaimer

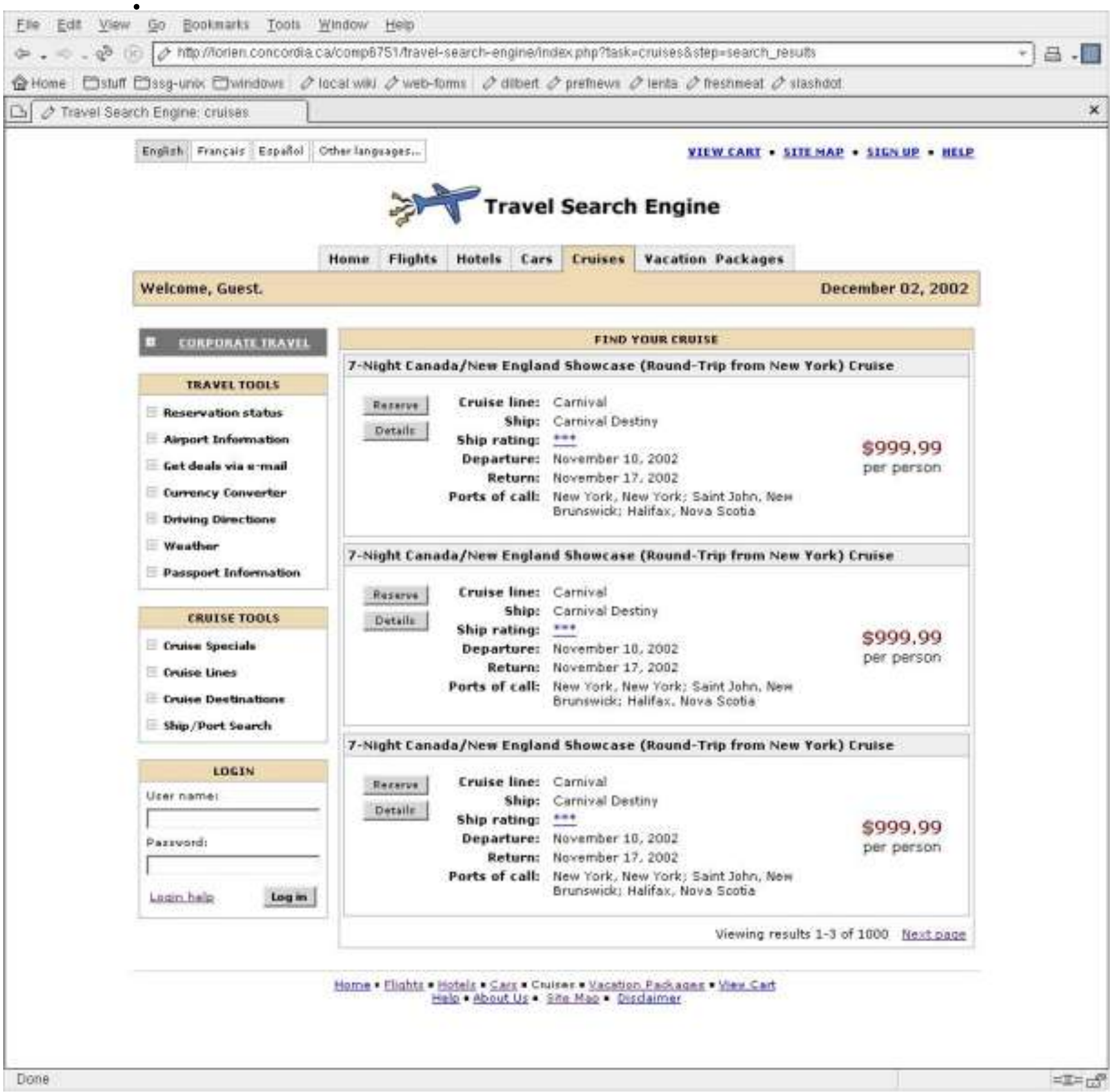

File Edit View Go Bookmarks Tools Window Help
http://lorien.concordia.ca/comp6751/travel-search-engine/index.php?task=cruises&step=search_results
Travel Search Engine: cruises
English Français Español Other languages...
VIEW CART • SITE MAP • SIGN UP • HELP
Travel Search Engine
Home Flights Hotels Cars Cruises Vacation Packages
Welcome, Guest.
December 02, 2002
CORPORATE TRAVEL
TRAVEL TOOLS
Reservation status
Airport Information
Get deals via e-mail
Currency Converter
Driving Directions
Weather
Passport Information
CRUISE TOOLS
Cruise Specials
Cruise Lines
Cruise Destinations
Ship/Port Search
LOGIN
User name:
Password:
Login help
Log in
FIND YOUR CRUISE
7-Night Canada/New England Showcase (Round-Trip from New York) Cruise
Reserve
Details
Cruise line: Carnival
Ship: Carnival Destiny
Ship rating: ***
Departure: November 10, 2002
Return: November 17, 2002
Ports of call: New York, New York; Saint John, New Brunswick; Halifax, Nova Scotia
$999.99 per person
7-Night Canada/New England Showcase (Round-Trip from New York) Cruise
Reserve
Details
Cruise line: Carnival
Ship: Carnival Destiny
Ship rating: ***
Departure: November 10, 2002
Return: November 17, 2002
Ports of call: New York, New York; Saint John, New Brunswick; Halifax, Nova Scotia
$999.99 per person
7-Night Canada/New England Showcase (Round-Trip from New York) Cruise
Reserve
Details
Cruise line: Carnival
Ship: Carnival Destiny
Ship rating: ***
Departure: November 10, 2002
Return: November 17, 2002
Ports of call: New York, New York; Saint John, New Brunswick; Halifax, Nova Scotia
$999.99 per person
Viewing results 1-3 of 1000 Next page
Home • Flights • Hotels • Cars • Cruises • Vacation Packages • View Cart
Help • About Us • Site Map • Disclaimer
Done

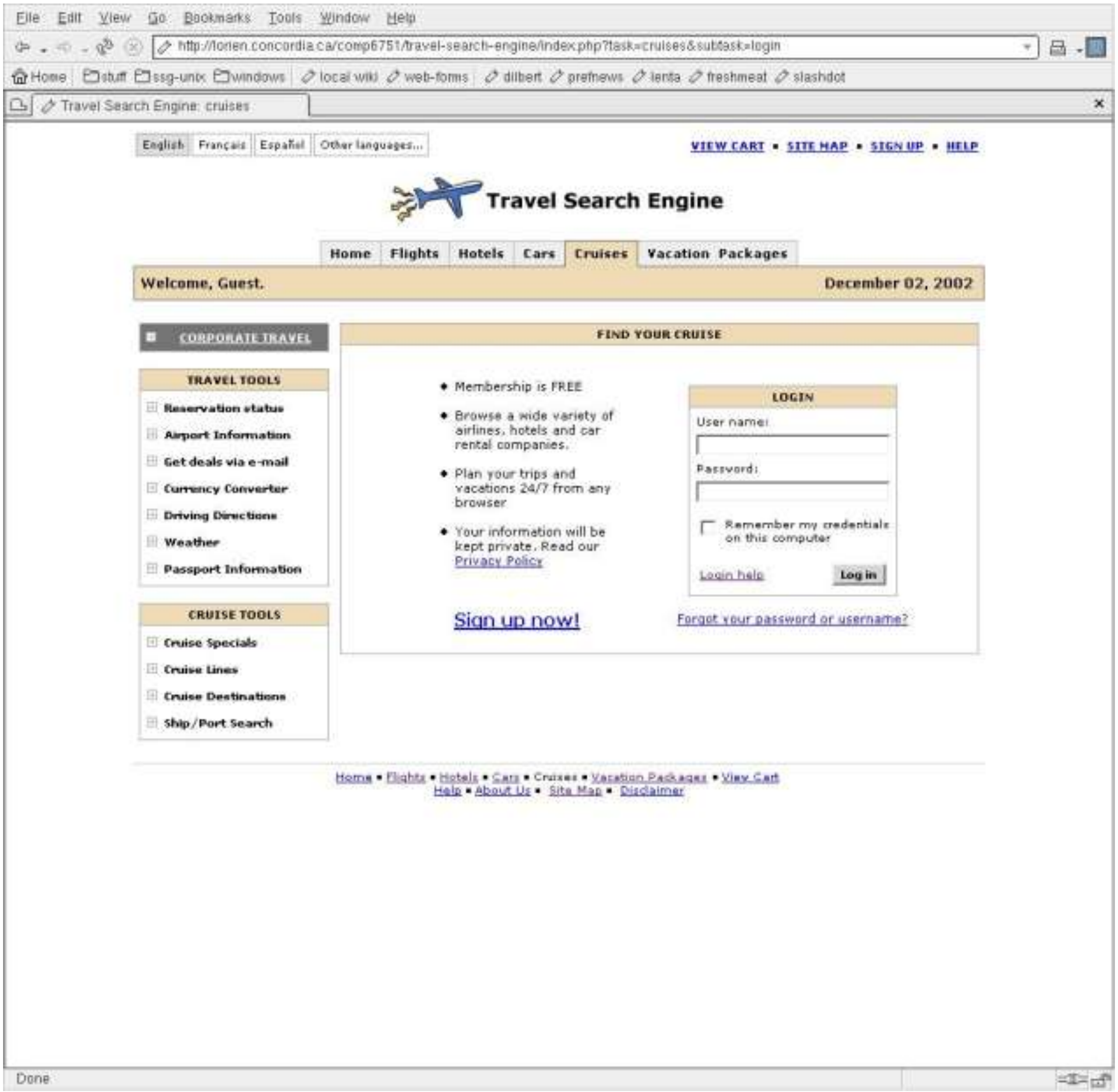
File Edit View Go Bookmarks Tools Window Help
http://lorien.concordia.ca/comp6751/travel-search-engine/index.php?task=cruises&subtask=login
Home stuff ssg-unix windows local wiki web-forms dilbert prefnews lenta freshmeat slashdot
Travel Search Engine: cruises
English Français Español Other languages...
VIEW CART • SITE MAP • SIGN UP • HELP
Travel Search Engine
Home Flights Hotels Cars Cruises Vacation Packages
Welcome, Guest.
December 02, 2002
CORPORATE TRAVEL
TRAVEL TOOLS
Reservation status
Airport Information
Get deals via e-mail
Currency Converter
Driving Directions
Weather
Passport Information
CRUISE TOOLS
Cruise Specials
Cruise Lines
Cruise Destinations
Ship/Port Search
FIND YOUR CRUISE
Membership is FREE
Browse a wide variety of airlines, hotels and car rental companies.
Plan your trips and vacations 24/7 from any browser
Your information will be kept private. Read our Privacy Policy
LOGIN
User name:
Password:
Remember my credentials on this computer
Login help
Log in
Sign up now!
Forgot your password or username?
Home • Flights • Hotels • Cars • Cruises • Vacation Packages • View Cart
Help • About Us • Site Map • Disclaimer
Done

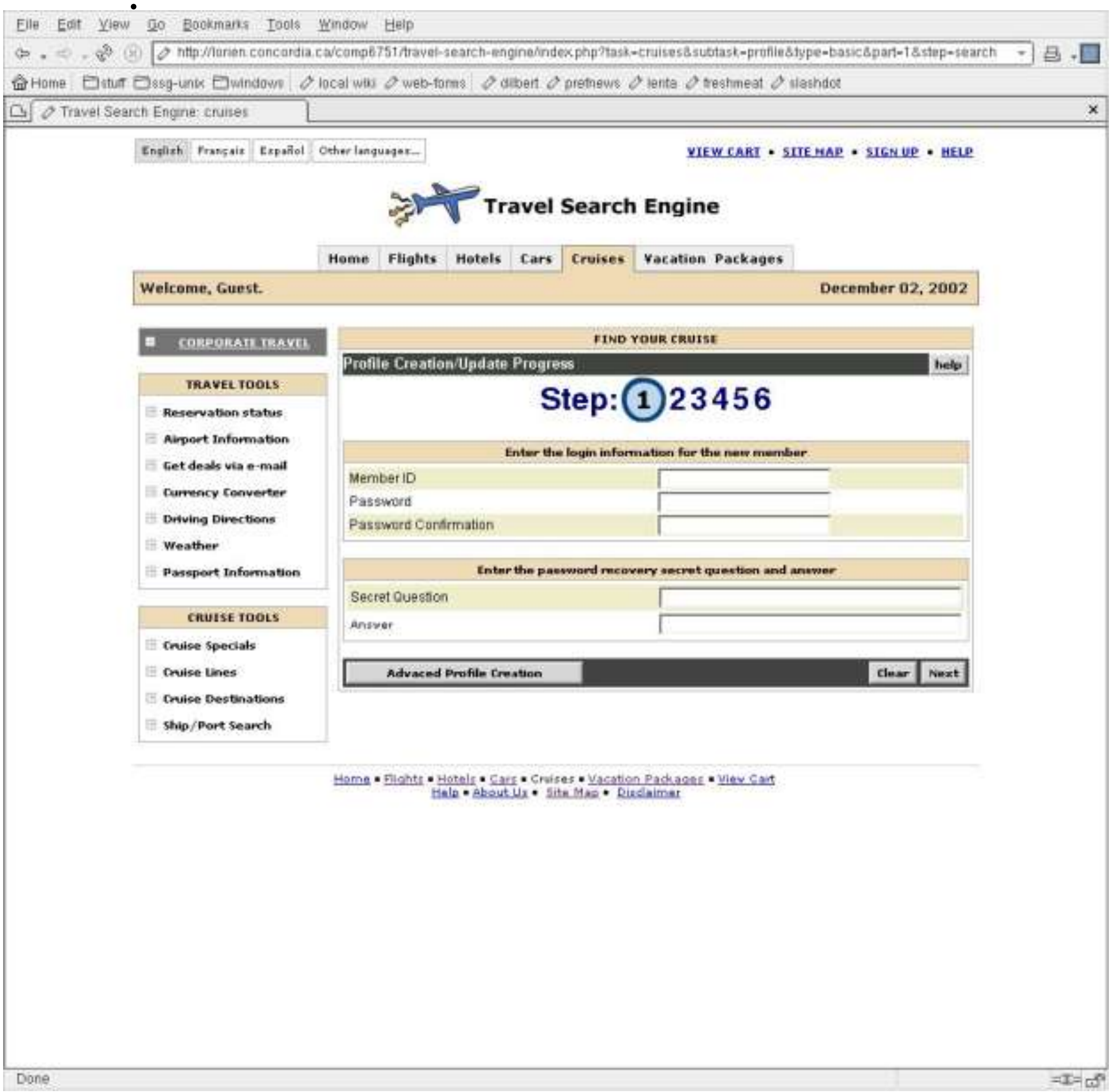

## Reserve a vacation:

- Reserve a golf vacation
- On the north pole
- Departing from Australia
- Departing on Nov. 10, 2002
- Returning on Nov. 19, 2002
- $999.99 per person
- ***** rating
- Two adults, two kids
- Children facility in hotel

- In-door pool in hotel
- $999.99 - $1999.99 price range
- Round-trip air-fair

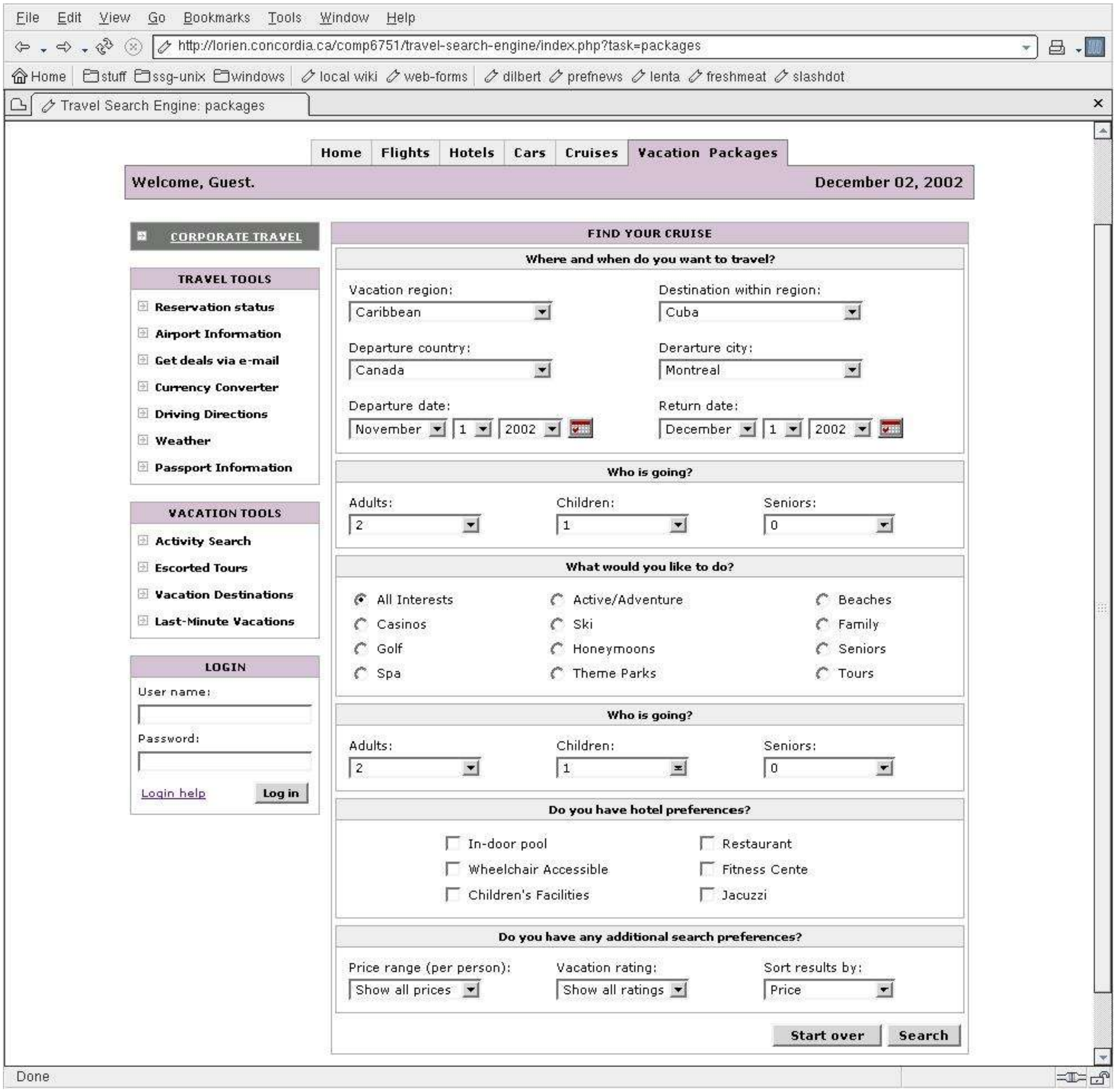

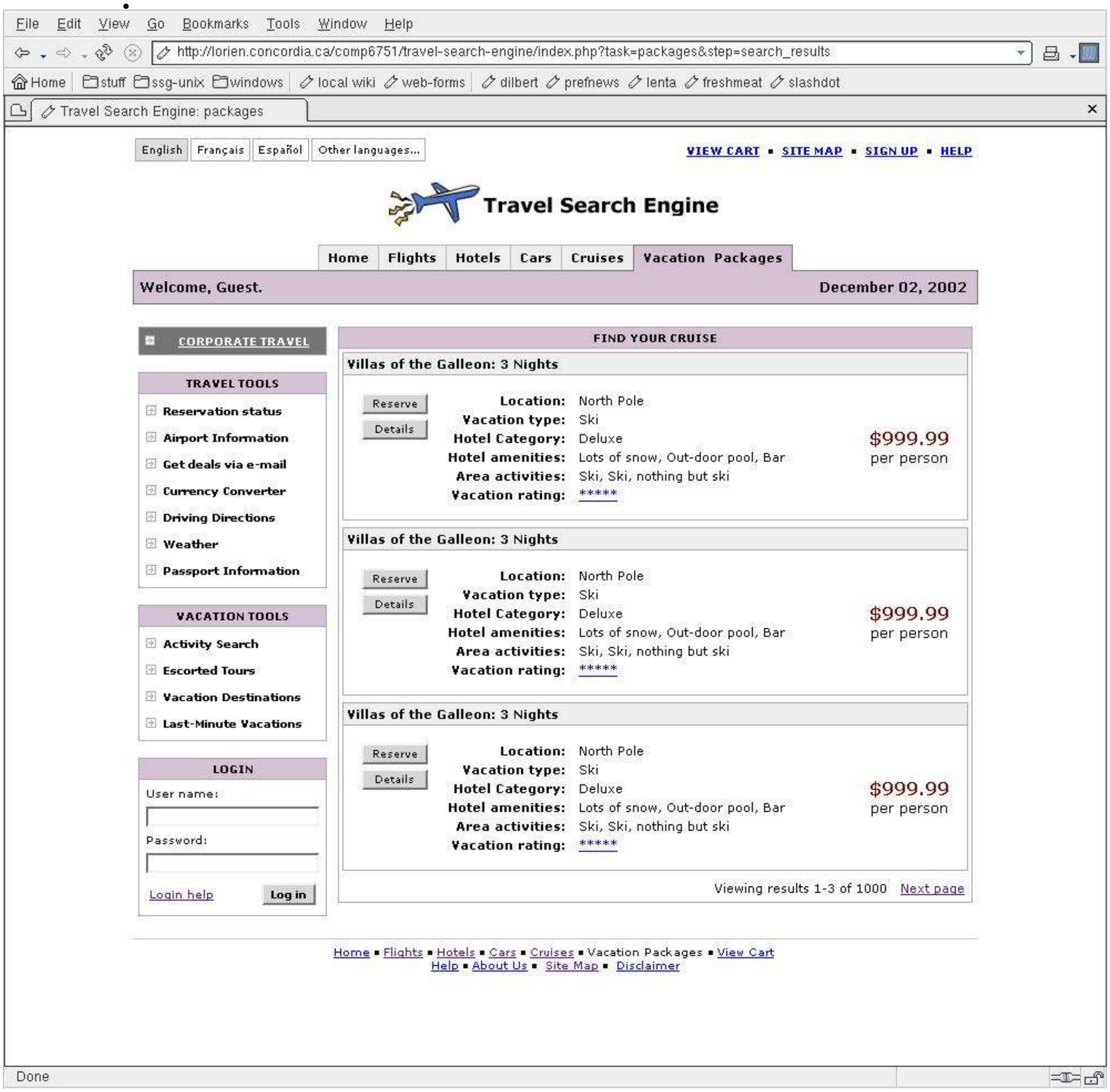
File Edit View Go Bookmarks Tools Window Help
http://lorien.concordia.ca/comp6751/travel-search-engine/index.php?task=packages&step=search_results
Home stuff ssg-unix windows local wiki web-forms dilbert prefnews lenta freshmeat slashdot
Travel Search Engine: packages
English Français Español Other languages...
VIEW CART ▪ SITE MAP ▪ SIGN UP ▪ HELP
Travel Search Engine
Home Flights Hotels Cars Cruises Vacation Packages
Welcome, Guest.
December 02, 2002
CORPORATE TRAVEL
TRAVEL TOOLS
Reservation status
Airport Information
Get deals via e-mail
Currency Converter
Driving Directions
Weather
Passport Information
VACATION TOOLS
Activity Search
Escorted Tours
Vacation Destinations
Last-Minute Vacations
LOGIN
User name:
Password:
Login help
Log in
FIND YOUR CRUISE
Villas of the Galleon: 3 Nights
Reserve
Details
Location: North Pole
Vacation type: Ski
Hotel Category: Deluxe
Hotel amenities: Lots of snow, Out-door pool, Bar
Area activities: Ski, Ski, nothing but ski
Vacation rating: *****
$999.99
per person
Villas of the Galleon: 3 Nights
Reserve
Details
Location: North Pole
Vacation type: Ski
Hotel Category: Deluxe
Hotel amenities: Lots of snow, Out-door pool, Bar
Area activities: Ski, Ski, nothing but ski
Vacation rating: *****
$999.99
per person
Villas of the Galleon: 3 Nights
Reserve
Details
Location: North Pole
Vacation type: Ski
Hotel Category: Deluxe
Hotel amenities: Lots of snow, Out-door pool, Bar
Area activities: Ski, Ski, nothing but ski
Vacation rating: *****
$999.99
per person
Viewing results 1-3 of 1000 Next page
Home ▪ Flights ▪ Hotels ▪ Cars ▪ Cruises ▪ Vacation Packages ▪ View Cart
Help ▪ About Us ▪ Site Map ▪ Disclaimer
Done

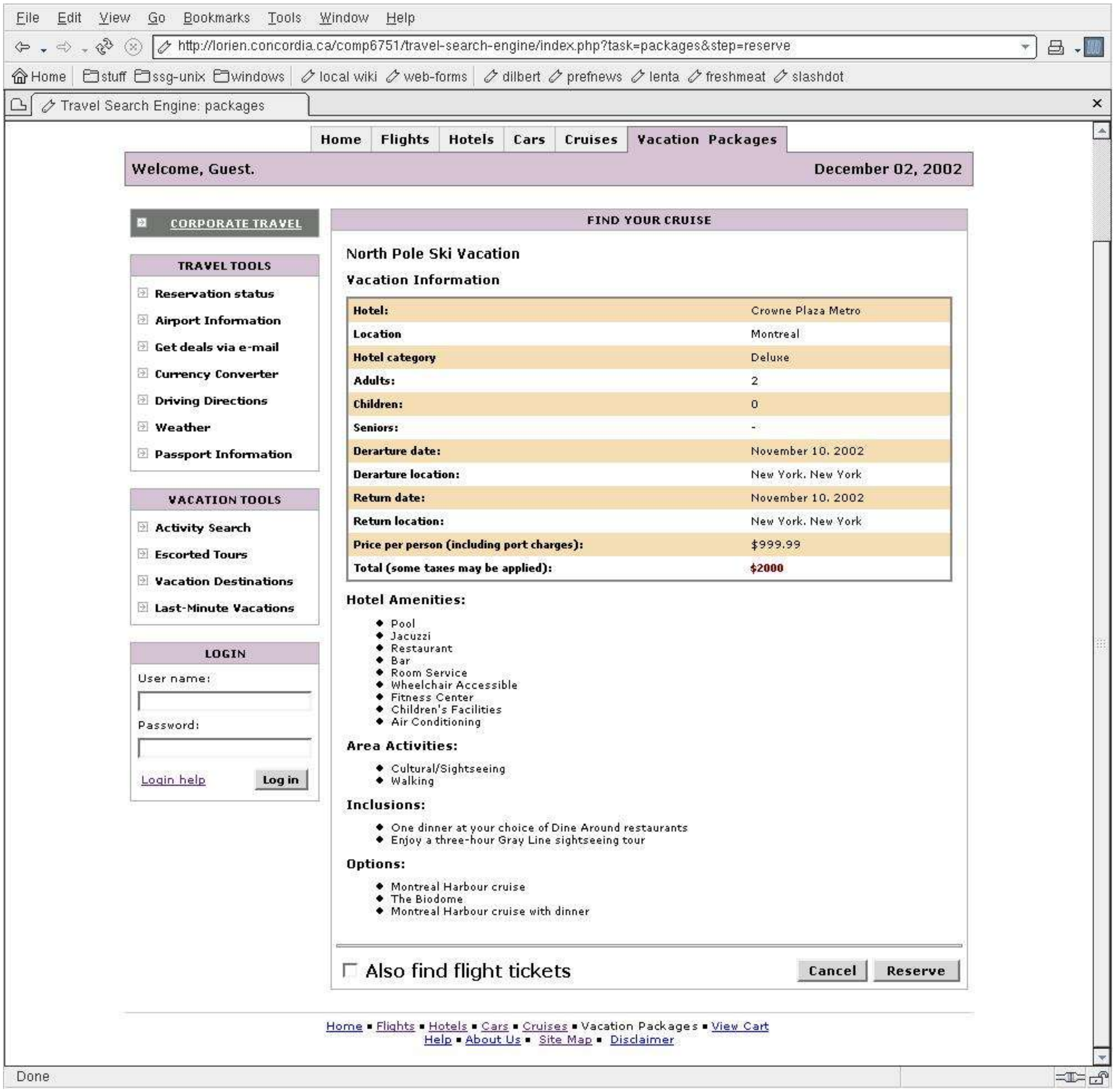
File Edit View Go Bookmarks Tools Window Help
http://lorien.concordia.ca/comp6751/travel-search-engine/index.php?task=packages&step=reserve
Home stuff ssg-unix windows local wiki web-forms dilbert prefnews lenta freshmeat slashdot
Travel Search Engine: packages
Home Flights Hotels Cars Cruises Vacation Packages
Welcome, Guest.
December 02, 2002
CORPORATE TRAVEL
TRAVEL TOOLS
Reservation status
Airport Information
Get deals via e-mail
Currency Converter
Driving Directions
Weather
Passport Information
VACATION TOOLS
Activity Search
Escorted Tours
Vacation Destinations
Last-Minute Vacations
LOGIN
User name:
Password:
Login help
Log in
FIND YOUR CRUISE
North Pole Ski Vacation
Vacation Information
Hotel: Crowne Plaza Metro
Location Montreal
Hotel category Deluxe
Adults: 2
Children: 0
Seniors: -
Derarture date: November 10, 2002
Derarture location: New York, New York
Return date: November 10, 2002
Return location: New York, New York
Price per person (including port charges): $999.99
Total (some taxes may be applied): $2000
Hotel Amenities:
Pool
Jacuzzi
Restaurant
Bar
Room Service
Wheelchair Accessible
Fitness Center
Children's Facilities
Air Conditioning
Area Activities:
Cultural/Sightseeing
Walking
Inclusions:
One dinner at your choice of Dine Around restaurants
Enjoy a three-hour Gray Line sightseeing tour
Options:
Montreal Harbour cruise
The Biodome
Montreal Harbour cruise with dinner
Also find flight tickets
Cancel
Reserve
Home ▪ Flights ▪ Hotels ▪ Cars ▪ Cruises ▪ Vacation Packages ▪ View Cart
Help ▪ About Us ▪ Site Map ▪ Disclaimer
Done

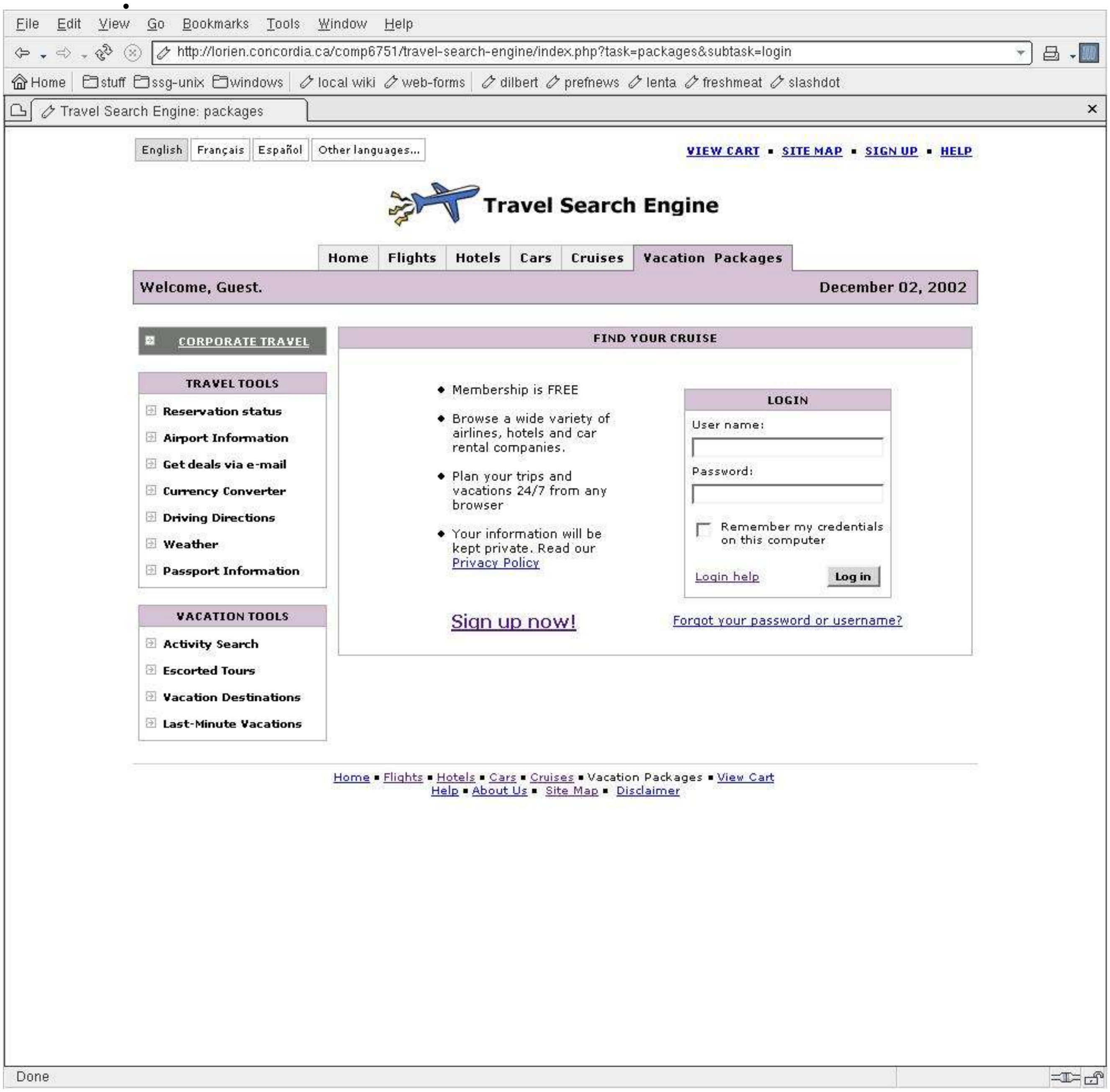
File Edit View Go Bookmarks Tools Window Help
http://lorien.concordia.ca/comp6751/travel-search-engine/index.php?task=packages&subtask=login
Home stuff ssg-unix windows local wiki web-forms dilbert prefnews lenta freshmeat slashdot
Travel Search Engine: packages
English Français Español Other languages...
VIEW CART ▪ SITE MAP ▪ SIGN UP ▪ HELP
Travel Search Engine
Home Flights Hotels Cars Cruises Vacation Packages
Welcome, Guest.
December 02, 2002
CORPORATE TRAVEL
TRAVEL TOOLS
Reservation status
Airport Information
Get deals via e-mail
Currency Converter
Driving Directions
Weather
Passport Information
VACATION TOOLS
Activity Search
Escorted Tours
Vacation Destinations
Last-Minute Vacations
FIND YOUR CRUISE
Membership is FREE
Browse a wide variety of airlines, hotels and car rental companies.
Plan your trips and vacations 24/7 from any browser
Your information will be kept private. Read our Privacy Policy
LOGIN
User name:
Password:
Remember my credentials on this computer
Login help
Log in
Sign up now!
Forgot your password or username?
Home ▪ Flights ▪ Hotels ▪ Cars ▪ Cruises ▪ Vacation Packages ▪ View Cart
Help ▪ About Us ▪ Site Map ▪ Disclaimer
Done

## Shopping Cart

### Browsing the Shopping Cart

At the end of all purchase customer can review his/her items in the shopping cart. Shopping cart's initial view is to provide brief information about the purchases. Customer can delete any item from the cart.

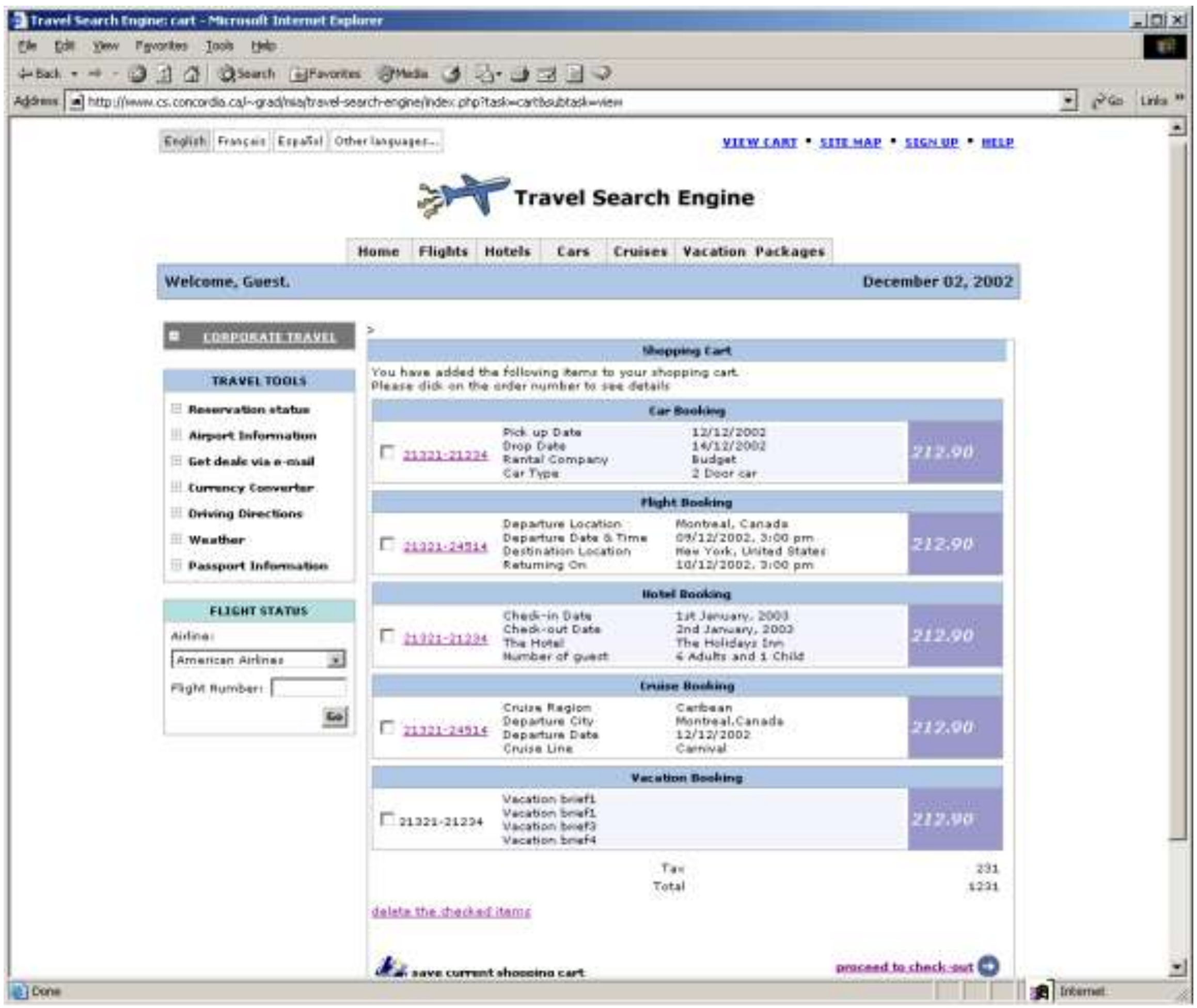

## Detailed Information from the Shopping Cart

From shopping cart customer can review detailed information from the cart for any item.

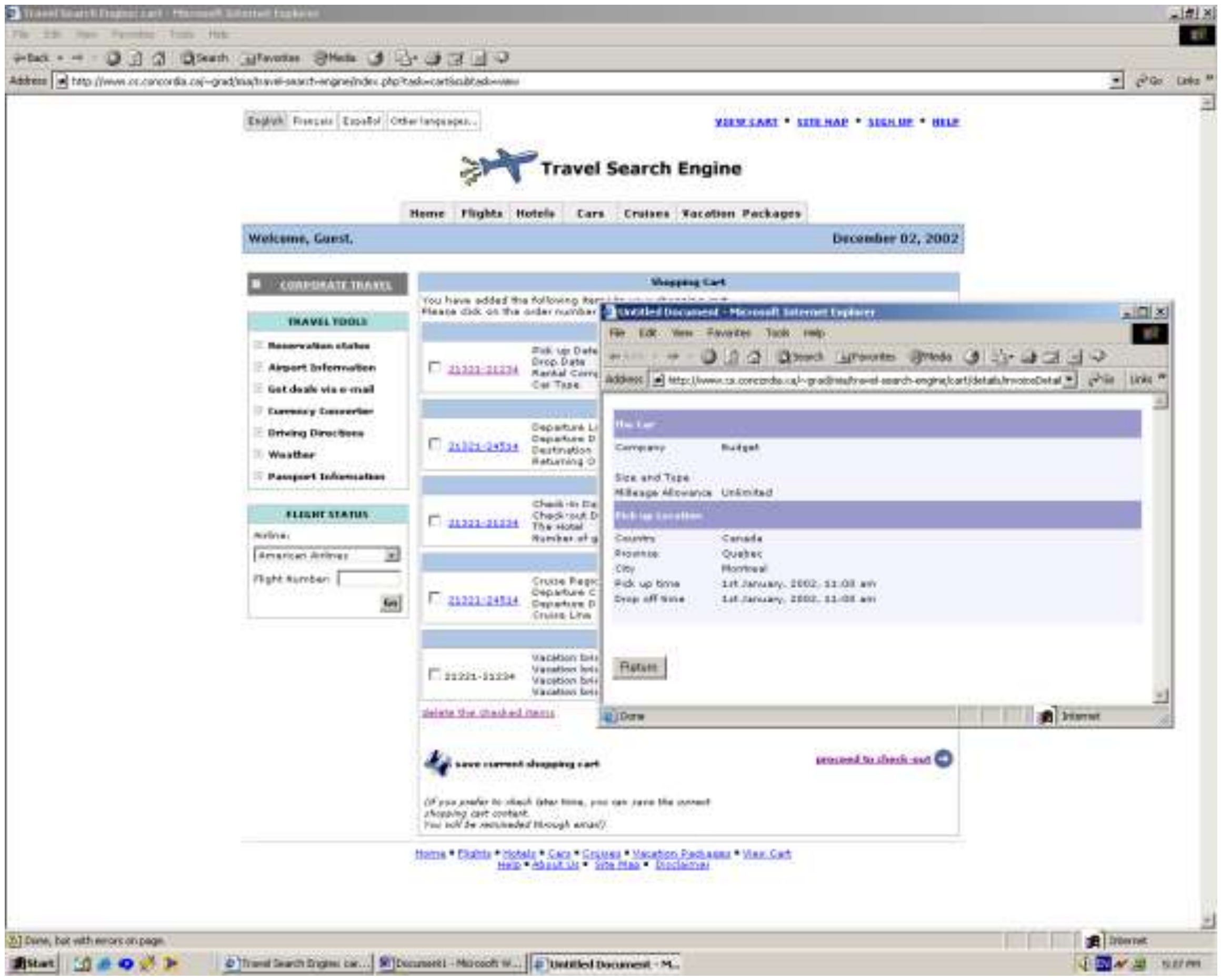

### Specifying Delivery Method

Flight tickets can be delivered three ways: e-ticket, delivery to an address or pick up from an nearby agent office.

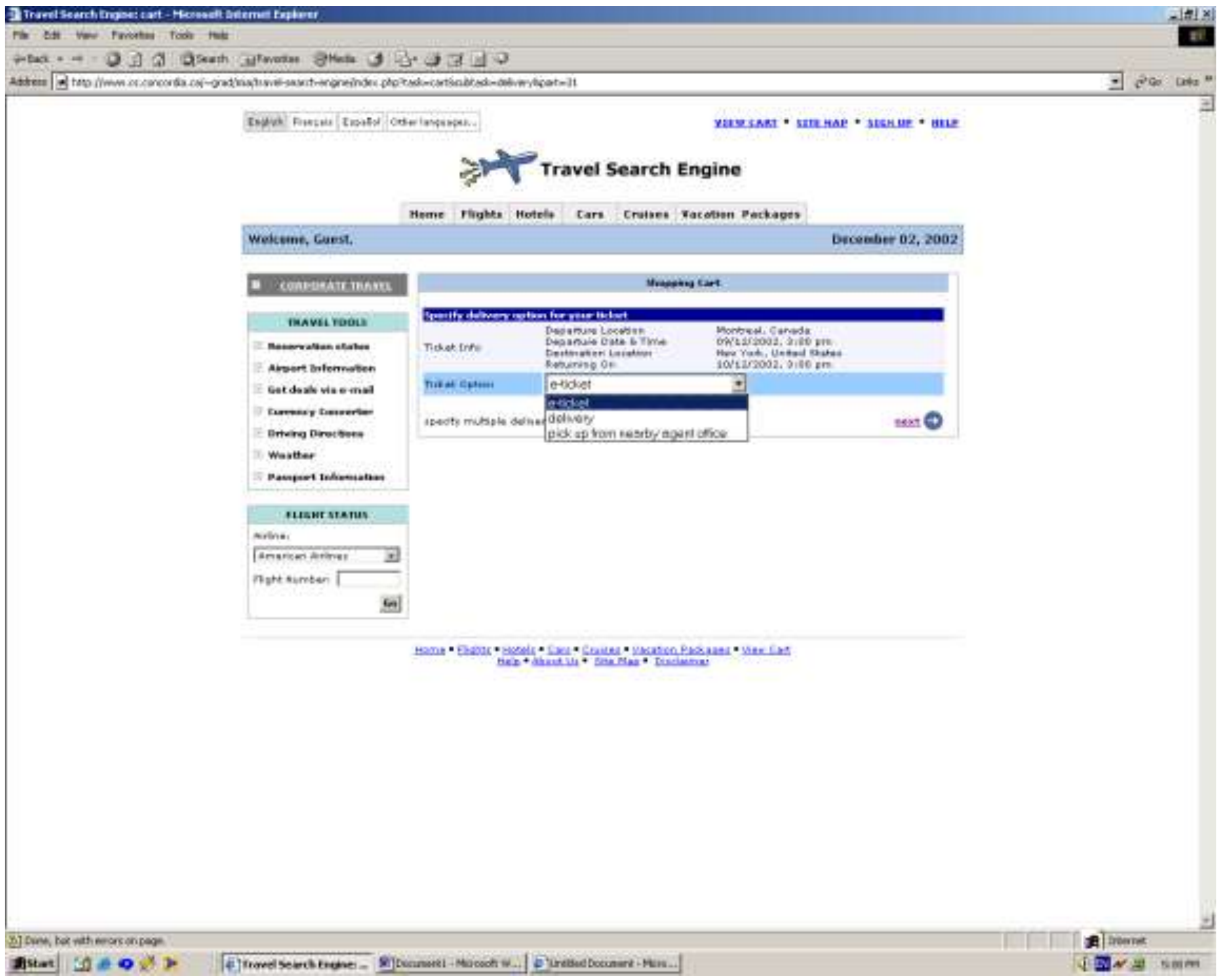

## Modifying Payment and Shipping Option

If desired a customer can have different payment method for payments.

Also, if desired the customer can have the purchased times delivered to different address than from the profile.

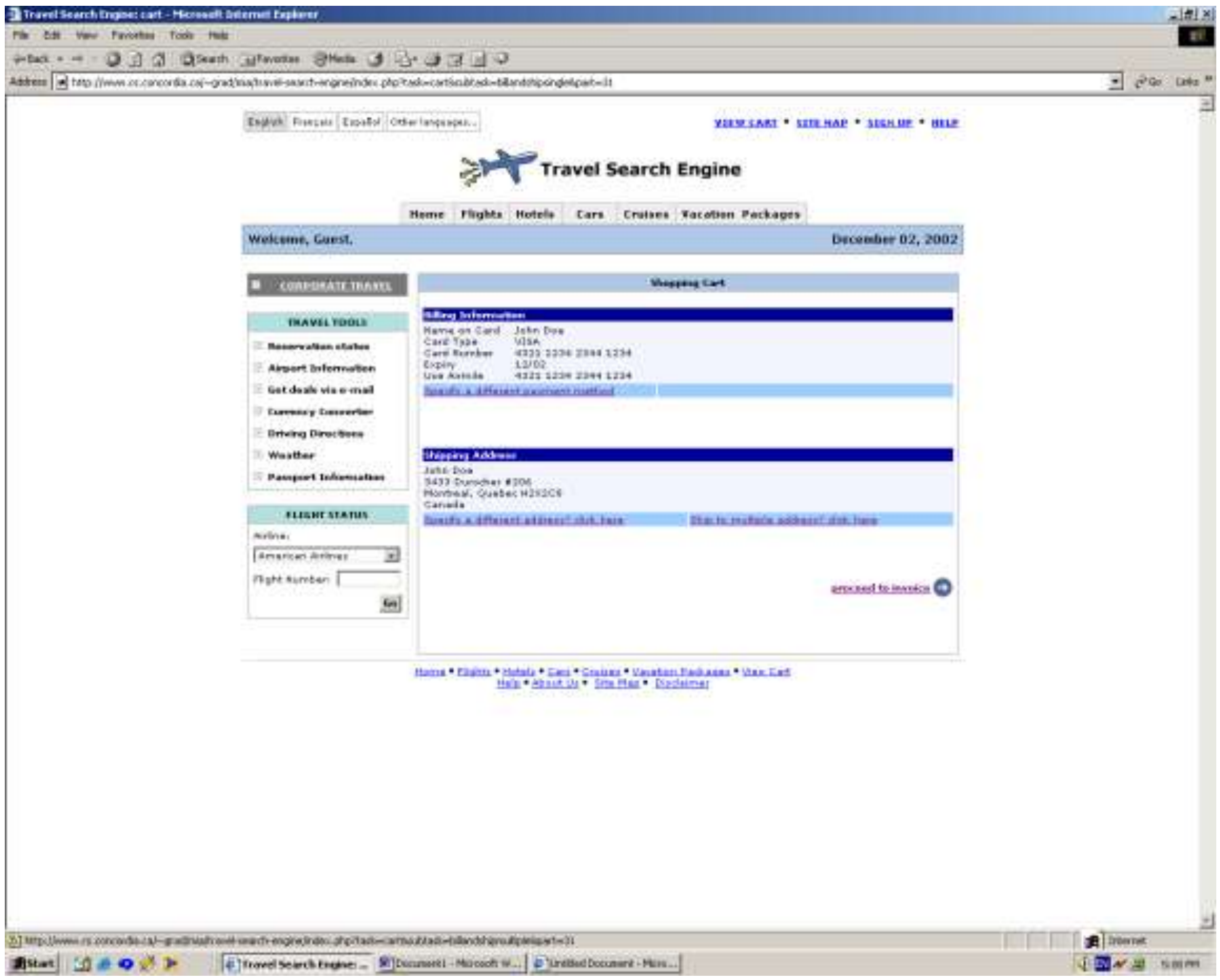

## Modifying Payment and Shipping Options

Different payment option

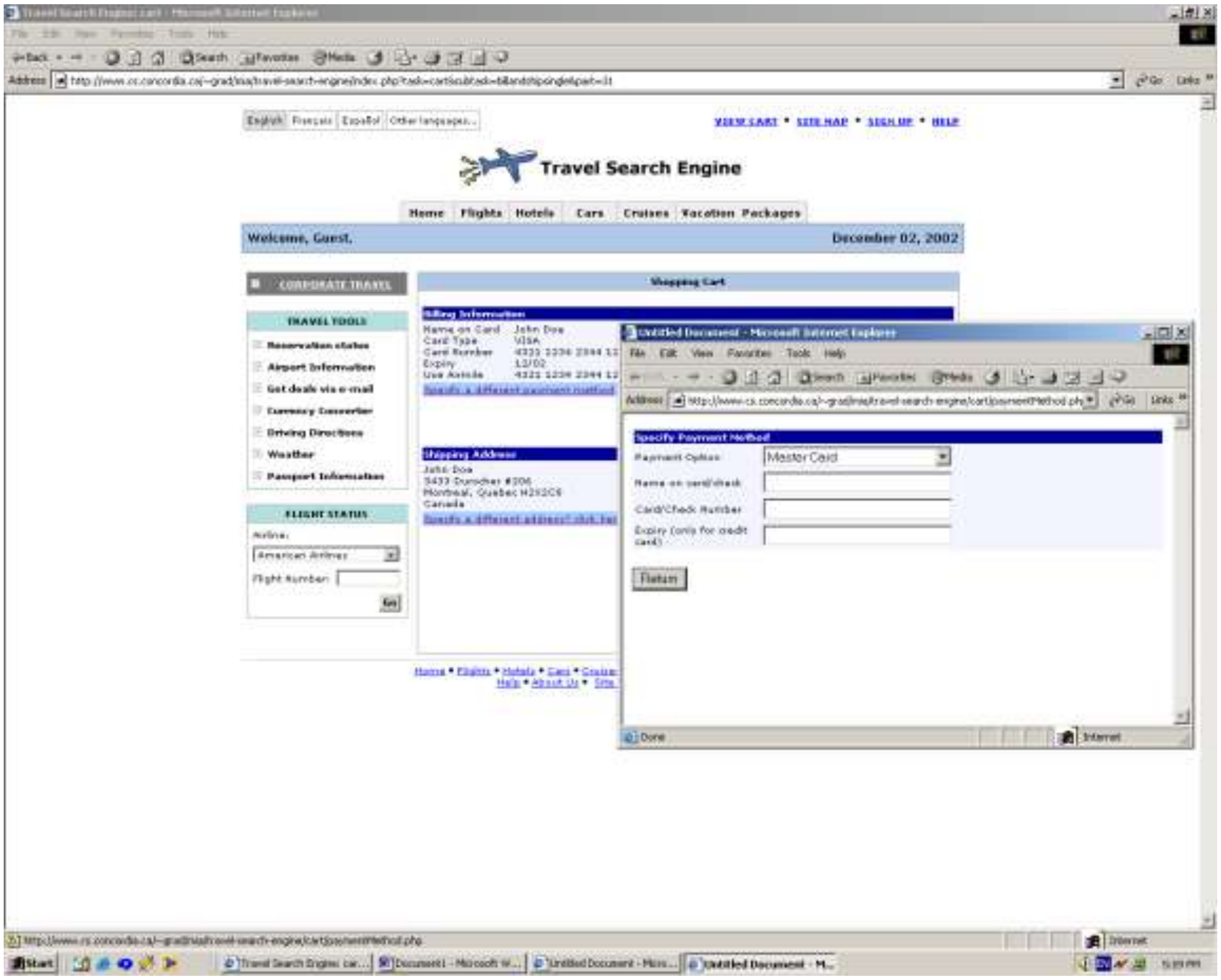

## Modifying Payment and Shipping Options

Different shipping address

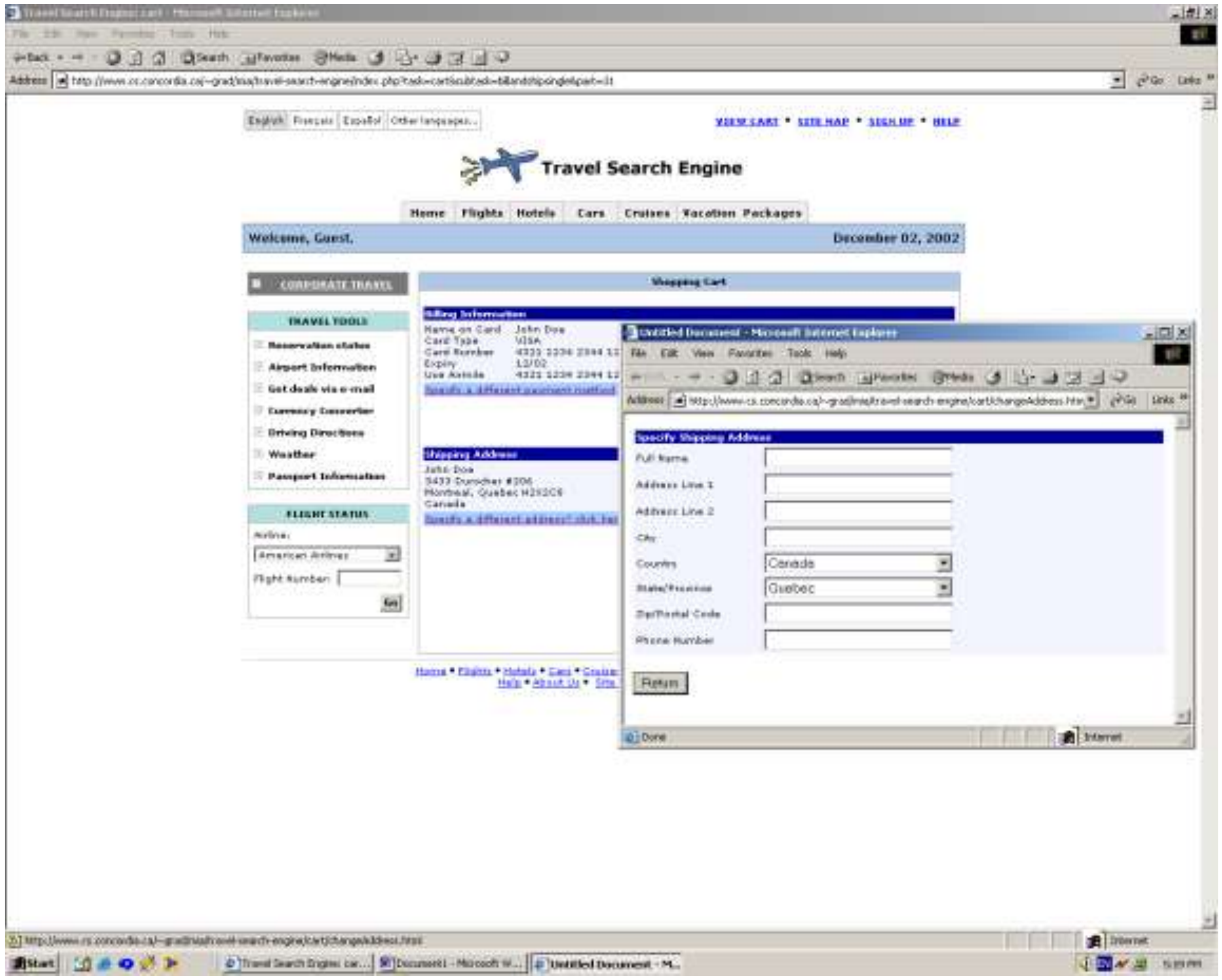

## Modifying Payment and Shipping Options

If desired, the customer can specify different shipping address for different items in the cart.

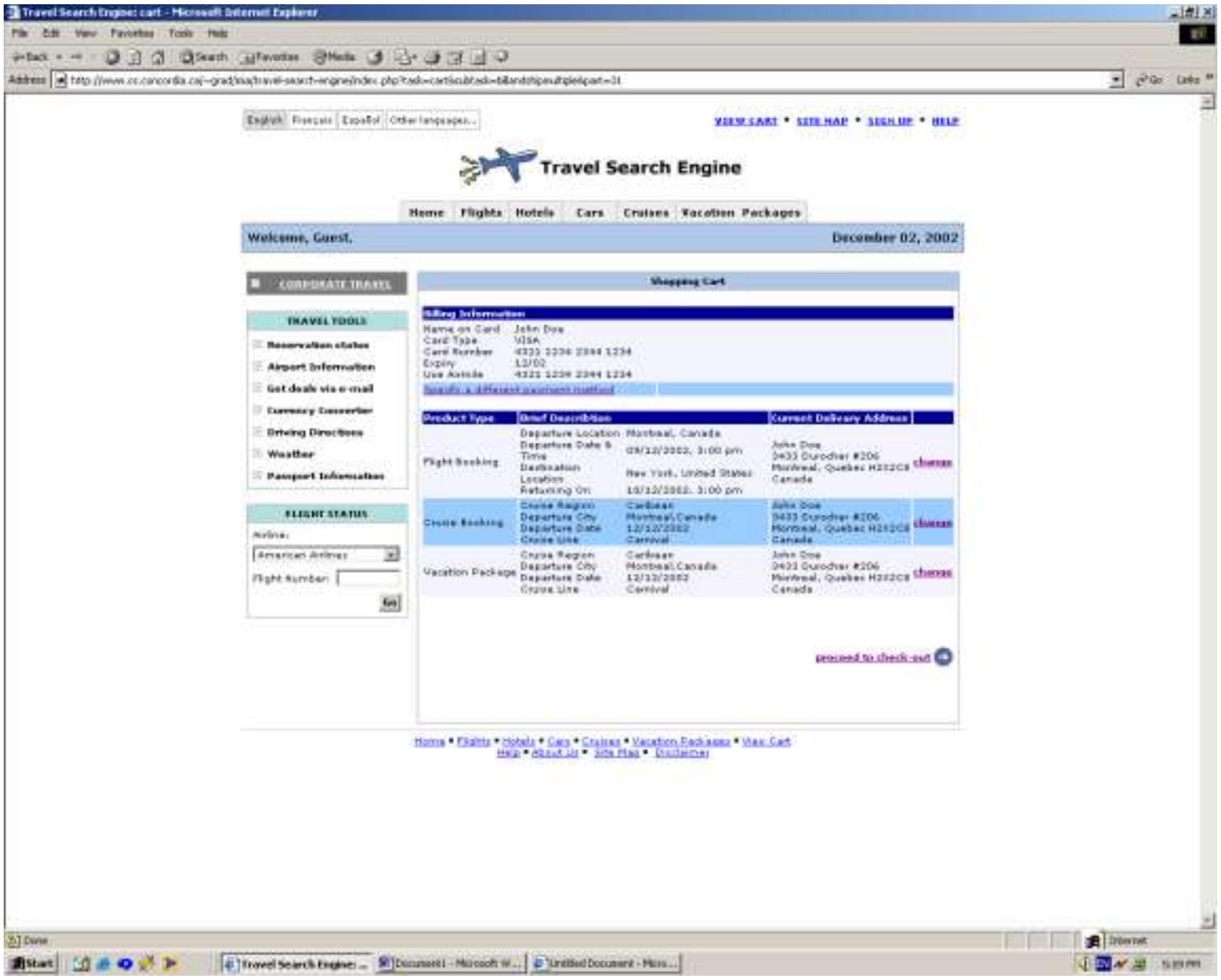

## Modifying Payment and Shipping Options

Different shipping addresses for different items in the cart.

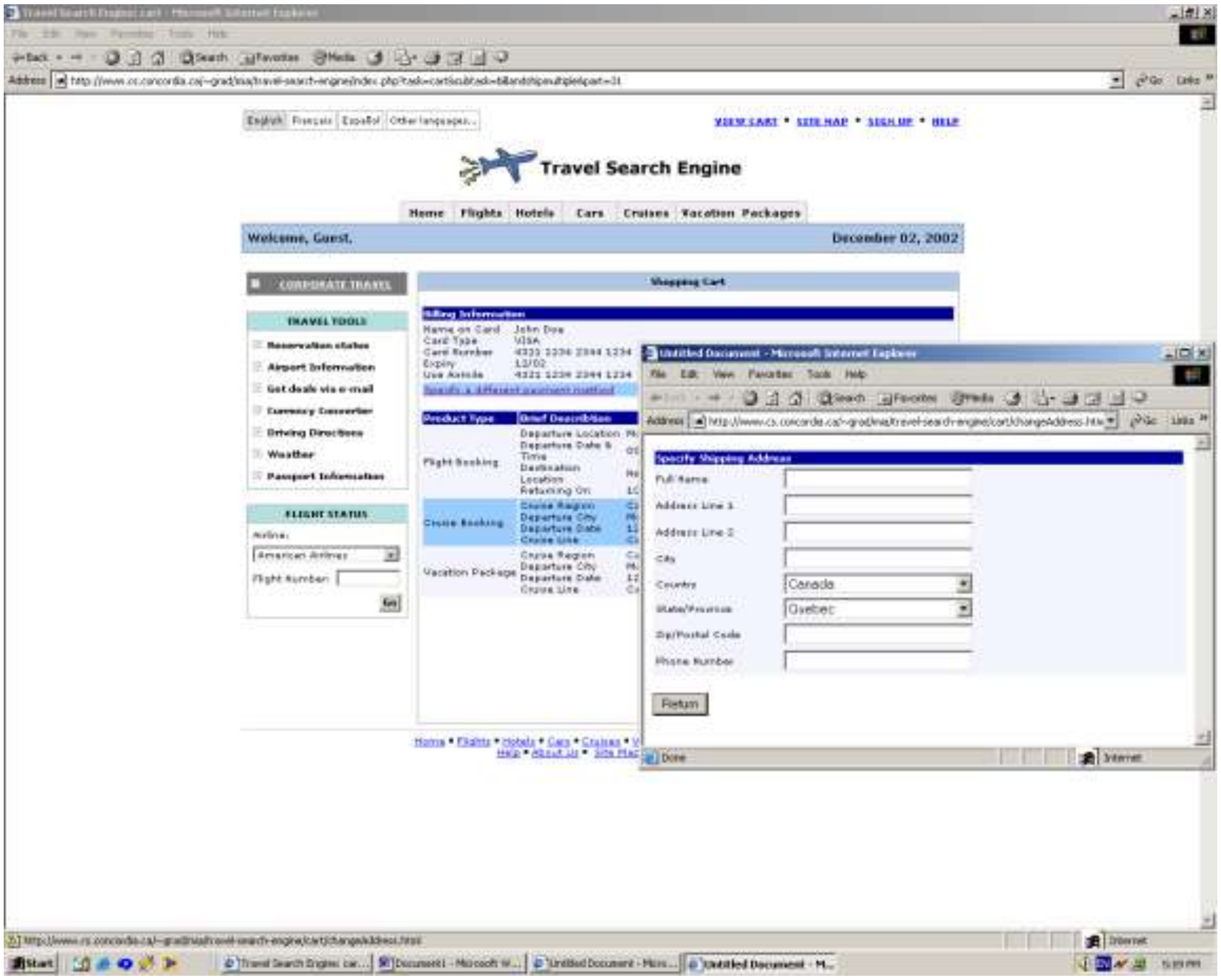

## Invoice

Finally customer can review the invoice.

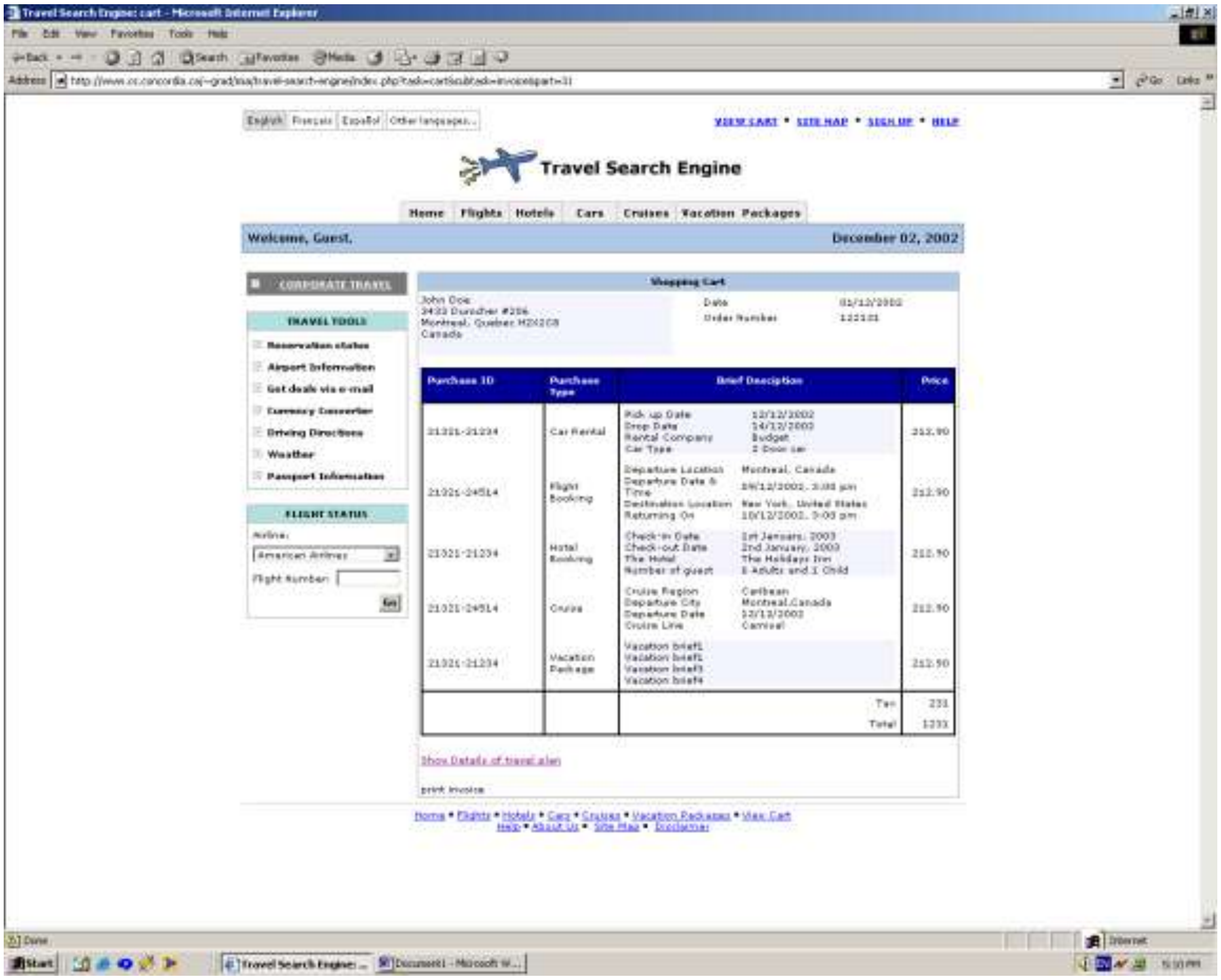

## Detailed Invoice

Customer can also access a detailed version of the invoice

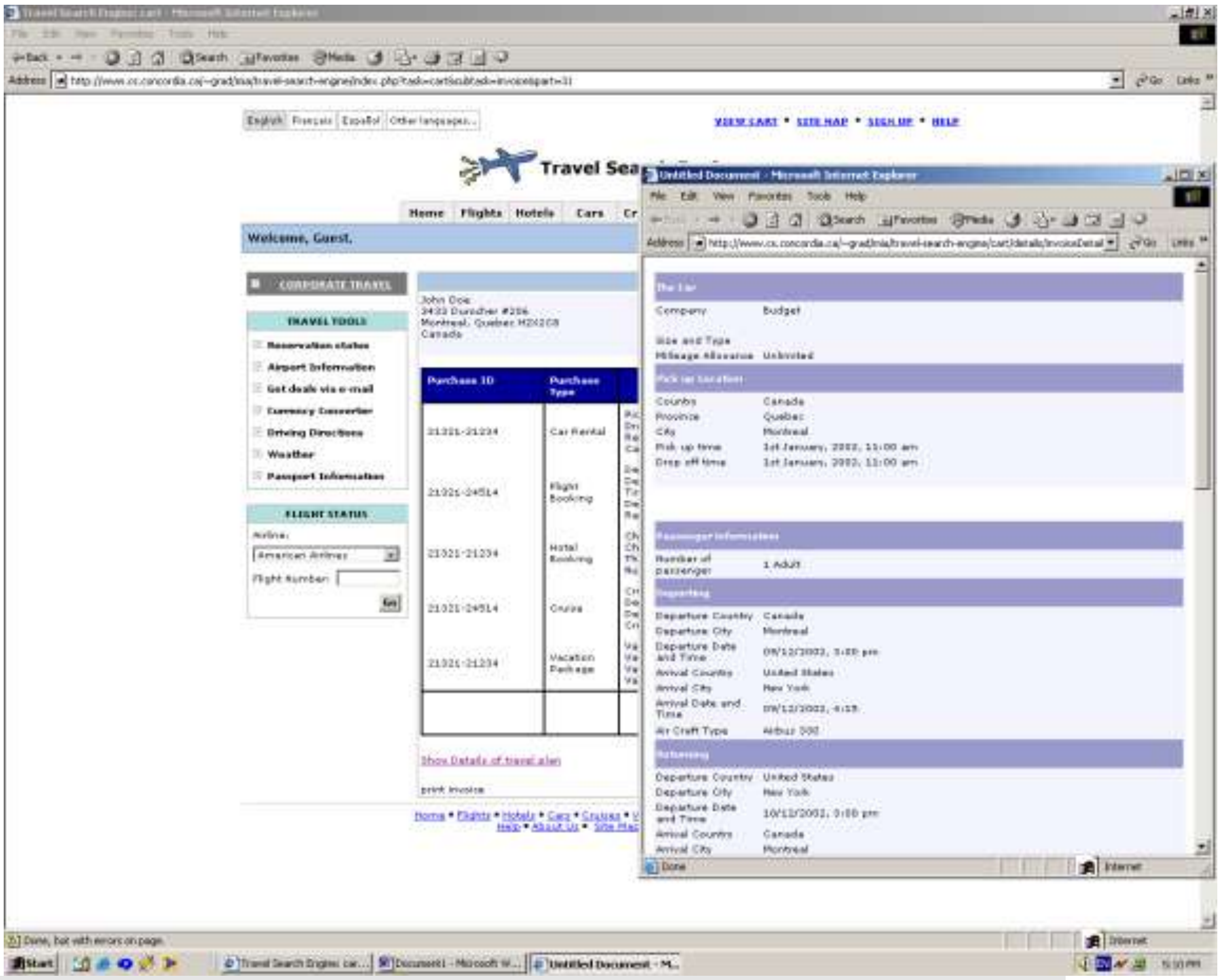

# References

Below is the list of references, web sites used to look up for research and comparative studies. It does include travel search engines as well because we had to design the features they have to add more on top of them.

## Flights

## Hotels

## Cars

## Search Engines

## Course-Related Materials and Others